\documentclass[12pt, a4paper]{article}
\pdfoutput=1
\usepackage[utf8]{inputenc}
\usepackage{fancybox}
\usepackage[T1]{fontenc}
\usepackage{lmodern, textcomp}
\usepackage{bm}
\usepackage[british]{babel}
\usepackage{csquotes}

\usepackage{eurosym}
\usepackage[nottoc]{tocbibind}
\usepackage{authblk}
\usepackage{fancyhdr}
\usepackage{xspace}
\usepackage{soul}

\usepackage{graphicx}
\usepackage{subcaption}
\usepackage{float}
\usepackage{makecell}
\usepackage{multirow}
\usepackage{multicol}
\usepackage{marginnote}
\usepackage[multiple]{footmisc}

\usepackage[usenames, dvipsnames, svgnames, table]{xcolor}

\usepackage[backend=bibtex8,
            style=ieee,
            dashed=false,
            citestyle=numeric-comp,
            maxbibnames=99,
			sorting=none,
            sortcase=false,
            sortcites=true,
            giveninits=true]{biblatex}

\usepackage[normalbf,normalem]{ulem}
\usepackage{amsmath, amsfonts, amssymb}
\usepackage{mathtools}
\usepackage{mathrsfs}
\usepackage{cancel}
\usepackage{braket}
\usepackage{tensor}
\usepackage{slashed}
\usepackage{siunitx}
\usepackage{listings}

\usepackage{stmaryrd}
\usepackage{caption}
\usepackage{relsize,exscale}
\usepackage{array}
\usepackage[toc,page]{appendix}

\usepackage{enumerate}

\DeclareSymbolFont{largesymbols}{OMX}{cmex}{m}{n}

\setcellgapes{1pt}
\makegapedcells
\newcolumntype{R}[1]{>{\raggedleft\arraybackslash }b{#1}}
\newcolumntype{L}[1]{>{\raggedright\arraybackslash }b{#1}}
\newcolumntype{C}[1]{>{\centering\arraybackslash }b{#1}}

\newtheorem{remark}{Remark}

\newcommand{\beq}{\begin{equation}}
\newcommand{\eeq}{\end{equation}}
\newcommand{\bea}{\begin{eqnarray}}
\newcommand{\eea}{\end{eqnarray}}

\definecolor{mygray}{gray}{0.3}

\newcommand{\lnorm}{\lvert \!\vert}
\newcommand{\rnorm}{\vert \! \rvert}

\newcommand{\bes}{\begin{eqnarray}}
\newcommand{\ees}{\end{eqnarray}}

\newcommand\restr[2]{{
  \left.\kern-\nulldelimiterspace 
  #1 
  \vphantom{\big|} 
  \right|_{#2} 
  }}

\usepackage{multicol}
\usepackage{amssymb}

\usepackage[heightrounded, top=3.5cm, bottom=3.5cm, left=2.5cm, right=2.5cm, headheight=14pt]{geometry}

\usepackage[pdftex, breaklinks=true, linktocpage,
	colorlinks=true, urlcolor=blue, linkcolor=blue, citecolor=red
]{hyperref}

\newcommand{\email}[1]{\href{mailto:#1}{\nolinkurl{#1}}}
\newcommand{\emailfoot}[1]{\thanks{\email{#1}}}

\usepackage{marginnote}
\newcounter{draftcommentcnt}
\NewDocumentCommand{\draftcomment}{s O{red} m}{%
	\def\margnote{\IfBooleanTF{#1}{\marginnote}{\marginpar}}%
	\stepcounter{draftcommentcnt}%
	\textcolor{#2}{#3}%
	\margnote{\textcolor{#2}{$\Leftarrow$ \arabic{draftcommentcnt}}}%
}

\fancypagestyle{preprint}{
	\fancyhf{}

	\fancyhead[R]{\textsf{\preprint}}
}

\numberwithin{equation}{section}
\usepackage{biblatex}
\hypersetup{
	pdftitle={Stochastic disordered random tensors, equilibrium renormalization group and beyond},
}
\title{Large time effective kinetics $\beta$-functions for quantum $(2 + p)$-spin glass}

\author[1]{Vincent Lahoche\emailfoot{vincent.lahoche@cea.fr}}
\author[1,2]{Dine Ousmane Samary\emailfoot{dine.ousmanesamary@cipma.uac.bj}}
\author[1]{Parham Radpay\emailfoot{parham.radpay@cea.fr}}

\affil[1]{%
	Université Paris-Saclay, \textsc{Cea}, Palaiseau, F-91120, France
}

\affil[2]{%
	Faculté des Sciences et Techniques (ICMPA-UNESCO Chair)
	\protect\\
	Université d'Abomey-Calavi, 072 BP 50, Benin
}

\begin{document}

\maketitle
\hrule

\hrule
\begin{abstract}
This paper examines the quantum $(2+p)$-spin dynamics of an $N$-vector $\textbf{x} \in \mathbb{R}^N$ through the lens of renormalization group (RG) theory. The RG approach is based on coarse-graining the eigenvalues of a matrix-like disorder, interpreted as an effective kinetic term whose eigenvalue distribution follows a deterministic law in the large $N$ limit. We focus our investigation on perturbation theory and vertex expansion for the effective average action, which proves more tractable than standard nonperturbative approaches due to the unique non-local temporal and replicative structures that arise in the effective interactions after disorder integration. Our work involves formulating rules to handle these non-localities within the perturbative framework, culminating in the derivation of one-loop $\beta$-functions. Our explicit calculations focus on the cases $p=3$ and $p=\infty$, with additional analytical material provided in the appendix.
\end{abstract}
\hrule
\hrule
\bigskip

\noindent \textbf{keywords:} Functional renormalization group, Stochastic field theory, random matrix theory, non-local field theory.

\setcounter{footnote}{0}
\newpage

\hrule
\pdfbookmark[1]{\contentsname}{toc}
\tableofcontents
\bigskip
\hrule

\clearpage


\section{Introduction}

Understanding the physics of glassy systems, characterized by the complexity of their energy landscape, remains an active research domain almost 60 years after their formal introduction in theoretical physics, mathematics, and data science \cite{Parisi,MezardInformation,Mehta2004,talagrand2010mean1,talagrand2010mean2,talagrand2003spin,ghio2023sampling}. Similar to the Ising model for ferromagnets, spin glass models emerge as an idealization of the behavior of certain alloys at low temperatures. In these models, the role of randomly distributed impurities in the crystal lattice significantly modifies the energy landscape, introducing an exponentially large number of metastable states in which the system remains trapped for extended periods \cite{cugliandolo2002dynamics,nishimori2001statistical,mezard1984nature}. Computationally, this implies that the typical problem of finding the ground state of the landscape becomes an NP-hard problem \cite{fan2023searching}. The slow dynamics of the spin glass phase are further characterized by non-equilibrium effects at low temperatures \cite{Dominicis,cugliandolo1995full,bouchaud1996mode,cugliandolo1993analytical}, with some explicitly solvable models, such as the spherical Sherrington-Kirkpatrick model, exhibiting this phenomenon. The origin of this complex energy landscape can be traced back to the phenomenon of frustration, which arises because random impurities introduce effective couplings between spins that cannot be satisfied simultaneously along a given closed loop (i.e., the product of the couplings along the loop is negative). Furthermore, the disorder is \textit{quenched}, meaning that the typical time evolution for the "spin" is much larger than the time evolution of the disorder (impurity configuration), leading thermodynamic quantities like free energy to \textit{self-average} in the large $N$ limit.

\medskip

\begin{figure} \begin{center} \includegraphics[scale=1]{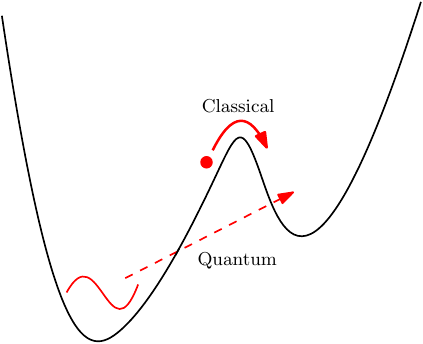} \end{center} \caption{Quantum tunneling versus thermal annealing.}\label{fig1} \end{figure}

Standard examples of spin glasses can be approached using classical tools, primarily the mean field techniques of statistical physics \cite{Parisi,Dominicis}. This is justifiable, as in ferromagnetic theory, because the transition temperature and the corresponding typical energy scale are far from the typical energy scale of quantum fluctuations. Therefore, quantum transition physics typically operates at zero temperature, where quantum fluctuations can disrupt long-range order \cite{vojta2003quantum}. In a spin glass system, it is experimentally possible to adjust external parameters such that the glassy transition temperature becomes comparable to the typical quantum energy fluctuations. This interplay can significantly influence the dynamics through the complex energy landscape of the glassy state. These effects are particularly relevant for the so-called quantum annealing (as opposed to thermal annealing), which involves the ability of quantum particles to escape from metastable states via quantum tunneling \cite{bapst2013quantum} (see Fig. \ref{fig1}).

\medskip

\begin{figure}
\begin{center}
\includegraphics[scale=1]{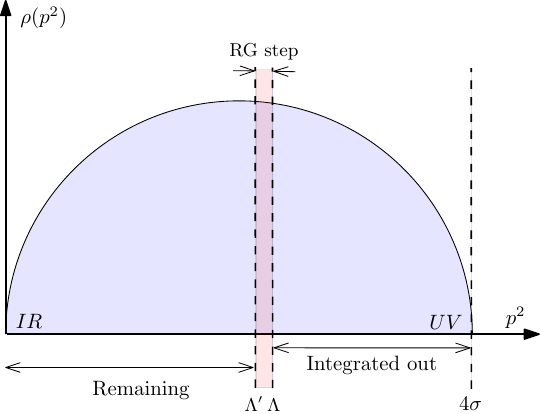}
\end{center}
\caption{Partial integration over Wigner modes.}\label{fig2}
\end{figure}

There are many approaches to describing a quantum spin glass system \cite{baldwin2017clustering,cugliandolo2004effects,chowdhury2022sachdev,rosenhaus2019introduction,biroli2001quantum,rokni2004dynamical}. In this paper, we consider an $N$-dimensional quantum particle moving through a random energy landscape realized by two real Gaussian tensors $(K,J)$ with ranks 2 and $p$, respectively, a quantum model inspired by \cite{cugliandolo2001imaginary,biroli2001quantum}. Typically, renormalization group (RG) techniques are not prominently featured among the preferred tools in the literature on this topic. Here, we propose to study the problem by adopting the effective kinetics arising from the rank 2 tensor, whose eigenvalues are distributed according to Wigner's law as $N \to \infty$. These eigenvalues thus behave like effective moments, and an RG can be constructed by partially integrating over the corresponding modes (Fig. \ref{fig2}) \cite{Bradde,lahoche20241,lahoche20242,lahoche20243,lahoche20244,lahoche4,lahoche3}. We first focus on the case $p=3$ and then consider the limit as $p \to \infty$. Our goal for this initial work is to introduce the general theoretical framework and explore simple approximations in various limits. This will allow us to assess the feasibility of a program based on renormalization and identify potential areas for future research. In the long term, our objective is to uncover various criteria based on RG that will enable us to characterize the different phases of a quenched glassy quantum system. This research will have applications in situations where standard analytical methods become less effective, such as in the quantum detection of signals. This work is part of the bibliographical thread \cite{lahoche2022generalized,lahoche4,Lahoche2022functional,Lahoche2023functional,lahoche20241,lahoche20242,lahoche20243,lahoche20244,Finotello1}, aiming to consider RG as a powerful tool for investigating open problems in signal detection.
\medskip

The manuscript is organized as follows: In the section \eqref{sec2} we define the quantum model by its Hamiltonian, in which the potential is an operator that contains all the information about the disorder. We also construct a path integral approach adapted to our study, particularly in Fourier modes. Averaging around the disorder and in the large N approximation, we use the replica method to compute the partition function and then the classical action.
In the section \eqref{sectionPert} we introduce the functional renormalization group through the Wetterich equation. It also has the advantage of regularizing the divergences that can appear in the infrared regime. In section \eqref{secWard} we study the model's symmetry through the Ward identities and discuss their implications in the functional renormalization group. In section \eqref{sectionscaling} we discuss the scaling and dimensions of the couplings regarding the specific nature of the Wigner spectrum. In section \eqref{sec6} the large $N$ $\beta$-functions are computed using perturbation theory and investigated numerically. In the section \eqref{sec7} we provide the same analysis non-perturbatively using the vertex expansion at the leading order of the derivative expansion. Finally, in section \eqref{largeN} we study the large $N$-limit from the closed equation of the self-energy in the deep infrared regime, and we consider a first look for the $2PI$ formalism. This last point shows that singularities seem to blind a first-order phase transition. 
We conclude our work in section \eqref{conclusion}. We provide four appendices to clarify the results throughout the manuscript.

\section{The model}\label{sec2} 

\paragraph{Definition of the model.} We consider a quantum particle over $\mathbb{R}^N$, whose  wave function $\Psi(\textbf{x},t)$, satisfies the Schr\"odinger equation:
\begin{equation}
i \hbar \frac{\partial }{\partial t}\Psi(\textbf{x},t)= \hat{H}_{\text{SG}}\, \Psi(\textbf{x},t)\,,\label{schro}
\end{equation}
where $\textbf{x}\in \mathbb{R}^N$  is the classical position of the particle and the Hamiltonian $\hat{H}_{\text{SG}}$ is:
\begin{equation}
 \hat{H}_{\text{SG}}= \frac{\hat{\textbf{p}}^2}{2m_0}+U_{J,K}(\hat{\textbf{x}})+V(\hat{\textbf{x}}^2)\,.\label{Hamiltonian Model}
\end{equation}
Note that in this Hamiltonian, we denoted momentum and position components operators with a hat, to distinguish them from their classical counterparts. In particular, they satisfy the quantum commutation relations:
\begin{equation}
[\hat{p}_k,\hat{x}_j]=i \hbar \delta_{kj}\,, \quad [\hat{p}_k,\hat{p}_j]=[\hat{x}_k,\hat{x}_j]=0\,,\forall (j,k)\,.
\end{equation}
The potential $U_{J,K}(\hat{\textbf{x}})$ is the disorder term, which is given explicitly by:
\begin{align}
U_{J,K}&:=\frac{1}{2}\sum_{i,j=1}^N K_{ij}\hat{x}_{i}\hat{x}_{j}+ \sum_{1\le i_1<\cdots <i_p\le N}\, J_{i_1\cdots i_p} \, \hat{x}_{i_1} \cdots  \hat{x}_{i_p}\,,\label{defUJK}
\end{align}
where both $J$ and $K$ have random Gaussian entries, with zero mean and variance defined as follows:
\begin{align}
\overline{J_{i_1\cdots i_p}J_{i_1^\prime \cdots i_p^\prime}}&=\left(\frac{\lambda p!}{N^{p-1}}\right) \,\prod_{\ell=1}^p \delta_{i_\ell i_\ell^\prime}\,,\label{averageJ}\\
\mathbb{E}[{K_{i_1i_2}K_{i_1'i_2'}}]&=\frac{\sigma^2}{N}(\delta_{i_1i_1'}\delta_{i_2i_2'}+\delta_{i_1i_2'}\delta_{i_2i_1'})\,.\label{averageK}
\end{align}
Note that throughout the rest of this paper, we keep the notation $\overline{X}$ (with overbar) for averaging over the disorder tensor ${J}$ only. Moreover, the absence of a factor $2$ in \eqref{averageK} is due to the presence of a factor of $1/2$ in \eqref{defUJK}, under the assumption that the matrix $K$ is \textit{symmetric}. Note that with the definition \eqref{averageK}, the probability distribution for $K$ is $O(N)$ invariant\footnote{$K$ is a random matrix from the Gaussian orthogonal ensemble (GOE).}. This will be relevant later for non-trivial Ward identities arising because of the Gauge fixing (see equation \eqref{Gauge} and section \ref{secWard}). The potential $V(\textbf{x}^2)$, depending only on the square length $\textbf{x}^2$ of the classical coordinate, suppresses large $\textbf{x}$ configurations in the classical equation of motion:
\begin{equation}
m_0 \frac{d^2 x_i}{dt^2}= -\frac{\partial}{\partial x_i} \left(U_{J,K}({\textbf{x}})+V({\textbf{x}}^2)\right)\,.
\end{equation}
As we will see below, in the symmetric phase, the mass of the quantum particle $m_0$ does not renormalize, and we will set $m_0=1$ for simplicity. Furthermore, according to \cite{van2010second}, we consider a quartic potential:
\begin{equation}
V({\textbf{x}}^2)=\frac{1}{2} m^2 {\textbf{x}}^2+\frac{\mu_2}{4N}\, ({\textbf{x}}^2)^2\,,
\end{equation}
such that for $m^2<0$ and $\mu_2>0$, there are infinitely many degenerate stable minima such that:
\begin{equation}
{\textbf{x}}^2_0(t)=-\frac{m^2}{\mu_2}\,N\,,
\end{equation}
implementing a dynamic constraint resembling a spherical constraint. Alternatively, one could impose the constraint dynamically using the theory of Lagrange multipliers. However, our approach is better suited for investigation using the Renormalization Group (RG).  In this paper, we do not rely on Schrödinger formalism \eqref{schro} but rather on Feynman's path integral approach. This framework is based on the path integral definition of the partition function $\mathcal{Z}[K,J,\textbf{L}]$:
\begin{equation}
\mathcal{Z}[K,J,\textbf{L}]:=\int \, [d x]\, e^{\frac{i}{\hbar}S_{\text{cl}}[\textbf{x}]+i\int_{-\infty}^\infty\,dt\,\sum_{k=1}^N L_k(t)x_k(t)}\,,\label{pathintegralZ}
\end{equation}
where $S_{\text{cl}}$ is classical action:
\begin{equation}
S_{\text{cl}}[\textbf{x}]:=\int_{-\infty}^{+\infty} dt \left(\frac{1}{2}\dot{\textbf{x}}^2-U_{J,K}(\hat{\textbf{x}})-V(\hat{\textbf{x}}^2)\right)\,.
\end{equation}
Note that we use the notation $[d x]$ for the path integral ``measure'', to distinguish it from the standard Lebesgue measure $d x$. As is well known in quantum field theory (QFT), the path integral \eqref{pathintegralZ} allows computing vacuum-vacuum expectation value of fields correlations at different times \cite{ZinnJustinBook2}:
\begin{align}
 \langle \Omega \vert \hat{x}_{i_1}(t_1)\hat{x}_{i_2}(t_2)\cdots \hat{x}_{i_n}(t_n) \vert \Omega \rangle = \frac{(-i)^{n}}{\mathcal{Z}[K,J,0]} \frac{\partial^n \mathcal{Z}[K,J,\textbf{L}]}{\partial L_{i_1}(t_1)\partial L_{i_2}(t_2)\cdots \partial L_{i_n}(t_n)}\,\Bigg\vert_{J=0}\,.
\end{align}
As usual in QFT, the loop expansion of the path integral could be structured as a power series in $\hbar$, and the limit $\hbar \to 0$ corresponds to the \textit{classical limit}. 
\medskip

\paragraph{Large $N$ Gaussian theory.} As previously indicated, we will not treat the two disorders on the same footing. Unlike the disorder tensor $J$, which will be integrated out, in the case of the tensor $K$ we will take advantage of \textit{Wigner semicircle theorem} \cite{Mehta2004}. 
To this aim, we first note that the effective non-interacting, bare propagator can be expressed as the resolvent of the random matrix $K$. In fact, in the Fourier space we have,
\begin{eqnarray}
 S_{\text{cl,kin}}[\textbf{x}]&&=\frac12\int_{-\infty}^{\infty} d\omega (x(-\omega)\left[(\omega^2-m^2)I-K\right]x(\omega))\cr
&&=\frac{1}{2}\,\int_{-\infty}^{+\infty} d\omega \sum_{\mu=1}^N\, x'_{\mu}(-\omega)\left(\omega^2-\lambda_\mu -m^2\right)x'_{\mu}(\omega),\label{kineticeff}
\end{eqnarray}
where
\begin{equation}
    x'_{\mu}(t):=(x(t),v_\mu),\label{Gauge}
\end{equation}
$\{v_\mu\}$ denote the eigenvectors of $K$ associated to the eigenvalues $\{\lambda_\mu\}$ and $(x(t),v_\mu)$ is the scalar product of $x(t)$ with $v_\mu$. Assuming we are in this setting, the $K$-dependent effective bare propagator is 
\begin{equation}
C^{(K)}(\omega^2)=\frac{i}{(\omega^2-m^2+i\epsilon)I-K},
\end{equation}
in Feynman's prescription, which is the resolvent of $K$ evaluated outside the real line.
Now let $\lnorm K \rnorm_{op}$ denote the operator norm of $K$. Then, we can define the ``$K$-dependent generalized momenta'' (or ``background dependent generalized momenta'') - notations will have to change -
\begin{equation}
    p_\mu(K)^2:=\lambda_\mu+\lnorm K \rnorm_{op}
\end{equation}
which are positive semi-definite for each realization of the disorder $K$. Denote also the ``$K$-dependent'' physical mass
$$\mu_1(K):=m^2-\lnorm K\rnorm_{op}.$$ 
One notes that $\mu_1(K)$ can be negative with non-vanishing probability. This means that the action is not expressed in terms of the fluctuations around the physical vacuum.\\
In the basis provided by the $\{v_\mu\}$, the resolvent is diagonal. The diagonal elements write
\begin{equation}
C^{(K)}(p_\mu(K),\omega)_{\mu \mu}:=\frac{i}{\omega^2-p_\mu(K)^2-\mu_1+i\epsilon}\,.\label{barepropa-finite-N}
\end{equation}

\noindent
In the definition of the kinetic kernel \eqref{kineticeff}, we include explicitly the mass term, with coupling strength $m^2$. This contribution is essential for the \textit{free theory} to avoid IR singularities because the Wigner spectrum is not positive definite. We further require that the mass term $m^2>2\sigma$. In fact, the empirical eigenvalue distribution $\nu(\lambda):=\frac1N\sum_{\mu=1}^N\delta(\lambda-\lambda_\mu)$ (almost surely) converges to the semicircle distribution $\mu(\lambda):=\frac1{2\pi \sigma^2}\sqrt{4\sigma^2-\lambda^2}\mathbf{1}_{[-2\sigma, 2\sigma]},$ where $\mathbf{1}_{[-2\sigma, 2\sigma]}$ denotes the indicator function for the interval $[-2\sigma, 2\sigma]$. Moreover, the largest and smallest eigenvalues of $K$ are also known to (almost surely) converge to the edges of the limiting spectrum. This justifies the large $N$ limit heuristics that the $K$-generalized momenta do not depend on the realization $K$. In this regime, they can be replaced by the \emph{generalized momenta},
\begin{equation}
p^2_\mu:=\lambda_\mu+2\sigma\,,
\end{equation}
which are positive semi-definite. Additionally, our condition $m>2\sigma$ ensures that the physical mass does not depend on $K$, is positive, and simply becomes,
\begin{equation}
\mu_1:=m^2-2\sigma\,,
\end{equation}
\medskip
such that, at large $N$, the effective (bare) propagator becomes \cite{peskin}:
\begin{equation}
\boxed{C(p,\omega):=\frac{i}{\omega^2-p^2-\mu_1+i\epsilon}\,.}\label{barepropa}
\end{equation}
Recall that we use the Feynman convention to define the propagator, corresponding to the time-ordered propagator of the free scalar field, which is compatible with path integral quantization. Furthermore, this convention avoids crossing singularities in the region $(\mathrm{Im}(\omega)>0,\mathrm{Re}(\omega)>0$ on one hand, and in $(\mathrm{Im}(\omega)<0,\mathrm{Re}(\omega)<0)$ on the other, and Wick rotation toward Euclidean theory is well-defined.
\medskip

In this form, the propagator takes a suggestive form, reminiscent of propagators appearing in relativistic quantum field theory, where $p$ would play the role of a ‘‘spatial momenta''. However, this reminiscence remains artificial, and the theory is essentially non-relativistic. Space, defined by these generalized momenta and time, therefore, does not have to be treated symmetrically as required by a relativistic theory. Hence, it is suitable and physically relevant to construct a coarse-graining over-generalized momenta only, disregarding frequencies. Such a strategy, for time-dependent phenomena, was considered for instance in \cite{duclut2017frequency}, in the context of time-dependent critical phenomena and \cite{Lahoche2023functional} for classical $p$-spin dynamics. 
\medskip

The large $N$ distribution of generalized momenta $\rho(p^2)$ deduces from the distribution of eigenvalues $\lambda_\mu$, we get:
\begin{equation}
\boxed{\rho(p^2)=\frac{\sqrt{p^2(4\sigma-p^2)}}{2\pi \sigma^2}\,,}\label{distribution_p}
\end{equation}
which is the distribution shown in Fig \ref{fig2}. Finally, note that for an ordinary Euclidean QFT over $\mathbb{R}^d$, the distribution of momentum is $\rho(\vec{p}\,^2) \sim (\vec{p}\,^2)^{\frac{d-2}{2}}$. Then, from the point of view of spatial interpretation, $p$ would have a strange appearance, and from the coarse-graining point of view discussed before, it is as if the effective dimension of space changed as the scale changed. Furthermore, asymptotically, as $p\to 0$, the distribution matches with what we expect for a 3D standard QFT, i.e. $\rho(\vec{p}\,^2)\sim (\vec{p}\,^2)^{\frac{1}{2}}$, and the effective kinetic theory behaves well in that limit, as an ordinary QFT. 
\medskip

\paragraph{Averaging over the disorder $J$.} Now let us address the issue of the disorder $J$. In this paper we are aiming to construct RG, not for a given sample $J_{i_1\cdots i_p}$ but for its quenched average. This is entirely justified in the large $N$ limit due to the expected self-averaging property of the functional self-energy $W[K,J,\textbf{L}]:=-i\ln \mathcal{Z}[K,J,\textbf{L}]$:
\begin{equation}
\lim_{N\to \infty} W[K,J,\textbf{L}] \to \overline{W[K,J,\textbf{L}]}\,.
\end{equation}
The standard trick to compute the average logarithm is the \textit{replica method} \cite{castellani2005spin,Parisi,Dominicis}, coming from the elementary observation that:
\begin{equation}
\overline{W[K,J,\textbf{L}]}=-i\lim_{n\to 0} \frac{\overline{\mathcal{Z}^n[K,J,\textbf{L}]}-1}{n}\,.\label{averagedfree}
\end{equation}
The interest lies in the fact that, once we have constructed the replicated partition function $\mathcal{Z}^n[K,J,\textbf{L}]$,  the averaging can be easily performed from \eqref{averageJ} (see appendix \ref{AppA}). Explicitly:
\begin{align}
 \mathcal{Z}^n:=\int \, \prod_{\alpha=1}^n[d x_\alpha]\, \exp \Bigg(\frac{i}{\hbar}\sum_{\alpha=1}^nS_{\text{cl}}[\textbf{x}_\alpha]+i\int_{-\infty}^\infty\,dt\,\sum_{k=1,\alpha=1}^{N,n} L_{k,\alpha}(t)x_{k,\alpha}(t)\Bigg)\,,
\end{align}
where Greek indices $\alpha,\beta,\delta \cdots$ denote replica. Note that each replica corresponds to the same sample of the disorder ${J}$, and we get:
\begin{equation}
\overline{\mathcal{Z}^n}:=\int \, \prod_{\alpha=1}^n[d x_\alpha]\, e^{\frac{i}{\hbar}\overline{S_{\text{cl}}}[\{\textbf{x}\}]+i \mathcal{J}}\,,\label{partition1}
\end{equation}
where the source term is:
\begin{equation}
\mathcal{J}:=\int_{-\infty}^\infty\,dt\,\sum_{k=1,\alpha=1}^{N,n} L_{k,\alpha}(t)x_{k,\alpha}(t)\,,
\end{equation}
and the \textit{classical averaged action} reads, in the large $N$ limit:

\begin{align}
\nonumber\overline{S_{\text{cl}}}[\{\textbf{x}\}]&:=\sum_{\alpha}S_{\text{cl,kin}}[\textbf{x}_\alpha] \\
&+\int_{-\infty}^{+\infty} dt \left(i\frac{\lambda N}{2\hbar}\int_{-\infty}^{+\infty} dt^\prime\sum_{\alpha,\beta}\,  \left(\frac{\textbf{x}_\alpha(t)\cdot \textbf{x}_\beta(t^\prime)}{N}\right)^p -\sum_{\alpha} \frac{\mu_2}{4! N} (\textbf{x}_\alpha^2)^2\right)\,.\label{classicalaveraged}
\end{align}

\noindent
\textbf{Remark.}
\textit{Obviously, this theory is for the moment very formal, and the convergence of the path integral is not guaranteed since the sextic interaction (with $p=3$) is not positive definite. Strictly speaking, the integration measure should be completed in order to ensure the convergence of the integral. We will return to this point in future work.}
\medskip

Usually, the replica trick requires: (1) using the same source field for all replicas and (2) constructing a suitable analytic continuation to define the limit $n\to 0$. This last point in particular introduces subtleties regarding the construction of the saddle point solutions, opening the possibility of a \textit{replica symmetry breaking}. In this paper, we disregard this issue, and adopt a different procedure, assuming different sources for replica. These sources break explicitly the replica symmetry but allow characterizing the random distribution for $W[K,J,\mathbf{L}]$ from its cumulants, agreeing with the strategy of the authors in \cite{Tarjus2}. Hence, in the rest of this paper, $n$ is an integer $n\in \mathbb{N}^*$. Finally, note that our analysis will essentially focus on non-local moments of order $2$ and order $6$ for the approximations that we will consider in this article. A limitation that will be overcome in the subsequent work.

\section{RG Perturbation theory}\label{sectionPert}

Since Wilson pionner's work, RG \cite{ZinnJustinBook2,Wilson} has been based on the idea that long-range physics depends in principle only marginally on microscopic physics. Concretely it relies on the observation that small wavelength phenomena can be integrated out gradually, to provide effective couplings for long wavelength degrees of freedom which do not depend on the details of the microscopic physics.
\medskip

In the literature, there are many realizations of the general Wilson idea, and in this paper, we essentially focus on the effective average action (EAA) formalism due to Wetterich and Morris \cite{Delamotte_2012}. This choice will have the advantage of setting up a unified formalism with our subsequent nonperturbative studies. Furthermore, this formalism has the advantage of regularizing the infrared divergences that inevitably appear in the case $\mu_1<0$. This allows for a more precise discussion of the behavior of the flow in a regime where the development around the vacuum ‘‘zero" becomes unstable, pushing back the singularity of the propagator due to zero modes by a finite gap \cite{Litim}. However, a comprehensive study would require exploring beyond the symmetric phase, which we will discuss in a later article.
\medskip 

According to \cite{Berges}, we add to the classical averaged action $\overline{S}$ a scale dependent mass called \textit{regulator}:

\begin{equation}
\Delta S_k[\{\textbf{x}\}]:=-\frac{1}{2}\,\int dt \sum_{\alpha=1}^n\sum_{\mu=1}^N\, x_{\mu\alpha}(t) R_k(p_\mu^2)x_{\mu\alpha}(t)\label{defregulatorD}
\end{equation}
Note that the regulator is expected to be diagonal both in time and in the replica space. This is equivalent to focusing on large-time behavior (see Appendix \ref{AppB}). We introduce the \textit{effective average action} $\Gamma_k$ defined by the following relation:
\begin{equation}
\Gamma_k[\mathcal{M}]+\Delta S_k [\mathcal{M}]={W}_{k}^{(n)}[\mathcal{L}]-\sum_{\alpha=1}^n \int\, dt \, \textbf{L}_\alpha(t) \cdot \textbf{M}_\alpha(t)\,,\label{defGammak}
\end{equation}
where ${W}_{k}^{(n)}[\mathcal{L}]$ is again the free energy of the modified partition function $\overline{\mathcal{Z}^n_k}$ and  $\mathcal{L}:=\{\mathbf{L}_\alpha \}$,  $\mathcal{M}:=\{\mathbf{M}_\alpha \}$ are the set of sources and classical fields (a $N$-dimensional vector per replica). The classical fields are given by:
\begin{equation}
\frac{\delta }{\delta L_{i\alpha}}{W}_{k}^{(n)}[\mathcal{L}]=M_{i\alpha}\,.
\end{equation}
The role of the regulator is to make $\Gamma_k[\mathcal{M}]$ a smooth interpolation between the classical action $\overline{S_{\text{cl}}}$ for $k=4\sigma$ and the full effective action $\Gamma$ for $k=0$, which is the Legendre transformation of the free energy \eqref{averagedfree}. 
\medskip

More precisely, the regulator freezes out large-scale degrees of freedom ($p_\mu  \lesssim k$), which acquire a large mass, whereas microscopic degrees of freedom ($p_\mu  \gtrsim k$) essentially unaffected by the regulator are integrated out. More precisely, we require that:

\begin{enumerate}
    \item $\lim_{k\to 0} R_k(p_\mu^2)=0$, meaning that all the degrees of freedom are integrated out as $k\to 0$ (infrared (IR) limit).
    \item $\lim_{k\to 4\sigma} R_k(p_\mu^2)\to\infty$, meaning that no fluctuations are integrated out in the deep ultraviolet (UV) limit $k\to 4\sigma$.
\end{enumerate}

Note that, as usual, we define the IR regime for small $p_\mu$ and the UV regime for large $p_\mu$, on the right side of the spectrum (see Fig. \ref{fig2}). In this paper, we focus on a slightly modified version of the standard Litim regulator \cite{Litim}, inspired by \cite{lahoche20241}:
\begin{equation}
R_k(p_\mu^2):=\frac{4\sigma}{4\sigma-k^2}(k^2-p_\mu^2)\theta(k^2-p_\mu^2)\,.\label{LitimReg}
\end{equation}
\medskip

\begin{figure}
\begin{center}
\includegraphics[scale=1.2]{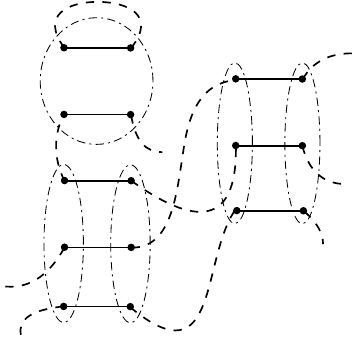}
\end{center}
\caption{A typical Feynman diagram for $p=3$, involving two non-local vertices, one local quartic vertex, and six external points. }\label{FeynmanDiag}
\end{figure}

Any vertex function $\Gamma_k^{(2n)}$ (with $2n$ external points) then expands in power series in perturbation theory:
\begin{equation}
\Gamma_k^{(2n)}(\omega_1,\cdots \omega_{2n})= \sum_{\mathcal{G}\in\mathbb{G}_{2n}}\,\frac{1}{s(\mathcal{G})} \mathcal{A}_{\mathcal{G}}(\omega_1,\cdots \omega_{2n})\,,
\end{equation}
where $\mathbb{G}_{2n}$ is the set of 1PI Feynman diagrams with $2n$ external points, with its corresponding Feynman amplitude being $\mathcal{A}_{\mathcal{G}}$. The additional factor, $s(\mathcal{G})$ called symmetry factor, depends on the dimension of the automorphism group of the graph\footnote{If the coupling constant is appropriately normalized, $s(\mathcal{G})$ equals the dimension of the automorphism group.}. A Feynman graph is essentially a set of edges and vertices, representing the bare propagator from Wick contractions and the interactions involved in the bare action, respectively. Here, we need to establish a notation that ensures easy readability of 1) the contractions of the fields, 2) the replicated structure, and, 3) the non-locality. \\
The solution we consider is inspired by \cite{Lahoche2022functional}. Vertices are realized by sets of black nodes, corresponding to the fields $\textbf{x}_\alpha$, linked together by solid edges representing Euclidean scalar products of vectors, in the components of the field degrees of freedom. Hence:
\begin{equation}
\vcenter{\hbox{\includegraphics[scale=1]{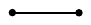}}}\equiv \sum_{i=1}^N \, x_{i\alpha} (t)x_{i\alpha}(t)\,.
\end{equation}
This takes care of point 1, but points 2 and 3 remain to be considered. For the latter, it is useful to note that the local time components are associated with the same replica index. We therefore only need one notation to specify the replica or the local time component. Concretely, we introduce the graphical convention that any set of dots surrounded by a dashed-dotted circle are local in time and with the same replica indices. For instance:
\begin{equation}
\vcenter{\hbox{\includegraphics[scale=1.2]{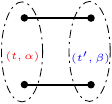}}}\equiv \int {\color{red}dt}\int {\color{blue}dt^\prime} \sum_{{\color{red}\alpha},{\color{blue}\beta}}\, (\textbf{x}_{{\color{red}\alpha}}({\color{red}t})\cdot \textbf{x}_{{\color{blue}\beta}}({\color{blue}t^\prime}))^2\,.
\end{equation}
Finally, we represent the propagator involved in the Wick contractions as a dashed edge (note that the propagator is diagonal in frequencies and different indices). A typical Feynman graph is pictured in Fig. \ref{FeynmanDiag} for $p=3$ model with $6$ external edges. We call \textit{external nodes} the black points connected with external edges. 
\medskip

\begin{figure}
\begin{center}
\includegraphics[scale=1.2]{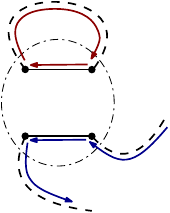}
\end{center}
\caption{Illustration of the concept of face, a closed face (red cycle) and an open face (blue cycle).}\label{figFace}
\end{figure}

Another relevant concept that has to be addressed in the large $N$ limit is the concept of face. We have the following definition: 
\medskip

\noindent
\textbf{Definition.} \textit{A face is a cycle made of an alternative sequence of dashed and solid edges. A face can be closed or open.}
\medskip

Fig. \ref{figFace} provides an illustration. Note that each closed face contributes a global factor $\sum_{i=1}^N\delta_{ii}=N$ to the diagram. Then, for the $p=2$ model, if we denote respectively by $V$ and $F$  the number of quartic vertices (local or not) and closed faces of some Feynman graph $\mathcal{G}$, the amplitude scale with $N$ as:
\begin{equation}
\mathcal{A}_{\mathcal{G}} \sim N^{-V+F}\,.
\end{equation}
In large $N$ limit, the leading order (LO) graphs are then those that maximize the number of created faces, taking into account all the other constraints imposed for the configurations of the internal dashed edges (not connected to external nodes). 
\medskip

An understanding of the LO sector provides moreover some powerful relations between different LO observables that in the large $N$ limit depend only on the 1PI $2$-point function $\gamma_{\mu\mu^\prime, \alpha, \beta}$. Standard reviews exist on this topic, and we refer the reader to them, for instance \cite{ZinnJustinReview,ZinnJustinBook2,Lahoche2018}, and we will only give a summary description here, referring the reader to these reviews and references therein. 
\medskip

To understand LO graphs, we will need some definitions and intermediate results.  In what follow we will mainly focus on the case $p=3$, assuming that only non-local interactions are sextic, the local ones being assumed quartic. However, we consider two types of propagator. This will give an additional generality to the construction and  allow us to refer directly to this section in our future work. We denote by $C_{\text{L}}$ and $C_{\text{NL}}$ these two kinds of propagators, in which  the subscript ‘‘L" and ‘‘NL" mean respectively ‘‘local" and ‘‘non-local". We define them from their Fourier transform; the Fourier transform of the propagator $C_{\text{L}}$ is proportional to $\delta(\omega+\omega^\prime) \delta_{\alpha\beta}\delta_{\mu\mu^\prime}$, whereas the non-local propagator is proportional to $\delta(\omega) \delta(\omega^\prime)\delta_{\mu\mu^\prime}$. Note that both propagators are diagonal with respect to the indices of generalized momenta because of the underlying \textit{global} $O(N)$ symmetry of the theory. This means the locality or non-locality is with respect to the indices of replica and time. We make use of the following graphical notations:
\begin{align}
C_{\text{L}} &\equiv \vcenter{\hbox{\includegraphics[scale=1]{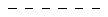}}}\,,\\
C_{\text{NL}} & \equiv  \vcenter{\hbox{\includegraphics[scale=1]{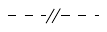}}}\,.
\end{align}
Now, let us define the concepts of contraction of some internal propagator edge and 
the sum of two $O(N)$ symmetric interaction vertices. 
\medskip

\noindent
\textbf{Definition.} \textit{Let $n_1\in u$ and $n_2 \in v$ be two nodes attached to the local components $u$, $v$, hooked to each other through some propagator edge, which can be local or non-local. Let $q=0,1$ be the number of solid edges between them. Furthermore, let  $p=1-q$ be the number of solid edges attached to the vertex $n_1$ (resp. $n_2$) but not to $n_2$ (resp. $n_1$). The contraction of the propagator edge is defined as:
\begin{enumerate}
\item The cancellation of the $q+1$ edges (including the dashed one) between the two nodes $n_1$ and $n_2$ and the nodes themselves,
\item The union of $p$ solid edges attached to these nodes.
\item The union of the local components attached to the nodes $n_1$ and $n_2$ if the propagator is local. 
\end{enumerate}
}

Fig. \ref{figcontraction} illustrates this definition. Another definition is that of the sum of two vertices:
\medskip

\begin{figure}
\begin{center}
\includegraphics[scale=0.9]{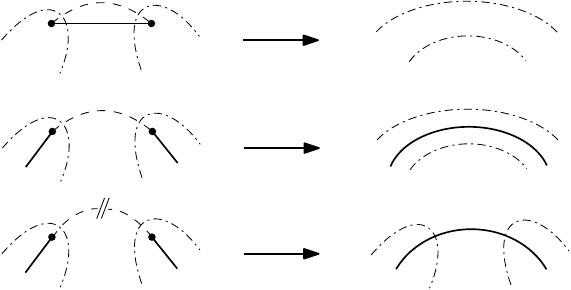}
\end{center}
\caption{Illustration of the local and non-local contraction.}\label{figcontraction}
\end{figure}

\begin{figure}
\begin{center}
\includegraphics[scale=1]{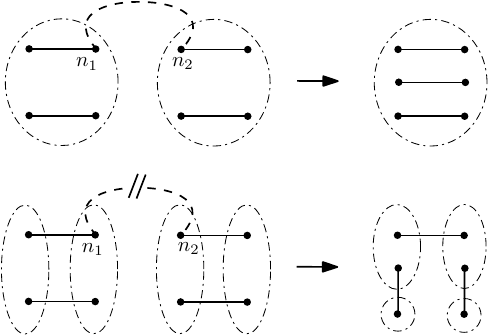}
\end{center}
\caption{The local sum of two local vertices (on the top) and the non-local sum of two non-local vertices (on the bottom).}\label{figsum}
\end{figure}

\begin{figure}
\begin{center}
\includegraphics[scale=1.2]{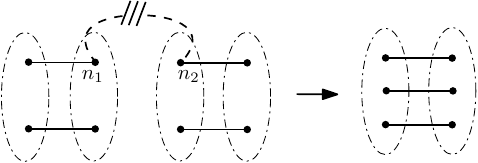}
\end{center}
\caption{Formal definition of the $C_0$ edge.}\label{figC0}
\end{figure}

\noindent
\textbf{Definition.} \textit{Let $v_1$ and $v_2$ two vertices, and let $n_1\in v_1$ and $n_2\in v_2$ two nodes. The local (resp. non-local) sum $v_1 \sharp_L v_2$ is obtained by:
\begin{enumerate}
\item Linked the two nodes $n_1$ and $n_2$ with some local propagator $C_{\text{L}}$ (resp. $C_{\text{NL}}$).
\item Contracting the edge $(n_1,n_2)$.
\end{enumerate}}
Fig. \ref{figsum} illustrates this definition, which enables the use of a purely quartic theory, where each sextic coupling is "broken" into two quartic vertices. However, it seems impossible to simply decompose the non-local sextic vertex due to disorder into a sum of two local vertices. For this we will add an \textit{ad hoc} rule and a new propagator\footnote{This propagator has not any explicit form, and is only defined by the contraction procedure} $C_0$,
\begin{equation}
C_{0} \equiv  \vcenter{\hbox{\includegraphics[scale=1.2]{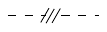}}}\,.
\end{equation}
which is assumed to be diagonal with respect to the field indices $i,j,\cdots$ and can only exist between two non-local quartic vertices. This propagator is defined such that the sum of two non-local quartic vertices gives a non-local sextic vertex for $p=3$ like disorder -- see Fig. \ref{figC0} -- and such that a given non-local quartic vertex cannot be hooked with more than one edge of type $C_0$. 
\medskip

\begin{figure}
\begin{center}
\includegraphics[scale=0.7]{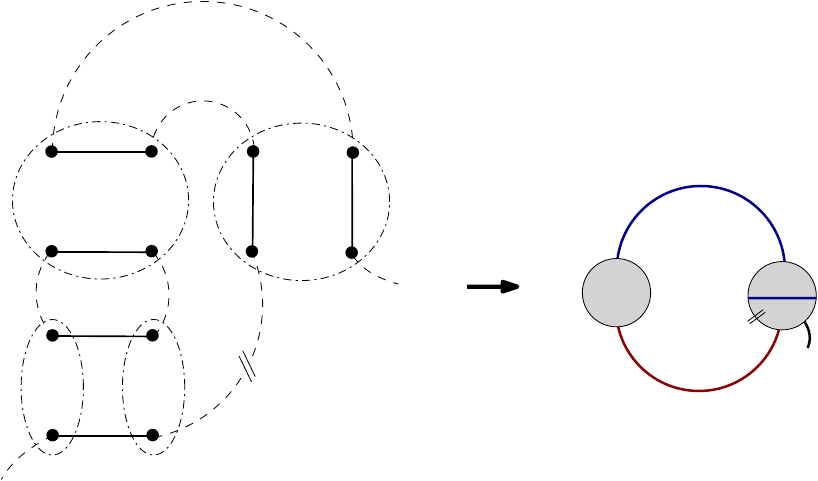}
\end{center}
\caption{illustration of the correspondence between original and intermediate field representation. Loop vertices are gray discs, and intermediate field propagators are solid (blue or red) edges. Blue edges (resp. red edges) are for local (resp. non-local vertices), and non-local propagators are marked by a double bar on the corresponding corner of the graph. Finally, the ‘‘cilium" on the loop vertex on the right-hand side represents the external edges in the original representation.}\label{figintermediate}
\end{figure}

\begin{figure}
\begin{center}
\includegraphics[scale=0.7]{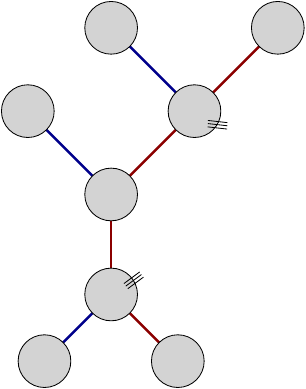}
\end{center}
\caption{A typical bicolored tree.}\label{figtree}
\end{figure}

We can then construct in a very general way the graphs corresponding to a quartic theory, including local and bi-local interactions and its three kind of propagators. The graphs of the theory that we are looking for will be those which will only include propagators of type $C$ and $C_0$.
Furthermore, for such a model, the construction of LO graphs is generally more transparent using intermediate field formalism and \textit{loop-vertex representation} (LVR) \cite{Riv2018}. Consider a vacuum-connected Feynman graph involving $L$ edges (local or not) and $V$ quartic vertices (local or not). Such a Feynman graph scales as $N^{1-(V-F+1)}$. In the intermediate field representation, vertices become intermediate field edges and faces are contracted as effective \textit{loops vertices}. The correspondence is illustrated graphically in Fig. \ref{figintermediate} for a Feynman diagram in the original representation, involving some local and non-local vertices and edges. Let us consider a vacuum diagram in the original representation, and let $\mathcal{E}$ and $\mathcal{V}$ be the number of intermediate field edges and loop vertices respectively. It can be shown that the graph scales as $N^{-\omega+1}$, such that\footnote{We refer  reader that, there is a typo in reference \cite{Lahoche2018} where $\mathcal{E}$ and $\mathcal{V}$ have the wrong sign in the corresponding formula}:
\begin{equation}
\omega=\mathcal{E}-\mathcal{V}+1 \geq 0\,.\label{omegaindex}
\end{equation}
$\omega$ is nothing but the number of loops in the intermediate field graph, and it vanishes only if the (connected) graph is a tree graph. Hence, the LO graphs are bicolored trees in the intermediate field representation (Fig. \ref{figtree}). Now, let us move on to the LO $1$PI $2$-points diagrams. General LO 1PI $2$-points diagrams can be obtained from a vacuum tree by cutting some loops on the leaves of the tree\footnote{If we cut edges which are not leafs, we break the 1PI condition.}.  Note moreover that the contraction of $C_0$ lines does not modify $\omega$, because it does not change the number of face, and if the number of vertices decrease by one, this is compensated by the fact that the sextic couplings scale as $N^{-2}$ i.e. has the product of two quartic vertices.
\medskip

\begin{figure}
\begin{center}
\includegraphics[scale=1.2]{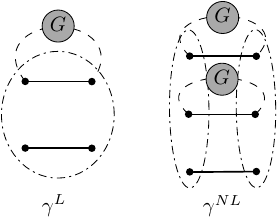}
\end{center}
\caption{Structure of the LO components for the 1PI 2-points function. The grey discs materialize LO 2-point function.}\label{figeffG}
\end{figure}

Let $\mathbb{G}_2$ be the set of $1$PI $2$-point graphs obtained by this method; it is obvious that $\mathbb{G}_2$ splits in two disjoint ensembles $\mathbb{G}_2=\mathbb{G}_{2\text{L}} \cup \mathbb{G}_{2\text{NL}}$, where $\mathbb{G}_{2\text{L}} $ (resp. $\mathbb{G}_{2\text{NL}}$) is the subset of graphs such that the external edges are attached to a local (resp. non-local) quartic vertex. Furthermore, for our sextic model, we must retain only graphs without non-local lines and such that $C_0$ lines contract to form a sextic interaction, without disturbing the connectivity of the graph\footnote{Note that we assume that only the sextic interactions are non-local from this point.}, and we call respectively $\mathbb{G}_{2\text{L,ph}}$ and $\mathbb{G}_{2\text{NL,ph}}$ the corresponding set of graphs. Then, the self-energy can be expressed as:
\begin{equation}
\Sigma_{\mu\mu^\prime, \alpha, \beta}= \underbrace{\sum_{\mathcal{G}\in\mathbb{G}_{2\text{L,ph}}}\, \mathcal{A}_{\mu\mu^\prime, \alpha, \beta}(\mathcal{G})}_{:=\gamma_{\mu\mu^\prime, \alpha,\beta}^{\text{L}}}\,+\,\underbrace{\sum_{\mathcal{G}\in\mathbb{G}_{2\text{NL,ph}}}\, \mathcal{A}_{\mu\mu^\prime, \alpha, \beta}(\mathcal{G})}_{:=\gamma_{\mu\mu^\prime, \alpha, \beta}^{\text{NL}}}\,,\label{closedzero}
\end{equation}
where $\mathcal{A}_{\mu\mu^\prime, \alpha, \beta}(\mathcal{G})$ denotes the Feynman amplitude for the graph $\mathcal{G}$. $\mu$, $\mu^\prime$ are the external generalized momenta, $\alpha,\beta$ the external replica indices. The external frequency indices in $\Sigma_{\mu\mu^\prime, \alpha, \beta}$ has been omitted. For  $\gamma_{\mu\mu^\prime, \alpha, \beta}^{\text{L}}$, further consideration shows that the summed Feynman graphs have to reconstruct effective (local) LO $2$-point function $G$, as shown on the left of Fig. \ref{figeffG}. For $\gamma_{\mu\mu^\prime, \alpha, \beta}^{\text{NL}}$, the two external points have to be connected to the same non-local sextic vertex, and the structure of the graphs has to be like on the right of Fig. \ref{figeffG} to be LO. 
\medskip

Recursively, it is straightforward to verify that the effective local 2-point function,  $G$, has to be proportional to $\delta(\omega+\omega^\prime)\delta_{\mu\mu^\prime}\delta_{\alpha\beta}$ i.e. diagonal with respect to generalized momenta indices and replica indices. Then, 
\begin{equation}
\Sigma_{\mu\mu^\prime, \alpha, \beta}=\gamma{(\omega)}\, \delta_{\mu\mu^\prime}\delta_{\alpha\beta}\,,
\end{equation}
with $\gamma(\omega)$ being a function of the external frequency and the relation \eqref{closedzero} takes the form (for $p=3$):
\begin{align}
\nonumber &\gamma(\Omega)=  \frac{A \mu_2}{N} \, \int d\omega\,\sum_{\mu} G_k(\omega,p_\mu,p_\mu,\alpha)\\
&+\frac{B \lambda}{N^2}\, \int d\omega\,\sum_{\mu,\nu} G_k(\omega,p_\mu,p_\mu,\alpha) G_k(-\omega-\Omega,p_\nu,p_\nu,\alpha)\,,
\end{align}
where:
\begin{equation}
G_k(\omega,p_\mu,p_\nu,\alpha)=\frac{\delta_{\mu\nu}}{\omega^2-p_\mu^2-\mu_1-\gamma(\omega)-R_k(p_\mu^2)+i\epsilon}\,,
\end{equation}
and where we reestablished the frequency dependency of $\gamma$. Constants $A$ and $B$ can be fixed by leading order perturbation theory, and we get finally in the continuum limit (we choose the units such that $\hbar=1$):
\begin{align}
\nonumber \gamma(\Omega)=& -i\int \frac{d\omega}{2\pi}\,\rho(p^2) dp^2 \bigg(\frac{\mu_2}{6}\, G_k(\omega,p,p,\alpha)\\
&+{12\lambda}\,\int  \rho(q^2)  dq^2 G_k(\omega,p ,p,\alpha) G_k(-\omega-\Omega,q,q,\alpha)\bigg)\,.\label{closedeqation}
\end{align}
Because of the dependency of $\gamma$ on the external frequency $\Omega$, this equation is difficult to solve exactly and requires approximations. The rest of this article essentially describes various approximation schemes, allowing us to understand the behavior of solutions to this equation in various regimes. Furthermore, note that in the case where $\lambda=0$, $\gamma$ becomes independent of frequency, and the equation takes the form of a transcendental equation.
\medskip

\noindent

\textbf{Remark.} \textit{The equation \eqref{closedeqation} can also be deduced from the so-called Schwinger-Dyson equation. The previous approach, however, has the advantage of being more transparent regarding the structure of the graphs.}
\medskip

\begin{figure}
\begin{center}
\includegraphics[scale=1.2]{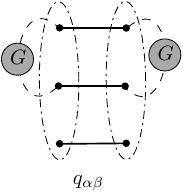}
\end{center}
\caption{Leading order contribution to $q_{\alpha\beta}$, where $G_k$ is the LO effective $2$-point function.}\label{Figqab}
\end{figure}

Furthermore, we will also need the following quantity under the condition ($\alpha \neq \beta$):
\begin{equation}
Q_{\alpha\beta}(t,t^\prime):=\frac{1}{N} \,\sum_{\mu=1}^N \, \langle x_{\mu,\alpha}(t)x_{\mu,\beta}(t^\prime) \rangle \,.
\end{equation}
Because of the previous analysis, it follows that $Q_{\alpha\beta}$ is next to the leading order with respect to N. It is convenient to decompose it in its 1PI components, and we denote as $q_{\alpha\beta}$ the 1PI component:
\begin{equation}
q_{\mu,\nu,\alpha\beta}(t,t^\prime):=\, \langle x_{\mu,\alpha}(t)x_{\nu,\beta}(t^\prime) \rangle_{\text{1PI}} \,,
\end{equation}
such that, formally (products are understood as matrix multiplication):
\begin{equation}
Q_{\alpha\beta}=\frac{1}{N}\, \mathrm{Tr}\, G_k\,q_{\alpha\beta} \bigg(1+ \sum_{n=1}^\infty (G_k\,q_{\alpha\beta})^n \bigg) G_k\,.
\end{equation}
Because it is next to the leading (NLO) quantity, $q_{\alpha\beta}\equiv\mathcal{O}(1/N)$, and in the large $N$ limit, the leading  order contribution comes from the first term of the expansion:
\begin{equation}
Q_{\alpha\beta}\approx \frac{1}{N}\, \mathrm{Tr}\, G_k\,q_{\alpha\beta} \,G_k\,,\label{Qab}
\end{equation}
We still have to find an equation for $q_{\alpha\beta}$. Because it is a non-local quantity, the external edges have to be attached to different local components. There are two possibilities: either the two edges are connected to the same vertex or never. But it is easy to check that the leading order contribution for $q_{\alpha\beta}$, of order $1/N$, is given by the configuration shown in Fig. \ref{Figqab}. Explicitly:
\begin{equation}
\boxed{q_{\alpha\beta} (\Omega,\Omega^\prime)=-\frac{12 \lambda}{N} \, \int \rho(p^2) dp^2\, \mathcal{L}_k(p,\alpha) \delta(\Omega)\delta(\Omega') \,,}
\end{equation}
where $\mathcal{L}_k(p,\alpha):=\left(\int d\omega \, G_k(\omega,p,p,\alpha)\right)^2$.

\section{Symmetries of the path integral}\label{secWard}

The path integral \eqref{partition1} has many symmetries, which
at the quantum level can be translated in terms of Ward identities. The Ward identities can be classified into two
categories:

\begin{itemize}
\item Type $1$ transformations, which left the path integral measure invariant but not the classical action.
\item Type $2$ transformations, which left both the path integral
measure and the classical action invariant.
\end{itemize}

Usually, these two kinds of transformation provide different kinds of Ward identities. Ward identities are the quantum version of the classical Noether's theorem, they provide relations between observables that are essential for the coherence of the quantum theory \cite{ZinnJustinBook2} and the underlying RG flow. In particular, we expect a complete nonperturbative program to take these relations into account, and this will be the subject of a forthcoming work. 
\medskip

For transformations of the second kind, the variation of the path integral reads:
\begin{align}
\nonumber &\langle \delta \mathcal{J} \rangle = \int_{-\infty}^\infty\,dt\,\sum_{k=1,\alpha=1}^{N,n} L_{k,\alpha}(t)\delta \langle x_{k,\alpha}(t) \rangle \\
&= \int_{-\infty}^\infty\,dt\,\sum_{k=1,\alpha=1}^{N,n} \frac{\delta \Gamma}{\delta \langle x_{k,\alpha}(t) \rangle } \delta\langle x_{k,\alpha}(t) \rangle= \delta \Gamma=0\,,\label{variationGamma}
\end{align}
where $\Gamma$ is the classical action, i.e. the Legendre transform of the free energy\footnote{$\Gamma:=\Gamma_k+ \Delta S_k$, where $\Gamma_k$ is defined by \eqref{defGammak}.} ${W}_{k}^{(n)}:=\ln \overline{\mathcal{Z}^n_k}$. Equation \eqref{variationGamma} implies that the effective average action must also be invariant under the symmetries that leave the bare action invariant. Time translation invariance is an example of such a symmetry. In this work, we will focus specifically on asymptotic states in the limit $t\to \infty$, with particular interest in the symmetries related to the internal and replica indices.
\medskip

The choice of gauge \eqref{Gauge} breaks the underlying global rotation invariance $O(N)$. Once the gauge is fixed, the interactions remain invariant under gauge transformations: 
\begin{equation}
x_{\mu\alpha} \to x_{\mu\alpha}^\prime=\sum_{\mu^\prime=1}^N\,O_{\mu\mu^\prime} x_{\mu^\prime\alpha}\,,
\end{equation}
for some orthogonal matrix $O$ which \textit{does not depend on the replica index}. Such a transformation is therefore described as local. On the other hand, kinetic action explicitly breaks this symmetry, leading to non-trivial relations between observables. Note that in a similar context, these relations would be used to close the hierarchy of flow equations \cite{lahoche20242,lahoche20244,Lahochebeyond}. For infinitesimal rotations, this identity is given by:
\begin{align}
\nonumber \int dt \sum_\alpha \Bigg(\frac{i}{\hbar}\delta E_k(p^2_{\mu_1},p^2_{\mu_2})\, &\frac{\partial^2}{\partial L_{\mu_1,\alpha}(t) \partial L_{\mu_2,\alpha}(t)}\\
 &-L_{\mu_1,\alpha}(t)\frac{\partial}{\partial L_{\mu_2,\alpha}(t)}+L_{\mu_2,\alpha}(t)\frac{\partial}{\partial L_{\mu_1,\alpha}(t)}\Bigg) e^{{W}_{k}^{(n)}[\mathcal{L}]}=0\,,
\end{align}
where:
\begin{equation}
\delta E_k(p^2_{\mu_1},p^2_{\mu_2})=p_{\mu_2}^2-p_{\mu_1}^2+R_k(p_{\mu_2}^2)-R_k(p_{\mu_1}^2)\,.
\end{equation}
But the measure of the path integral has a larger invariance $(O(N))^n$ for different rotations depending on the replica index $O_{\mu\mu^\prime}(\alpha)$. These transformations leave the local interactions in the replica space invariant, but not the bi-local terms, such as those generated by integrating out the disorder $J$; furthermore, the kinetic term is not invariant either. The computation of the variation of the non-local term:
\begin{equation}
V_J:=\frac{i\lambda }{2N^{p-1}\hbar}\int_{-\infty}^{+\infty} dt\, dt^\prime\sum_{\alpha,\beta}\,  \left({\textbf{x}_\alpha(t)\cdot \textbf{x}_\beta(t^\prime)}\right)^p\,,
\end{equation}
leads to the identity:
\begin{align}
\nonumber & \int dt \Bigg(\frac{i}{\hbar}\delta E_k(p^2_{\mu_1},p^2_{\mu_2})\, \frac{\partial^2}{\partial L_{\mu_1,\alpha}(t) \partial L_{\mu_2,\alpha}(t)}-L_{\mu_1,\alpha}(t)\frac{\partial}{\partial L_{\mu_2,\alpha}(t)}+L_{\mu_2,\alpha}(t)\frac{\partial}{\partial L_{\mu_1,\alpha}(t)}\Bigg) e^{{W}_{k}^{(n)}[\mathcal{L}]} \\\nonumber
&\nonumber -\frac{i p \lambda }{\hbar^2} \times \Big\langle \int dt dt^\prime\sum_\beta (X_{\alpha\beta})^{p-1} ( x_{\mu^\prime,\alpha}(t)x_{\mu,\beta}(t^\prime)- x_{\mu,\alpha}(t)(t)x_{\mu^\prime,\beta}(t^\prime)) \Big\rangle\\
&=0\,,\label{localWard}
\end{align}
where the notation $\langle X \rangle$ means averaging with respect to $x_{\mu\alpha}$, and:
\begin{equation}
X_{\alpha\beta}(t,t^\prime):=\frac{1}{N}\,\sum_{\mu=1}^N \, x_{\mu,\alpha}(t)x_{\mu,\beta}(t^\prime)\,.
\end{equation}
If we assume that $X_{\alpha\beta}(t,t^\prime)$ self averages,  always expected in the high temperature regime:
\begin{equation}
\langle (X_{\alpha\beta})^{p-1} \rangle \simeq  (Q_{\alpha\beta})^{p-1}\,,
\end{equation}
and define the connected $2$-point function:
\begin{equation}
\langle x_{\mu,\alpha}x_{\nu,\beta} \rangle_c:=\langle x_{\mu,\alpha}x_{\nu,\beta} \rangle-\langle x_{\mu,\alpha} \rangle \langle x_{\nu,\beta} \rangle\,.
\end{equation}
From what we saw above, following the "naive" power counting, the correlations between replicas vanish in the limit $N\to \infty$, and the Ward identities essentially reduce to what they would be if the interactions were locally symmetric.

\section{Scaling and dimensions}\label{sectionscaling}

Following \cite{lahoche20241,lahoche3,Bradde}, a Wilsonian RG \cite{Wilson} can be constructed by partially integrating out degrees of freedom labelled by $p_\mu$ in the partition function \eqref{partition1}, from UV scales (large $p_\mu$) to IR scales (small $p_\mu$). Here, it should be noted that the particular nature of the spectrum over which we integrate changes the way of defining the rescaling of the couplings after each step of the RG. Indeed, let us first note that the action \eqref{classicalaveraged} must have the same dimension as $\hbar$, i.e. the dimension of an energy by a time. In the units convention where $\hbar=1$, one can set $ [\omega]=-[t]=1$, and this fixes the dimension of the couplings involved in $\overline{S_{\text{cl} }}[\{\textbf{x}\}]$:
\begin{equation}
[\mu_1]=2\,,\qquad [\lambda]=4\,,\qquad [\mu_2]=3\,.
\end{equation}
The equation that describes how the EAA changes as $k$ changes is the Wetterich equation \cite{Delamotte_2012}
\begin{equation}
\boxed{\dot{\Gamma}_k=-\frac{i}{2}\int \frac{d\omega}{2\pi}\, \sum_{p_\mu,p_\nu,\alpha} \dot{R}_k(p_\mu^2) G_k(\omega,p_\mu,p_\nu,\alpha)\,,}\label{Wett}
\end{equation}
where the dot denotes differentiation with respect to $t:=\ln(k/4\sigma)$, and $G_k$ is the effective $2$-point function (diagonal with respect to the replica and generalized momenta indices), such that:
\begin{equation}
[G_k^{-1}(\omega, p_\mu,p_\nu,\alpha)+R_k(p_\mu^2)\delta_{\mu\nu}]\delta(\omega+\omega^\prime)\delta_{\alpha\beta}:=\frac{\delta^2 \Gamma_k}{\delta M_{\mu\alpha}\delta M_{\nu\beta}}\,.
\end{equation}
This definition assumes that the right-hand side is diagonal both in frequencies and in replica indices. As seen explicitly above, this assumption is justified in the large $N$ limit, where the bare propagator \eqref{barepropa} is diagonal. Then in the large $N$ limit, the flow equation \eqref{Wett} admits the continuum expression:
\begin{equation}
\dot{\Gamma}_k=-\frac{iN}{2}\int \frac{d\omega}{2\pi}\,\int_0^{4\sigma} dp^2 \, \rho(p^2) \sum_{\alpha} \dot{R}_k(p^2\,) G_k(\omega,p,p,\alpha)\,.\label{Wett2}
\end{equation}
In this paper, we focus on perturbation theory. It is well known that, in the symmetric phase, the anomalous dimension $\eta=\mathcal{O}(\mu_2^2/N)$, and it is easy to check that these perturbative effects are relevant only up to order $\mu_2^3/N$. Hence, as soon as we project it along the truncation in the theory space, we assume the second derivative matrix to be of the form:
\begin{equation}
\frac{\delta^2 \Gamma_k}{\delta M_{\mu\alpha}\delta M_{\nu\beta}}=(\omega^2-p_\mu^2-\Sigma+i\epsilon)\delta(\omega+\omega^\prime)\delta_{\alpha\beta}\,,
\end{equation}
where according to the large $N$ analysis, the \textit{self-energy} is proportional to the identity $\delta_{ij} \delta_{\alpha\beta}$, with entry $\Sigma$ such that:
\begin{equation}
\Sigma=\mu_1+\mathcal{O}(\mu_2)\,.
\end{equation}
Note that in this order, the self-energy reduces to the effective mass and does not depend on external momenta, as suggested by the notations. Flow equations for $2n$-point functions $\Gamma_k^{(2n)}$ can be obtained by taking the $2n$th functional derivative on both sides of the flow equation \eqref{Wett2}. To begin, let us only consider local interactions. It is easy to see that the right-hand side of the flow equation for $\Gamma_k^{(2n)}$ is written as a sum of effective one-loop diagrams, involving products of vertex functions, as illustrated by Fig. \ref{figflow}. 
\medskip

\begin{subequations}
\begin{figure}
\begin{center}
\includegraphics[scale=1.1]{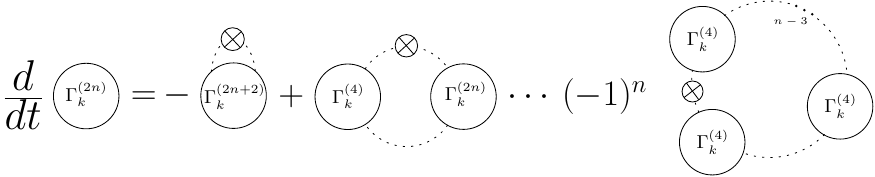}
\end{center}
\caption{Diagramatic illustration of the general structure of the flow equation for $\Gamma^{(2n)}_k$. Dotted edges materialize effective propagator $G_k$ and crossed circle materializes $\dot{R}_k$.}\label{figflow}
\end{figure}
\end{subequations}

Consider the flow equation for $\Gamma_k^{(2)}$. If we focus on the local sector, the effective $2n$-point functions can be written as: 
\begin{align}
\Gamma_k^{(2n)}(\{p_i, \omega_i\})&:=-\frac{u_{2n}}{(2n)! N^{n-1}} \delta\left(\sum_{i=1}^{2n}\omega_i\right) \sum_{\pi} \delta_{p_{\pi(1)}p_{\pi(2)}}\cdots \delta_{p_{\pi(2n-1)}p_{\pi(2n)}}\,,\label{localsectorGamma2n}
\end{align}
where $\pi$ denotes some permutation of the $2n$ external indices. Let us first focus on the case $k\ll1$, which corresponds to the IR limit of the theory. In this case, we can replace the regulator \eqref{LitimReg} by $(k^2-p^2)\theta(k^2-p^2)$, and if we consider only local interactions, the flow equation for $\Gamma_k^{(2)}$ involves the loop integral:
\begin{equation}
I_2(k):=\int d\omega \int dp^2\, \rho(p^2) \frac{\dot{R}_k(p^2)}{(\omega^2-k^2-\Sigma+i\epsilon)^2} \,.\label{definitionI2}
\end{equation}
The integral over frequencies factors out, and can be explicitly performed using the residue theorem:
\begin{equation}
\int \frac{d\omega}{(\omega^2-E^2+i\epsilon)^n}=\frac{i(-1)^n\sqrt{\pi }\, \Gamma \left(n-\frac{1}{2}\right) }{\left(E^2\right)^{n-\frac{1}{2}}\Gamma (n)}\,,
\end{equation}
thus we have:
\begin{equation}
I_2=\frac{i\pi\,\Omega(k)}{2 \left(k^2+\Sigma\right)^{3/2}} \,,
\end{equation}
where:
\begin{equation}
 \Omega(k):=\int dp^2\, \rho(p^2) \dot{R}_k(p^2)\,.
\end{equation}
Because all the flow equations involve only a single loop, the same factor appears indeed systematically, and a loop involving $n$ propagator involves the integral:
\begin{align}
 I_n:=\int_{-\infty}^{+\infty} d\omega \int dp^2\, \rho(p^2) \frac{\dot{R}_k(p^2)}{(\omega^2-k^2-\Sigma+i\epsilon)^n}=\frac{i(-1)^n\sqrt{\pi }\, \Gamma \left(n-\frac{1}{2}\right) }{\left(k^2+\Sigma\right)^{n-\frac{1}{2}}\Gamma (n)} \Omega(k)\,.
\end{align}
\medskip

If we set external momenta to zero in the equation for $\Gamma_k^{(2)}$, the left-hand side identifies with the derivative of the effective mass, and the flow equation is explicitly given by:
\begin{equation}
\dot{\Sigma}=\frac{i K}{2\pi} u_4 I_2 = -K u_4 \frac{\Omega(k)}{4 \left(k^2+\Sigma \right)^{3/2}}\,,
\end{equation}
for some numerical constant $K$. Generally, the explicit dependency on $k$ is canceled by a suitable rescaling of couplings, by $k$ to the power of their respective canonical dimension\footnote{This corresponds to the critical exponents around the Gaussian fixed point.}. With this rescaling, the flow equations form autonomous systems, and fixed-point solutions may exist. In the deep IR, for $k$ small enough, the Wigner distribution \eqref{distribution_p} can be approached by:
\begin{equation}
\rho(p^2)\approx \frac{\sqrt{p^2}}{\pi \sigma^{3/2}}
\end{equation}
which as pointed out before corresponds to the momenta distribution for square momentum $\vec{p}^2$ for standard field theory in dimension $3D$. Within this approximation, the factor $\Omega(k)$ reduces to a power law:
\begin{equation}
\Omega(k)\approx \frac{4 k^5}{3 \pi  \sigma ^{3/2}}\,. 
\end{equation}
Because $\mu_1$ is translated by $2\sigma$, it is natural to assume that $\mu_1$ scale as $k^2$ i.e. $\mu_1$ scales as eigenvalues under a global dilatation of the spectrum, according to \cite{lahoche4, Bradde, lahoche2022generalized}. We thus define the rescaled effective mass $\bar{\Sigma}$ as:
\begin{equation}
\bar{\Sigma}:=k^{-2} \Sigma\,,
\end{equation}
which reduces to $\bar{\mu}_1:=k^{-2} \mu_1$ at the order zero of the perturbation theory. With this rescaling the explicit dependency of the equation to $k$ is canceled, and we get, up to some numerical constant $K^\prime$:
\begin{equation}
\dot{\bar{\Sigma}}=-2\bar{\Sigma}- \frac{K^\prime \bar{u}_4 }{\left(1+\bar{\Sigma }\right)^{\frac{3}{2}}}\,,
\end{equation}
where $\bar{u}_4:=u_4$ i.e. the quartic coupling does not require to be rescaled; in the ordinary field theory language, this means that the quartic coupling is dimensionless. Now let us move to the flow equation for the quartic coupling itself. In the local approximation the flow equation for $u_4$ involves a term linear with respect to $u_6$, proportional to $I_2$, and a second term involving a product of two $\Gamma_k^{(4)}$ and the integral $I_3$:
\begin{equation}
\dot{{u}}_4=-K_1\, u_6  \frac{ \Omega(k)}{\left(k^2+\Sigma \right)^{3/2}}+K_2\, u_4^2  \frac{ \Omega(k)}{\left(k^2+\Sigma \right)^{5/2}}\,.
\end{equation}
This relation enforces the definition:
\begin{equation}
\bar{u}_6:=u_6\, k^2 \,,
\end{equation}
and the equation for the rescaled coupling $\bar{u}_4$ is:
\begin{equation}
\dot{\bar{u}}_4=- \frac{K_1^\prime\, \bar{u}_6 }{\left(1+\bar{\Sigma }\right)^{3/2}}+ \frac{K_2^\prime \, \bar{u}_4^2}{\left(1+\bar{\Sigma }\right)^{5/2}}\,.
\end{equation}
It is easy to check recursively that higher-order flow equations become autonomous with the rescaling:
\begin{equation}
\bar{u}_{2n}=u_{2n}\, k^{2n-4}\,.\label{rescaling1}
\end{equation}
Equation \eqref{rescaling1} is reminiscent of the power counting of  Euclidean field theory in dimension space $D=4$. This is not a surprise because as mentioned before $\rho(p^2)$ behaves as a $3D$ momentum distribution for $p$ small enough, and the integration over frequencies reduces the power $k$ of denominators by $1$. To summarize: 
\medskip

\noindent
\textbf{Remark. }\textit{In the deep IR, the renormalization group equations correspond to the ones of a non-local field theory in dimension $D=4$. In particular, only the quartic coupling is relevant in that limit.}
\medskip

We can attempt to probe the scaling behavior of the flow a little further into the UV regime. However, this will lead to some undesirable complications. The integral over $p$ would depend on the mass since the denominator of $G_k$ can no longer be factorized. We will still be able to define a rescaling, but this factor is no longer a simple power law (because Wigner's law \eqref{distribution_p} is not one), we will find dimensions that will depend not only on the scale but also on the mass $\Sigma$. However, the presence of this term is purely due to the method used in constructing this coarse-graining, and alternative methods can be considered to avoid this issue.
\cite{lahoche20241}. 
\medskip

To avoid this discussion and gain a little more insight into the scaling behavior of the couplings around the Gaussian point, let us focus on a theory in the vicinity of the critical point, i.e. $\mu_1=0$. We can then, at first order, neglect the $\Sigma$-dependencies in the denominator, and we find for example for the mass:
\begin{equation}
\dot{\Sigma}=\frac{i K}{2\pi}u_4 I_2 \approx -K u_4 \frac{\Omega^\prime(k)}{4 k^3}\,,
\end{equation}
where here, setting $\sigma=1$:
\begin{equation} 
\Omega^\prime(k):=k^3\int dp^2\, \frac{\rho(p^2) \dot{R}_k(p^2)}{(p^2+R_k(p^2))^{3/2}}=\frac{8 k^5}{3 \pi  \sqrt{4-k^2}}\,.
\end{equation}
As expected $\Omega^\prime(k)$ is not a simple power law. As before, it is always suitable to assume that $\Sigma$ scale as $k^2$ and in accordance with \cite{lahoche4, Bradde, lahoche2022generalized}, it is suitable to define the dimensionless coupling $\bar{u}_4$ as:
\begin{equation}
\bar{u}_4:=u_4\, \frac{\Omega^\prime(k)}{k^{5}}\,. 
\end{equation}
Recursively, and as we remains close enough to the IR regime, it is easy to check that $\bar{u}_{2n}$ must be defined as:
\begin{equation}
\bar{u}_{2n}=u_{2n}\, \frac{1}{k^2}\, \left(\frac{\Omega^\prime(k)}{k^3}\right)^{n-1}\,.\label{rescalingGaussian}
\end{equation}
Note that as we move away from the IR regime, some numerical factors forbids a simple power law. For instance, the contribution of order $u_4^2$ in the flow equation for the quartic coupling involves the additional factor:
\begin{equation}
R(k):= k^2 \Omega^{-1}(k) \int dp^2\, \frac{\rho(p^2) \dot{R}_k(p^2)}{(p^2+R_k(p^2))^{5/2}}\,,\label{Rfactor}
\end{equation}
but it is easy to check that this factor remains small enough in the IR regime. 
Numerical analysis confirm that these factors are indeed irrelevant until we reach the deep UV regime, and when this is the case we will systematically take them into account.\\

\begin{figure}
\begin{center}
\includegraphics[scale=0.7]{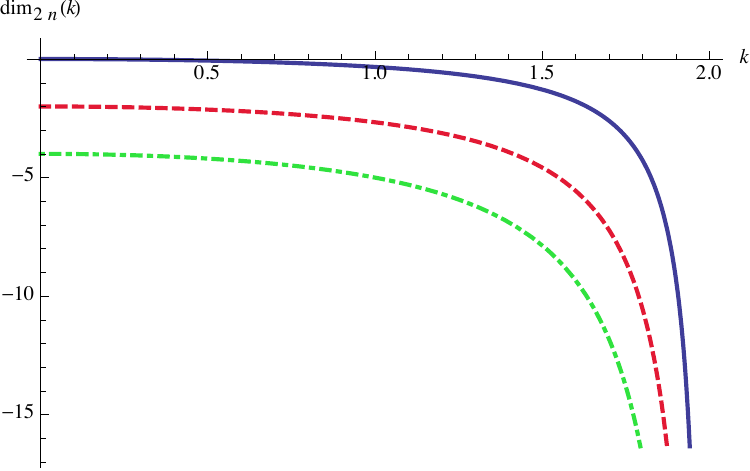}
\end{center}
\caption{Dependency on critical canonical dimensions ($\sigma=1$) for $n=2$ (solid blue curve), $n=3$ (dashed red curve) and $n=4$ (dashed-dotted green curve).}\label{figcanonical}
\end{figure}

Within this definition, the linear part of flow equations reads:
\begin{equation}
\dot{\bar{u}}_{2n}=-\mathrm{\dim}_{2n}(k) \bar{u}_{2n}+\cdots\,,
\end{equation}
where:
\begin{equation}
\boxed{\mathrm{\dim}_{2n}(k):= (n-1)\mathrm{\dim}_4(k)+2(2-n) \,,}
\end{equation}
and:
\begin{equation}
\boxed{\mathrm{\dim}_4(k):=\frac{d}{dt}\, \ln \,\left(k^5(\Omega^\prime)^{-1}(k)\right)=\frac{k^2}{k^2-4}\,.}
\end{equation}
For $k\to 0$, we recover the previous result: 
\begin{equation}
\mathrm{\dim}_{2n}(k)=(4-2 n)+\frac{k^2}{4} \left(1-n\right)+\mathcal{O}(k^3)\,,
\end{equation}
but this differs significantly from the $4D$ power counting in the deep UV (see Fig. \ref{figcanonical}). Note that $\mathrm{\dim}_{2}(k)=2$, as expected. 
\medskip

\noindent
\textbf{Remark.} \textit{The power counting is asymptotically the same as for a local Euclidean field theory in dimension $D=4$, despite the non-local nature of the interactions. This is linked to the specific nature of non-localities, particularly the fact that each loop only creates one face. This remains true for matrix theories \cite{lahoche20241}, but becomes wrong for tensor theories, for which the power counting is modified \cite{Lahochebeyond}.}
\medskip

To conclude, let us consider the case where the sextic coupling is non-local. In that case, the one-loop integral $I_2$ is no longer given by \eqref{definitionI2} but rather:
\begin{align}
\nonumber J_2(k)&:=\int d\omega \int dp^2\, \rho(p^2) \frac{\dot{R}_k(p^2)}{(\omega^2-p^2-R_k(p^2)+i\epsilon)^2}\, \delta(\omega)\\\nonumber
&= \int dp^2\, \rho(p^2) \frac{\dot{R}_k(p^2)}{(p^2+R_k(p^2)-i\epsilon)^2}\\
&= 16\pi \frac{4 \sin ^{-1}\left(\frac{k}{2}\right)-k \sqrt{4-k^2}}{4\pi^2  k^2}  \,.\label{definitionI22}
\end{align}
Asymptotically, we get $J_2(k) = 4k/3\pi + \mathcal{O}(k^3)$, and the asymptotic dimension for $u_6$ is $-1$ rather than $-2$. Explicitly, we find the dimension of the non-local coupling:
\begin{equation}
\boxed{\mathrm{dim}_{6,n.l}=\frac{k \left(k^2-8\right) \sqrt{4-k^2}+4 \left(8-3 k^2\right) \sin ^{-1}\left(\frac{k}{2}\right)}{\left(k^2-4\right) \left(k \sqrt{4-k^2}-4 \sin ^{-1}\left(\frac{k}{2}\right)\right)}\,.}
\end{equation}

\section{Perturbative RG flow}\label{sec6}

Due to the dependency of the canonical dimension on the scale $k$, the flow is more difficult to study compared to ordinary field theories. In particular, this dependence makes the notion of a global fixed point obsolete. Here, it is further emphasized, at the leading order, by the fact that the function $\beta$ of the sextic coupling does not vanish except for $u_6=0$ and because canonical dimension depends on the infrared cut-off $k$.

\subsection{Perturbative $\beta$-functions}

Let us move on to the computation of $\beta$ functions using perturbation theory. In this way, we assume that both $u_4$ and $\lambda$ are of order $\varepsilon$, and we focus on the leading order in the $\varepsilon$ and $1/N$ expansions. It is also useful to reintroduce temporarily the $\hbar$ parameter explicitly and to focus on the dominant order in $\hbar$. In standard field theory, the expansion corresponds to an expansion in the number of loops, but note that the effective sextic interaction being of order $\hbar^{-1}$ (see \eqref{classicalaveraged}), we will have to go to the order of two loops for these interactions (before setting again $\hbar=1$).
\medskip

To begin, let us write the leading order contributions to the 1PI $2n$-point functions $\Gamma^{(2n)}_k$ graphically, we have:

\begin{equation}
\Gamma^{(2)}_k= \omega^2-p^2_\mu-\mu_1 \,+\, \vcenter{\hbox{\includegraphics[scale=0.7]{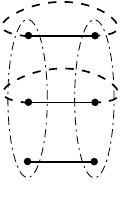}}}\,+\, \vcenter{\hbox{\includegraphics[scale=0.7]{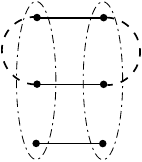}}}\,+\, \vcenter{\hbox{\includegraphics[scale=0.7]{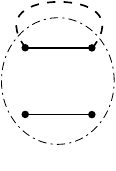}}}\,+\,\mathcal{O}(\hbar^2\varepsilon^2)
\end{equation}

\begin{align}
\nonumber &\Gamma^{(4)}_k= \vcenter{\hbox{\includegraphics[scale=0.7]{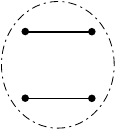}}}\,+\, \vcenter{\hbox{\includegraphics[scale=0.7]{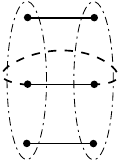}}}\,+\, \vcenter{\hbox{\includegraphics[scale=0.7]{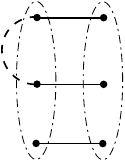}}}\,+\, \vcenter{\hbox{\includegraphics[scale=0.7]{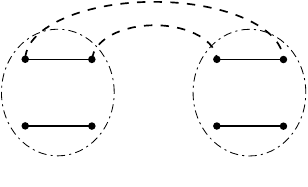}}}\,+\,\vcenter{\hbox{\includegraphics[scale=0.7]{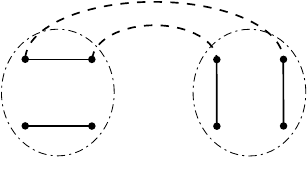}}}\\\nonumber
&\,+\, \vcenter{\hbox{\includegraphics[scale=0.7]{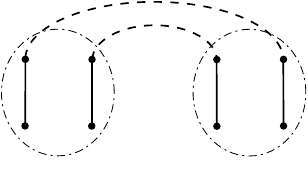}}}\,
+\,\vcenter{\hbox{\includegraphics[scale=0.7]{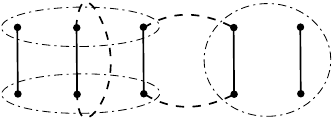}}}\\\nonumber
&\,+\,\vcenter{\hbox{\includegraphics[scale=0.7]{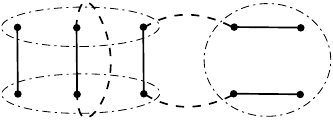}}}\,+\,\vcenter{\hbox{\includegraphics[scale=0.7]{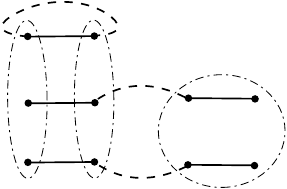}}}\\
&\,+\,\vcenter{\hbox{\includegraphics[scale=0.7]{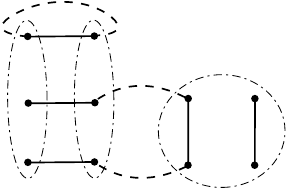}}}\,+\,\mathcal{O}(\hbar^2\varepsilon^2,\varepsilon^3)
\end{align}

\begin{align}
\nonumber &\Gamma^{(6)}_k=\vcenter{\hbox{\includegraphics[scale=0.7]{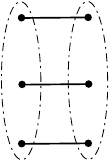}}}\,+\,\mathcal{O}(\varepsilon^3)\,.
\end{align}

In this paper, we assume $N$ is large enough, and contributions that do not maximize the number of faces must be neglected. However, it is necessary to take into account the dependence on $N$ intrinsic to the interaction, required by the existence of a thermodynamic limit when $N\to \infty$. We get:

\begin{equation}
\Gamma^{(2)}_k= \omega^2-p^2_\mu-\mu_1 \,+\, \vcenter{\hbox{\includegraphics[scale=0.7]{Gamma2Six.pdf}}}\,+\, \vcenter{\hbox{\includegraphics[scale=0.7]{Gamma2Four1.pdf}}}\,+\,\mathcal{O}(\hbar^2\varepsilon^2)
\end{equation}

\begin{align}
\nonumber \Gamma^{(4)}_k&= \vcenter{\hbox{\includegraphics[scale=0.7]{Gamma40.pdf}}}\,+\, \vcenter{\hbox{\includegraphics[scale=0.7]{Gamma4Six.pdf}}}\,+\, \vcenter{\hbox{\includegraphics[scale=0.7]{Gamma4Four2.pdf}}}
\,+\, \vcenter{\hbox{\includegraphics[scale=0.7]{Gamma4SixFour1.pdf}}}\,+\,\mathcal{O}(\hbar^2\varepsilon^2,\varepsilon^3)
\end{align}

\begin{align}
\nonumber &\Gamma^{(6)}_k=\vcenter{\hbox{\includegraphics[scale=0.7]{Gamma6Six1.pdf}}}\,+\,\mathcal{O}(\varepsilon^3,N^{-2})\,.
\end{align}

We want to construct flow equations for the effective couplings, and it should then be noted that certain effects must be summed. For example, the contributions for $\Gamma_k^{(2)}$ reveal, that at the dominant order in $\hbar$, the $4$-point function:
\begin{equation}
\Gamma^{(2)}_k= \omega^2-p^2_\mu-\mu_1 \,+\, \underbrace{\vcenter{\hbox{\includegraphics[scale=0.5]{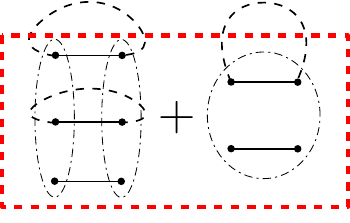}}}}_{\approx\, \Gamma_k^{(4)}}\,+\,\mathcal{O}(\hbar^2\varepsilon^2)\,.
\end{equation}
As we have seen in the previous section, the flow equation for $\Gamma_k^{(2)}$ involves only $\Gamma_k^{(4)}$ and $\Gamma_k^{(2)}$ in the symmetric phase (i.e. assuming the expansion around the zero classical field makes sense). Thus, the flow equation sums up these effects automatically, and we will therefore use it to construct our approximations. The simpler equation concern the sextic coupling $u_6$, defined by the condition \eqref{localsectorGamma2n}. Because $\Gamma_k^{(6)}$ involves only the initial coupling, the flow equation in the vicinity of the Gaussian fixed point reduces to its dimensional part:
\begin{equation}
\boxed{\dot{\bar{u}}_6=-\mathrm{dim}_{6,n.l}(k)\, {\bar{u}}_6\,,}
\end{equation}
where $\mathrm{dim}_{6,n.l}(k) = -1+\mathcal{O}(k^2)$. For the rest, part of the work has already been done in the previous section, and we mainly have to calculate the constants $K$, $K_1$, and $K_2$. We get\footnote{Recall that within the rescaling \eqref{rescalingGaussian}, some numerical factors are included in the definition of the dimensionless couplings.} for $\sigma=1$:
\begin{equation}
K=\frac{1}{12}\,,\quad K_1=\frac{1}{80}\,,\quad K_2=\frac{1}{16}\,.
\end{equation}
At the leading order around the Gaussian fixed point, assuming the sextic coupling to be non-local, we get:
\begin{equation}
\boxed{\dot{\bar{\Sigma}}\approx-2\bar{\Sigma}-\frac{\bar{u}_4}{48}+\frac{\bar{u}_4}{32} \bar{\Sigma}\,,}
\end{equation}
\begin{equation}
\boxed{\dot{\bar{u}}_4\approx -\mathrm{dim}_4(k)\, {\bar{u}}_4+ \frac{\bar{u}_6 }{40\pi}- \frac{\bar{u}_6 }{20\pi}\bar{\Sigma}+ \frac{\bar{u}_4^2}{16}\,.}
\end{equation}
\medskip

\noindent
\begin{remark}
Let us recall that in this equations we neglected some numerical factors, which are numerically of order $1$ for $k$ small enough. For instance, the contribution proportional to $\bar{u}_4^2$ involves the factor $R(k)$ discussed in \eqref{Rfactor}, and we have for instance $R(0.4)\approx 0.98$, $R(1)\approx 0.89$. Taking into account these factor does not change our conclusions qualitatively. 
\end{remark}

\noindent
\begin{remark}We derived the flow equation in the real-time formalism. We could also consider the imaginary time formalism, making contact with a thermal interpretation (see also Appendix \ref{AppC}). The long-time equations we constructed are then equivalent to the low-temperature limit in the Euclidean formalism, up to a Wick rotation.
\end{remark}

\subsection{Asymptotic infrared flow}

In this subsection, we investigate the behavior of the RG flow in the deep IR limit, assuming $k/4\sigma \ll 1$.
\medskip

\begin{figure}
\begin{center}
\includegraphics[scale=0.6]{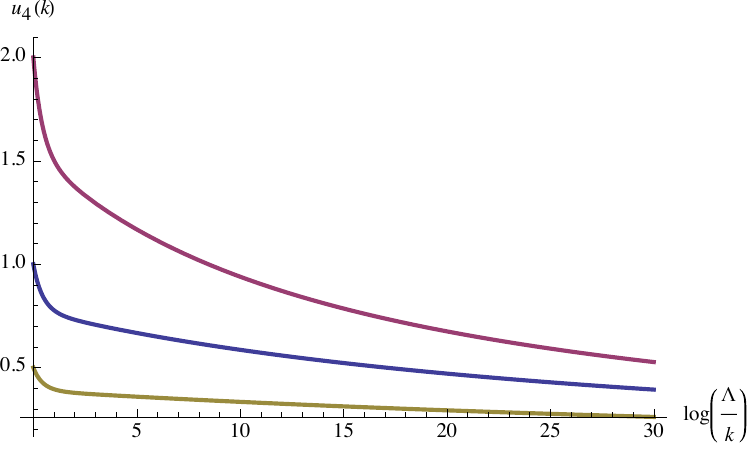}
 \includegraphics[scale=0.6]{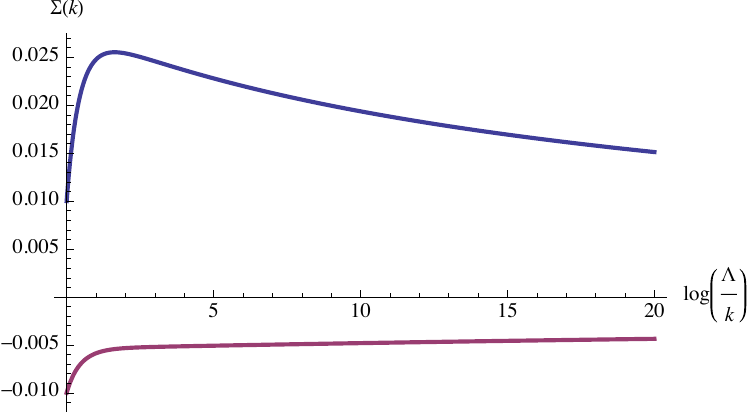}
\end{center}
\caption{Behavior of the effective coupling $u_4(k)$ and of the effective mass $\Sigma$ in the vicinity of the Gaussian fixed point.}\label{figflowu60}
\end{figure}

\begin{figure}
\begin{center}
\includegraphics[scale=1]{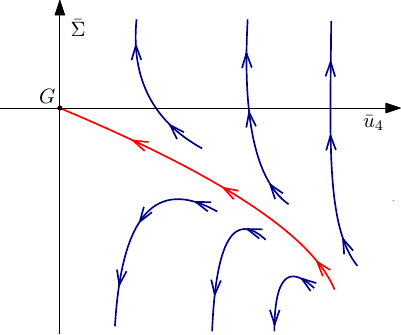}
\end{center}
\caption{The asymptotic behavior of the RG flow for $p=0$.}\label{figflowu602}
\end{figure}

\paragraph{Case $p=0$, mean field theory.} Setting $u_6=0$ and expanding quartic coupling's dimension at the leading order $k^2$, we get $\mathrm{dim}_4=-k^2/4+\mathcal{O}(k^4)$, and the flow equation reduces to:
\begin{align*}
&\dot{\bar{\Sigma}}=-2\bar{\Sigma}-\frac{\bar{u}_4}{48}+\frac{\bar{u}_4}{32}\bar{\Sigma}\\
&\dot{\bar{u}}_4= \frac{k^2}{4}\, {\bar{u}}_4+ \frac{\bar{u}_4^2}{24}\,.
\end{align*}
Note that no replicas are required in that case. The behavior of the flow is reminiscent of the behavior of the standard Euclidean scalar $\phi^4$ field theory in dimension $D=4+\epsilon$, with the difference that the parameter $\epsilon$ is now fixed by the nontrivial spectrum of the effective kinetics: $\epsilon\approx -k^2/4$.  Figs. \ref{figflowu60} and \ref{figflowu602} shows the behavior of mass and quartic coupling in the vicinity of the Gaussian fixed point, a picture which is again very reminiscent of ordinary $\phi^4$ field theory in dimension $D>4$: We find two regions, separated by a boundary resembling a critical line (the red line on Fig.  \ref{figflowu602}), one toward negative mass region (above the critical line), and one toward positive mass region (below the critical line). 
Then, asymptotically, the theory is expected to be well approximated by the Landau approximation: 
\begin{equation}
\Gamma[M] \approx \frac{\mu_1}{2}\, M^2+ \frac{\mu_2}{4!}\,M^4\,,
\end{equation}
where here $M$ is the $(p=0)$-component of the classical field $M_\mu \equiv M(p_\mu)$:
\begin{equation}
M_\mu:= \frac{\partial W_k}{\partial L_\mu}\,,
\end{equation}
where $M_\mu:= \sum_i M_i u_i^{(\mu)}$, with $u_i^{(\mu)}$ being the (normalized) eigenvector for $K_{ij}$ corresponding to the eigenvalue $\mu$. We then discover that, asymptotically, a phase transition occurs for the component $p_\mu=0$, as soon as $\mu_1<0$, for the value:
\begin{equation}
M^2=-\frac{6 \mu_1}{\mu_2}\,.
\end{equation}
In that case, we can go beyond this approximation by noticing that, for large $N$, the mass $\Sigma$ can be fixed by a closed equation (see \cite{lahoche3,ZinnJustinBook2} and Section \ref{sectionPert})
\begin{align}
\nonumber \Sigma-\mu_1&= -\frac{i \mu_2}{6} \, \int \frac{d\omega}{2\pi} \int \rho(p^2) dp^2\, \frac{1}{\omega^2-p^2-\Sigma+i\epsilon}\\
&=- \frac{\mu_2}{12} \, \int \rho(p^2) \frac{dp^2}{\sqrt{p^2+\Sigma}}\, \,,
\end{align}
where we set $R_k=0$. The integral can be computed by formally summing up the expansion in power of $p^2$, for $\Sigma >0$, we get a complicated closed equation involving some hypergeometric Gauss function:
\begin{equation}
\Sigma = \mu_1\,- \frac{\mu_2}{12}\, \frac{_2F_1\left(\frac{1}{2},\frac{3}{2};3;-\frac{4 \sigma }{\Sigma}\right)}{\sqrt{\Sigma}}\,.
\end{equation}
Close to the phase transition, the effective mass $\Sigma$ has to be arbitrarily close to zero, and we get the one-loop correction:
\begin{equation}
\Sigma \approx  \mu_1\,-\,  \frac{2\mu_2}{9 \pi \sqrt{\sigma }}\,.
\end{equation}
Because this is the physical mass $\Sigma$ and not the bare mass $\mu_1$ which truly controls the transition ($\Gamma$ is the asymptotic Gibbs potential), we get:
\begin{equation}
M^2=-\frac{6 (\mu_1-\mu_{1c})}{\mu_2}\,,\label{formulacompmacroBIS}
\end{equation}
where the critical value $\mu_{1c}$ is:
\begin{equation}
\mu_{1c}:= \frac{2\mu_2}{9 \pi \sqrt{\sigma }}>0\,.
\end{equation}
Equation \eqref{formulacompmacroBIS} should be compared with the exact result \eqref{formulacompmacro} of Appendix \ref{AppC}, to which it agrees qualitatively. The quantitative difference (a factor $2$) between the two approaches is expected to be produced by the fact that we expanded around different vacuums; in the symmetric phase in this section and around the true vacuum in Appendix \ref{AppC}, and the same phenomenon has been noticed in \cite{lahoche3} for the computation of the critical temperature in the classical case. 
\medskip

\noindent
\paragraph{Case $p=3$.} 

Now let us consider the case $p=3$, including the non-local interaction arising due to the averaging of the disorder. As stated before, there are no global fixed-points because of the dependence of canonical dimension on the scale $k$. However, there should be fixed trajectories, along which the $\beta$ functions vanish. In that case, we find such a trajectory:
\begin{equation}
\bar{\Sigma}_*(k)=\frac{u_6}{2 (480 \pi k+u_6)}\,,\quad \bar{u}_{4_*}(k)=-\frac{48u_6}{480 \pi k+u_6}\,.
\end{equation}
For $u_6$ small enough, the perturbation theory holds until $480 \pi k \gg u_6$. In the deep IR, the fixed point is however inevitably pushed outside the perturbative domain, and our approach then ceases to be valid. Furthermore, the coupling constant $\bar{u}_4$ has the wrong sign, and we have reason to doubt the reliability of such a fixed point.
\medskip

\begin{figure}
\begin{center}
\includegraphics[scale=0.55]{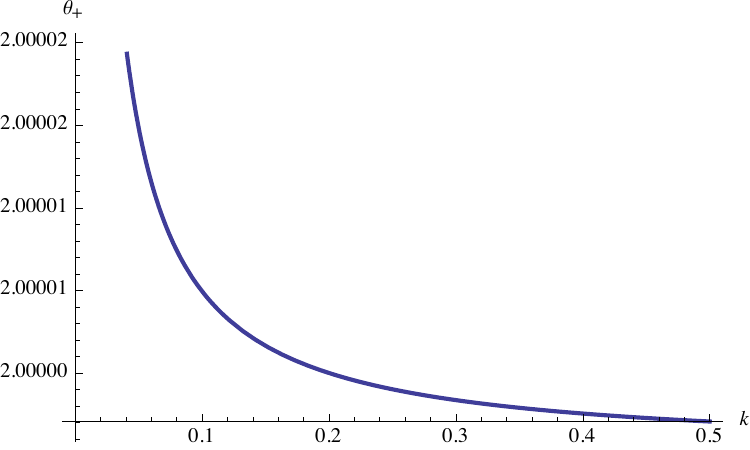}\qquad 
\includegraphics[scale=0.55]{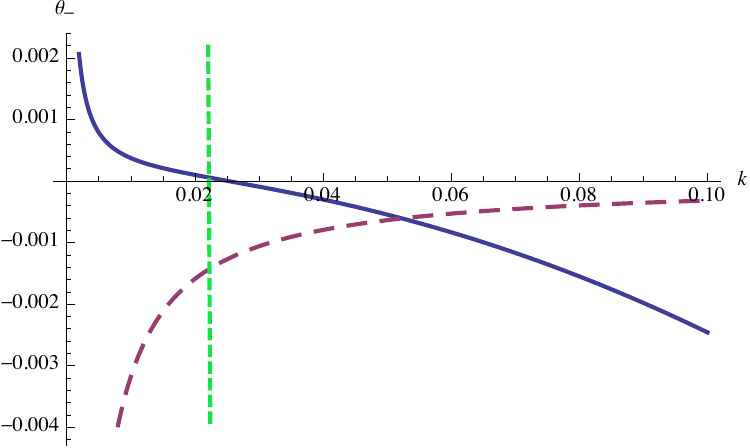}
\end{center}
\caption{Behavior of the critical exponents $(\theta_+,\theta_-)$ (solid blue lines) which are the opposite of the eigenvalues of the stability matrix along the fixed trajectory. In the figure on the bottom, the purple dashed line corresponds to the behavior of $\bar{u}_{4_*}(k)$ and the green dashed line is for $k=k_d$.}\label{Figexpo}
\end{figure}

It can however be expected that this is because we limit ourselves to the lowest order. Also, the sextic interactions, not taken into account by our truncation, restore the stability of the theory -- as is the case usually in the description of first-order phase transitions \cite{Nagy}.  Fig. \ref{Figexpo} shows the behavior of critical exponents for $u_6=0.001$ (we recall that, within our approximation scheme, $u_6$ does not renormalize). 
\medskip

When $k$ is still small enough, our approximation is still valid, and we then understand that an unexpected effect happens. At a sufficiently large scale, the fixed point has the properties of a Wilson-Fisher fixed point\footnote{i.e. one relevant and one irrelevant direction.}, but this behavior suddenly changes at a smaller scale $k_d$, and the fixed point then transforms into a \textit{UV attractor}. This mechanism evokes the appearance of a discontinuity in the flow, generally marked by a singularity, which can easily be evaluated numerically. In Fig. \ref{Figureflowbehavior}, we show the behavior of the RG flow in the vicinity of the non-Gaussian fixed point and for different initial conditions, with and without disorder. It seems that the presence of disorder induces the appearance of singularities “on a finite scale”, a phenomenon already observed classically \cite{Tarjus,Lahoche2022functional}. These divergences are expected to modify the power count (specifically, the power count of $N$) of the theory, and cause perturbatively irrelevant effects (like the correlations between replicas discussed in section \ref{secWard}) to become significantly amplified. However, it is difficult to describe this kind of divergence in a perturbative formalism (essentially because the quartic coupling becomes very large a long time after the divergence), and their rigorous study requires the use of a more advanced formalism. In this article, we will limit ourselves to vertex expansion, which happens to be unique with regard to the specific non-localities of the interactions. 
\medskip

 To summarize, we conclude that the presence of a fairly significant disorder in the IR influences the flow and induces singularities in the vicinity of the Gaussian point. Furthermore, these singularities have the particularity of sending the mass into the negative region, which corresponds to the region where the global symmetry $O(N)$ is spontaneously broken (see Appendix \ref{AppC}).

\begin{subequations}
\begin{figure}
\begin{center}
\includegraphics[scale=0.5]{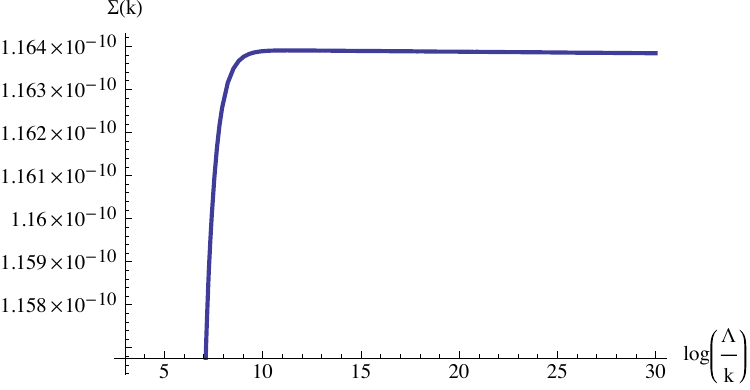}\qquad \includegraphics[scale=0.5]{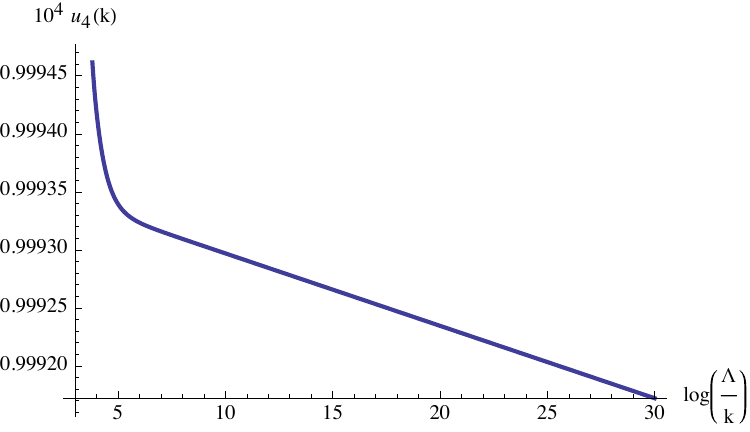}
\includegraphics[scale=0.5]{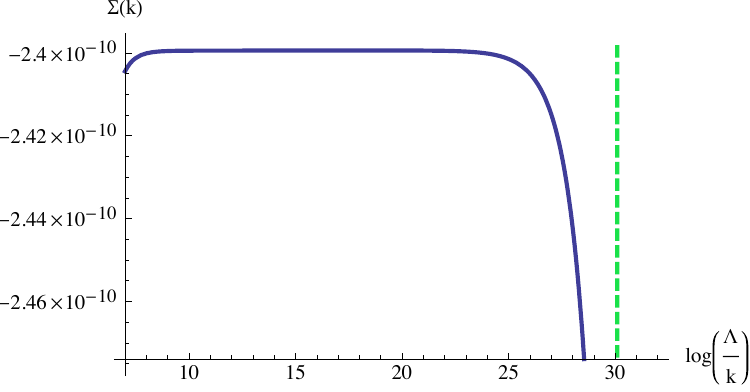}\qquad \includegraphics[scale=0.5]{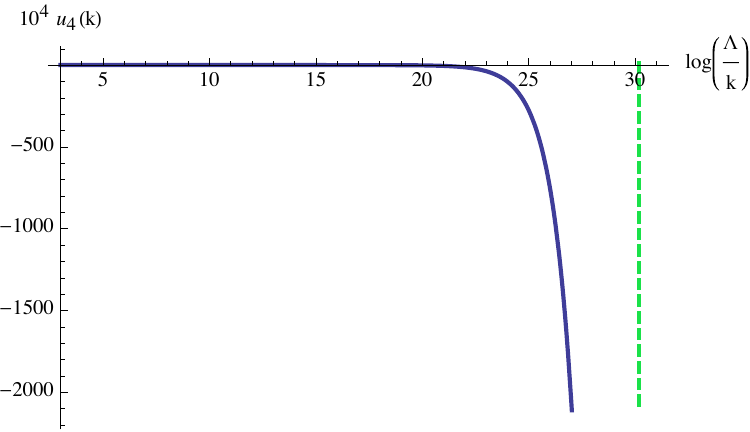}
\end{center}
\caption{On the top, behavior of the RG flow for $(\bar{\Sigma}(k_0),\bar{u}_4(k_0),u_6)=(-10^{-5},10^{-3},0)$. On the bottom, the same behavior for $u_6=0.1$.}\label{Figureflowbehavior}
\end{figure}
\end{subequations}

\subsection{Global flow}

\begin{figure}
\begin{center}
\includegraphics[scale=0.6]{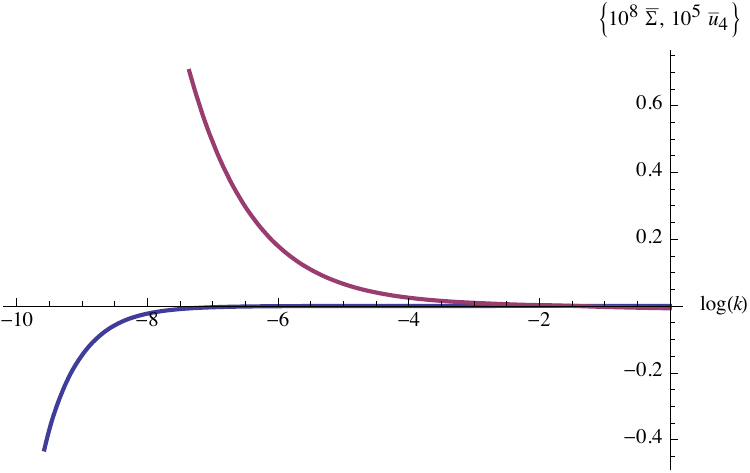}\,
\includegraphics[scale=0.6]{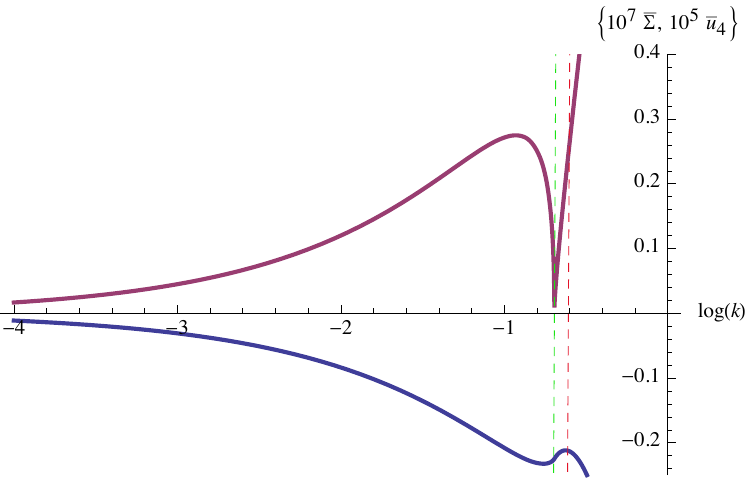}
\end{center}
\caption{On the top, global behavior of the RG flow in the vicinity of the Gaussian fixed points for $(\bar{\Sigma}(k_0),\bar{u}_4(k_0),\bar{u}_6)=(-10^{-7},10^{-7},0)$, $k_0=e^{-10}$. On the bottom, the same behavior for $\bar{u}_6(k_0)=10^{-7}$. The red curve is for $\bar{u}_4$ and blue curve for $\bar{\Sigma}$.}\label{Figureflowbehaviorfull}
\end{figure}

Now, let us consider the behavior of the RG flow globally. For perturbation theory, we computed the canonical dimensions in section \ref{sectionscaling}, and we expect the contributions that are irrelevant in the deep IR to play a significant when sufficiently far from the IR regime. Hence, even if we are arbitrarily close to the Gaussian point in the deep IR, $\bar{u}_6(k)$ grows with $k$, and high energy modes feel a large effective disorder, which significantly influences the RG trajectories. In Fig. \ref{Figureflowbehaviorfull}, we show the RG flow behavior from IR to UV, starting in the vicinity of the Gaussian fixed point. Once again, we show that as $u_6(k_0) \gtrsim 10^{-10}$, the RG flow develops a finite scale singularity (without symmetry breaking), which takes the form of a \textit{cusp non-analytic point} for $k_c^2 \approx 0.3$, reminiscent of the results in \cite{Tarjus} for random field Ising model. Note that this scale is numerically essentially independent of the initial conditions, as we remain close to the Gaussian regime. Furthermore, the cusp for mass is always located later than the cusp for couplings, which is the true origin of non-analyticity. 
\medskip

This non-analytic point furthermore becomes a true numerical singularity as soon as $u_6(k_0) \gtrsim 10^{-4}$, and below this point, sharp singularities appear along the flow. 
\medskip

Although these effects are difficult to discuss within the framework of perturbation theory, our interpretation of this local non-analyticity is different from that of the ``sharp" singularity, appearing at strong disorder. Non-analyticity prohibits the continuation of Ward identities, and therefore of symmetries, whether internal symmetries such as global invariance $O(N)$ or external symmetries such as invariance by translation over time. Conversely, a real singularity induces on a domain a strengthening of certain perturbatively neglected interactions and corresponding to an instability of the truncation.

\section{Nonperturbative vertex expansion}\label{sec7}

Within perturbation theory near the Gaussian fixed point, we see that the finite scale singularities seem to appear in the RG flow, meaning that long-range physics is separated from UV scales. One can suspect that these divergences arise because some couplings that do not appear in the perturbation theory could be generated by the quantum dynamics\footnote{Indeed, our parametrization of the truncated theory space is inspired by perturbation theory in the sense that expected couplings are those which can be generated perturbatively.}. Note that this is already exactly the interpretation that we give to this singularity of the flow for the simple Ising model: The singularity then simply means that we have chosen the “bad” vacuum. Moreover, such singular behavior has already been observed in subsequent work concerning glassy systems \cite{Tarjus,Lahoche2023functional,Lahoche2022functional}, but also concerning quantum chromodynamics at the confinement scale \cite{QCD,QCD2} or reaction-diffusion processes \cite{Delamotte2}. In this section, we will discuss these singularities in a nonperturbative formalism, focusing on standard vertex and derivative expansion. 
\medskip

\subsection{Vertex and derivative expansions}

Usually, the idea underlying vertex and derivative expansion is to expand the full EAA $\Gamma_k$ in the power of local vertices \eqref{localsectorGamma2n}, and to keep only the first term in the expansion of the $2$-point vertex in the power of the external momenta. For Euclidean scalar field theory in dimension $D$, we have for instance:
\begin{align}
 \Gamma_k[M]&=\frac{1}{2} \int d^D\vec{x} M(\vec{x}\,) (-Z(k)\partial_{\vec{x}}^2+\mu_1(k)) M(\vec{x}\,)+\sum_{n=2}^\infty\, \frac{u_{2n}(k)}{(2n)!}\, \int d^D\vec{x}\, M^{2n}(\vec{x}\,)\,,
\end{align}
where $M$ is the classical field, and $Z(k)$ is a constant called \textit{field strength renormalization}. In our case, however, we have to take into account the presence of non-local vertices in the original action. Formally, we then expect that $\Gamma_k[M]$ has local, bi-local, tri-local components, and so on, which are all perturbatively generated from the original action. However, the large $N$ limit drastically reduces the proliferation of non-local components, because the free propagator is diagonal. Within this limit, the sectors split, and non-local interactions do not renormalize to the dominant order. We will therefore pose as ansatz:
\begin{align}
\nonumber \Gamma_k[M]&=\frac{1}{2}\int dt \sum_{\mu,\alpha} M_{\mu,\alpha}(t)(-\partial_t^2-p_\mu^2-\mu_1)M_{\mu,\alpha}(t)\\\nonumber
&+\sum_{n=2}^\infty\int dt \sum_{\mu,\alpha} \frac{(2\pi)^{n-1}u_{2n}}{(2n)!N^{n-1}}\, \bigg(\sum_\mu M_{\mu,\alpha}^2(t) \bigg)^n\\
&+\frac{(2\pi)\tilde{u}_6}{6!N^2}\int dt dt^\prime \sum_{\alpha,\beta}\, \bigg(\sum_\mu M_{\mu,\alpha}(t) M_{\mu,\beta}(t^\prime) \bigg)^3\,.\label{truncationGamma}
\end{align}
Note that we set $Z(k)=1$, and it can be easily checked from perturbation theory that it is a consistent approximation in the large $N$ limit \cite{Lahoche2022functional}. Note that we do not renormalize the coefficient in front of $p^2_\mu$ in the propagator. This is justified for non-localities characterized by a global $O(N)$ symmetry, but would break down for theories whose global symmetry group is copies of $O(N)$ or $U(N)$ (see \cite{lahoche20241,Lahochebeyond,CarrozzaReview}). 
\medskip

We are aiming to compute the flow equation in the symmetric phase (assuming the vacuum ‘‘zero" good sufficiently well-behaved), and in a deep IR regime, where the power counting reduces to that of a Euclidean field theory in dimension 4. This study then ignores the specifics of power counting, in particular, regarding the intrinsic dependency of the canonical dimension on the scale; we will discuss these interesting aspects in a forthcoming work. 
\medskip

\begin{figure}
\begin{center}
\includegraphics[scale=0.8]{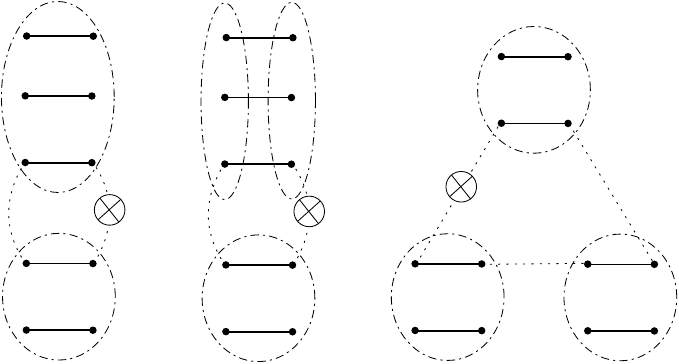}
\end{center}
\caption{Relevant contribution for the computation of $\dot{\bar{u}}_6$.}\label{figurephi6}
\end{figure}

The resulting flow equations can be obtained from \eqref{Wett}, taking the successive derivative with respect to the classical field and imposing, at the end on both sides the truncation \eqref{truncationGamma} for vanishing mean field. The procedure is standard \cite{Delamotte_2012, Lahochebeyond}, and we find for example, for a sextic truncation:
\begin{equation}
\boxed{\dot{\bar{\Sigma}}=-2\bar{\Sigma}-\frac{\bar{u}_4}{36\pi}\frac{1}{(1+\bar{\Sigma})^{\frac{3}{2}}}\,,}
\end{equation}
\begin{equation}
\boxed{\dot{\bar{u}}_4= \frac{\bar{\tilde{u}}_6 }{30\pi^2}\frac{1}{(1+\bar{\Sigma})^2}-\frac{\bar{u}_6 }{60\pi}\frac{1}{(1+\bar{\Sigma})^{\frac{3}{2}}}+ \frac{\bar{u}_4^2}{12\pi} \frac{1}{(1+\bar{\Sigma})^{\frac{5}{2}}}\,.}
\end{equation}
\begin{equation}
\boxed{\dot{\bar{{u}}}_6=2\, {\bar{{u}}}_6+\frac{72}{5\pi} \frac{\bar{u}_4\bar{u}_6}{(1+\bar{\Sigma})^{\frac{5}{2}}}+\frac{4}{5\pi^2} \frac{\bar{u}_4\bar{\tilde{u}}_6}{(1+\bar{\Sigma})^{3}}-\frac{5}{18\pi} \frac{\bar{u}_4^3}{(1+\bar{\Sigma})^{\frac{7}{2}}}\,,}
\end{equation}
\begin{equation}
\boxed{\dot{\bar{\tilde{u}}}_6=\, {\bar{\tilde{u}}}_6\,,}
\end{equation}
where relevant contributions in the large $N$ limit for the computation of $\dot{\bar{u}}_6$ is pictured on Fig. \ref{figurephi6}. 
\medskip

\subsection{Asymptotic infrared flow}

Initially, let us focus on the asymptotic fixed points in the deep IR. Setting $\bar{\tilde{u}}_6=0$ in this limit, no fixed point is found for the quartic truncation (i.e. setting $\bar{u}_6=0$ into the ansatz \eqref{truncationGamma}). For a sextic truncation, however, we get a single IR, reliable asymptotic fixed point for the values:
\begin{equation}
\bar{\Sigma} \approx 0.002\,,\quad \bar{u}_4\approx -0.44\,,\quad \bar{u}_6\approx 0.96\,,
\end{equation}
with critical exponents:
\begin{equation}
\theta_1\approx 2.0\,,\quad \theta_2\approx 0.02-0.15 i\,,\quad \theta_3\approx 0.02+0.15 i\,.
\end{equation}
This fixed point is IR-relevant and reminiscent of the fixed point that perturbation theory affected the flow of the perturbation theory. Furthermore, it appears to survive higher-order truncations and looks sufficiently reliable. 
\medskip

Let us investigate the behavior of the RG flow, setting $\tilde{u}_6=0$ to begin. In the positive region, i.e. where all couplings are positives, trajectories converge generally fast enough, as Fig \ref{figconvergence} show. This behavior is exactly what we expect from the Gaussian power counting, and the flow projected along the asymptotic just-renormalizable theory in the deep IR. We recover again the same kind of phenomenon in the negative region, as we remain sufficiently close to the Gaussian fixed point. However, farther away, we recover the finite scale singularity that we discussed in the previous section, and the flow is repelled fast far from the Gaussian region due to the repulsive eigendirections, as illustrated in the Fig. \ref{figdivergence}. Note that far enough from the interacting fixed point, the convergence seems to be recovered, and the finite scale singularities concern essentially the region between the Gaussian fixed point and the interacting one. This happens also at the boundary, along the eigendirection with critical exponent $\theta_1$. We can see numerically that trajectory is repelled from the mass axis and converges for a mass close, but different from the singular value\footnote{Once again, this singularity arises essentially because the symmetric phase assumption breaks down, and the zero field vacuum becomes a bad vacuum below this limit. Note furthermore that Litim's regulator corresponds to the larger extension of the validity domain of the symmetric phase.}  $\bar{\Sigma}=-1$ (Fig. \ref{conveigendirection}). Note that singularities are always related to the fact that the flow reaches the singular value $\bar{\Sigma}=-1$, forbidden by the choice of the truncation in the symmetric phase.
\medskip

In the region where the flow becomes singular, the mass reaches the negative region, and it is suitable to call it a ‘‘broken region". As Fig. \ref{figdivergencepositive} shows (the initial values are the same as in Fig. \ref{figconvergence}, but $\bar{\tilde{u}}_6(k_0)=1$), adding a disorder (or looking for the RG flow not exactly in the deep IR) modifies the behavior of the RG flow and extends the broken region in the positive region. As $\tilde{u}_6\neq 0$, the interpretation of the finite scale divergences is different from the ordinary broken symmetry mechanism. Indeed, equation \eqref{Qab} shows that $Q_{\alpha\beta}$, measuring correlation between replica and the failure of the local $O(N)$ symmetry in the Ward identity \eqref{localWard} becomes singular as well (Fig.\ref {Qabehavior}). The singularity is physically empty, but as the mass $\bar{\Sigma}$ reach the value $-1$, there is a small window in which $q_{\alpha\beta}$ becomes as relevant as the contributions that survive from the large $N$ power counting. In other words, we expect another phenomenon to add to the symmetry breaking as $\bar{\Sigma}\to -1$: correlations between replicas become large, and we expect the nature of the transition (favored by the disorder) to be different from ordinary phase condensation. At this stage, we conjecture that the physical meaning is that quantum tunneling is ‘‘frozen" in the broken region, and this freezing appears on a timescale smaller than the condensation phenomenon, in agreement with what we found within perturbation theory. However, it would still be necessary to verify that these singularities are not just a simple artifact of the truncation. In other words, for the moment our formalism cannot differentiate between a singularity coming from a development around a ‘‘bad" vacuum, and a truly physical singularity\footnote{This point in particular has been proved in \cite{Lahoche2022functional}, in the classical equilibrium dynamics framework.}, associated with the absence of certain interactions.
\medskip

\begin{subequations}
\begin{figure}
\begin{center}
\includegraphics[scale=0.4]{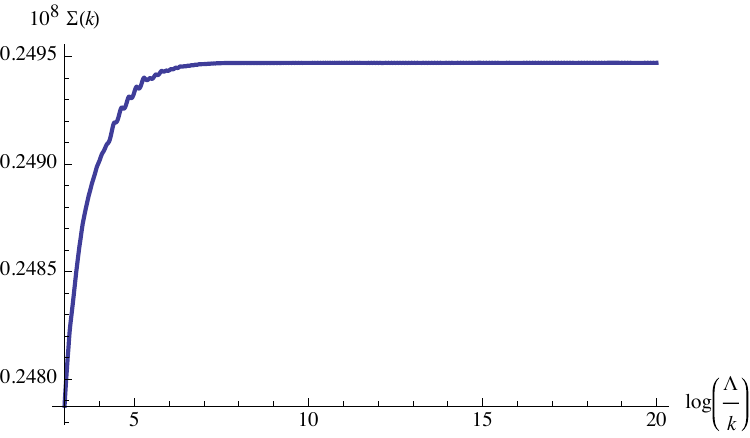}
\includegraphics[scale=0.4]{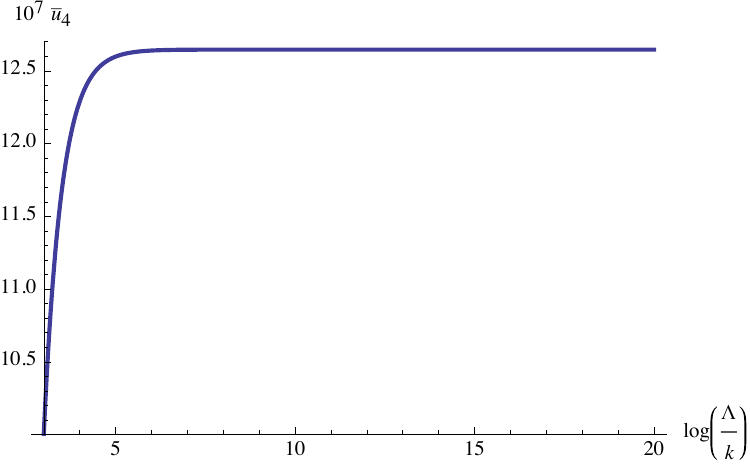}
\includegraphics[scale=0.4]{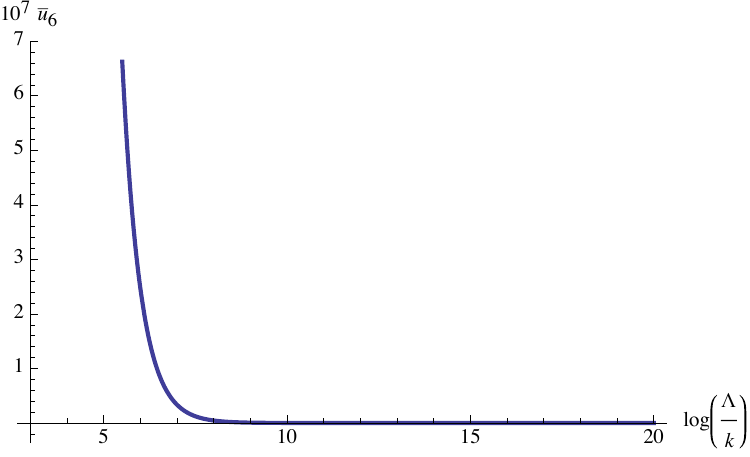}
\end{center}
\caption{Typical behavior of the RG flow in the positive region, near the Gaussian fixed point. Initial conditions for $k_0=e^{-3}$ are for $(\bar{\Sigma},\bar{u}_4,\bar{u}_6)=(10^{-5},10^{-5},10^{-3})$. }\label{figconvergence}
\end{figure}
\end{subequations}

Finally, let us note that  in the deep IR regime, the only singular directions of the flow are those which take it in the direction of the negative masses, in the region where the symmetry is broken. This differs from the results obtained for example in \cite{Lahoche2022functional}, and seems to mean that for the asymptotic flow, the glassy and condensed phases are superposed. 
\medskip

\begin{figure}
\begin{center}
\includegraphics[scale=0.5]{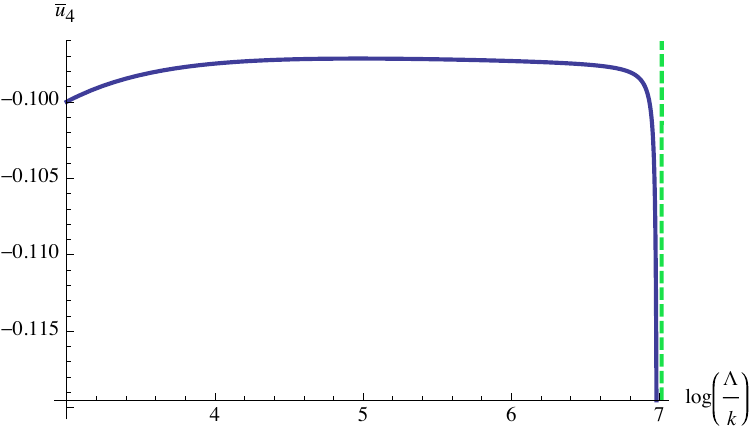}\qquad 
\includegraphics[scale=0.5]{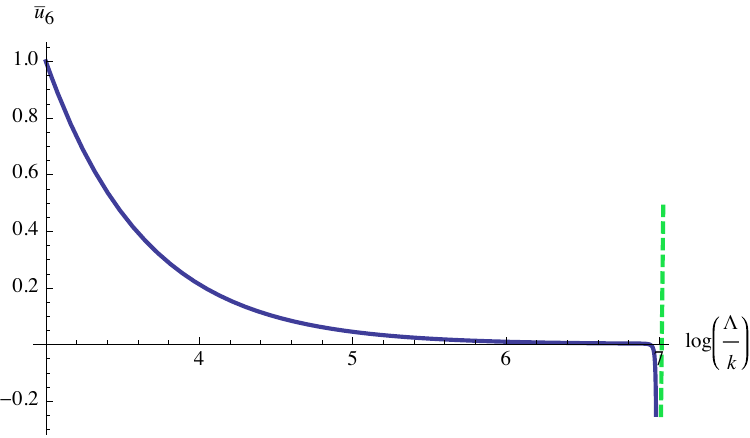}
\end{center}
\caption{Typical behavior of the RG flow in the negative region, far enough the Gaussian fixed point. Initial conditions for $k_0=e^{-3}$ are for $(\bar{\Sigma},\bar{u}_4,\bar{u}_6)=(10^{-4},-0.1,1)$ and $\tilde{\bar{u}}_6(k_0)=0$.}\label{figdivergence}
\end{figure}

\begin{figure}
\begin{center}
\includegraphics[scale=0.5]{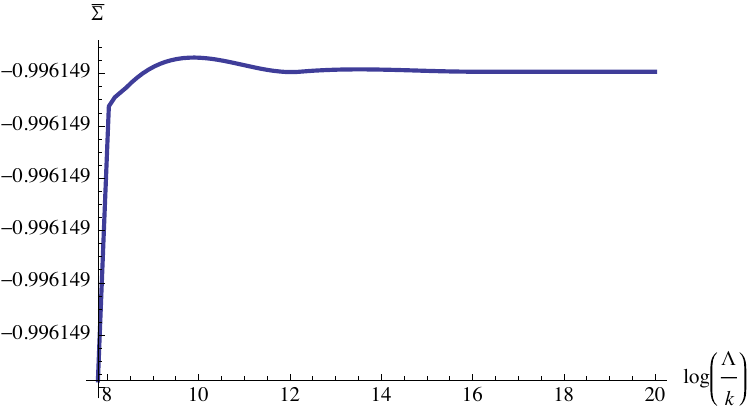}\qquad \includegraphics[scale=0.5]{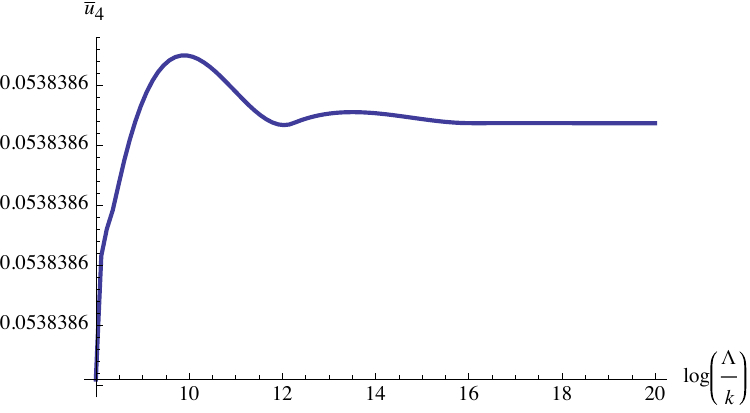}
\end{center}
\caption{Behavior of the RG flow from a small displacement along the mass axis around the interacting fixed point.}\label{conveigendirection}
\end{figure}

\begin{figure}
\begin{center}
\includegraphics[scale=0.5]{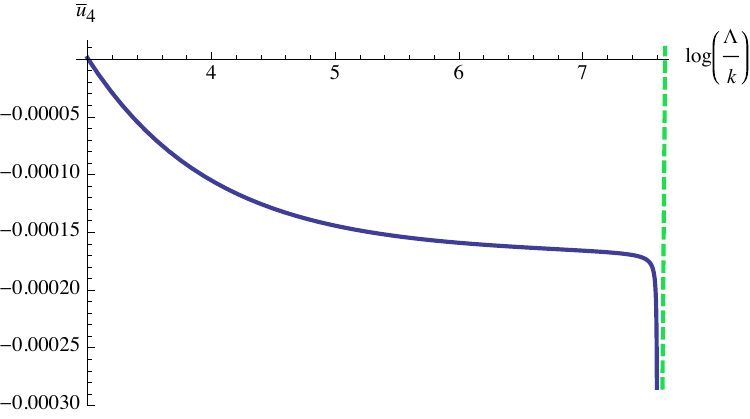}
\qquad \includegraphics[scale=0.5]{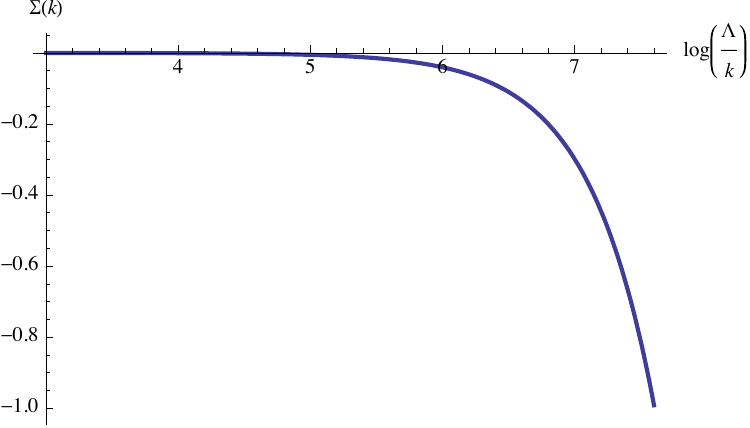}
\end{center}
\caption{Behavior of the RG flow in the positive region for $\bar{\tilde{u}}_6(k_0)=1$.}\label{figdivergencepositive}
\end{figure}

\begin{figure}
\begin{center}
\includegraphics[scale=0.5]{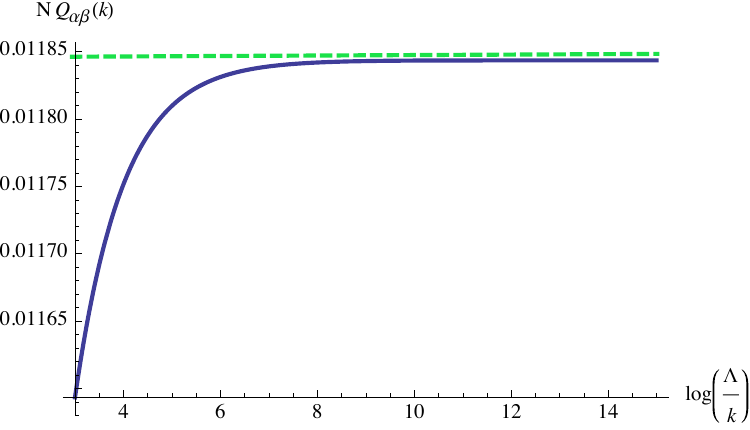}\qquad
\includegraphics[scale=0.5]{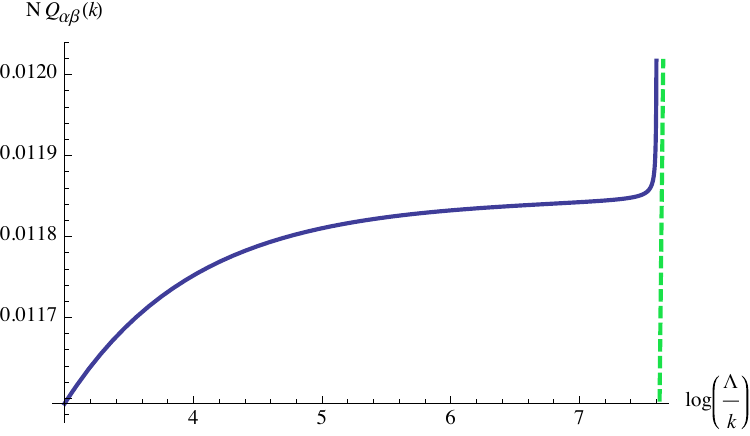}
\end{center}
\caption{Behavior of $N Q_{\alpha\beta}$ for $\tilde{u}_6=0$ (on the left) and for $\tilde{u}_6=1$ (on the right) in the positive region. }\label{Qabehavior}
\end{figure}

\subsection{Global flow near critical regime}

\begin{figure}
\begin{center}
\includegraphics[scale=0.5]{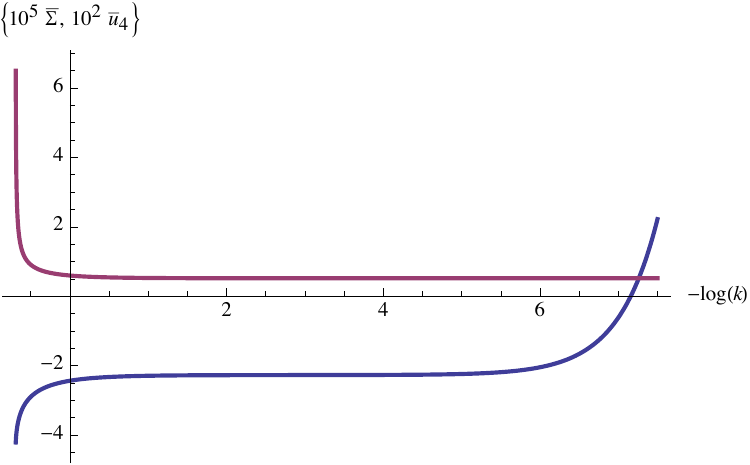}
\includegraphics[scale=0.5]{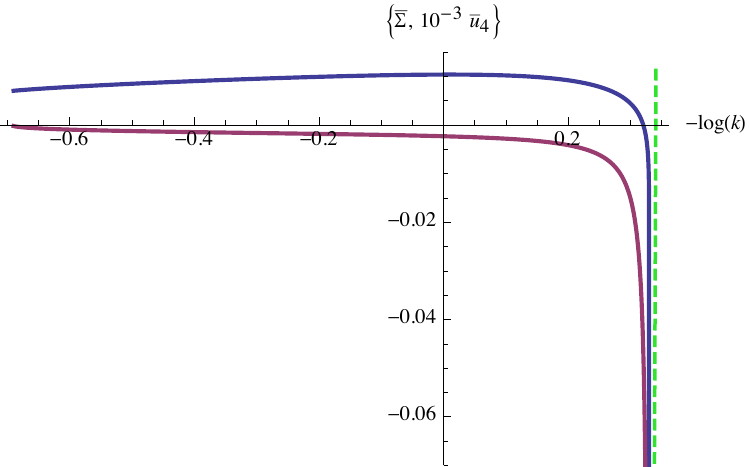}
\end{center}
\caption{On the left: behavior of the RG flow in the vicinity of the critical surface for $\tilde{\bar{u}}_6(k_0)=0$ for $(\bar{\Sigma}(k_0),\bar{u}_4(k_0),\bar{u}_6)=(-0.000045,0.3,1)$, $k_0=1.99$. On the right: behavior near the critical surface (adjusting the initial condition for $\bar{\Sigma}(k_0)$) for $\tilde{\bar{u}}_6(k_0)=10^5$. Red curve is for $\bar{u}_4$ and blue curve for $\bar{\Sigma}$.}\label{FigureflowbehaviorfullUV1}
\end{figure}

As we discussed in section \ref{sectionscaling}, the definition of the canonical dimension has to include mass or higher derivative coupling constants far from the IR regime. However, close enough to the critical surface, we expect the perturbative definition of the canonical dimension to hold. Using vertex expansion at the leading order of the derivative expansion, the criterion for the validity of this approximation is\footnote{Notice that this corresponds to the critical regime only near the Gaussian fixed point. Then, this is not the usual way the critical region is defined.}:
\begin{equation}
\bar{\Sigma} \ll 1\,. 
\end{equation}
and the flow equations are simplified as:
\begin{equation}
\boxed{\dot{\bar{\Sigma}}\approx -2\bar{\Sigma}-\frac{\bar{u}_4}{36\pi}\,,}
\end{equation}
\begin{equation}
\boxed{\dot{\bar{u}}_4\approx-\mathrm{dim}_{4}\bar{u}_4+ \frac{\bar{\tilde{u}}_6 }{30\pi^2}-\frac{\bar{u}_6 }{60\pi}+ \frac{\bar{u}_4^2}{12\pi}\,.}
\end{equation}
\begin{equation}
\boxed{\dot{\bar{{u}}}_6\approx -\mathrm{dim}_{6}\, {\bar{{u}}}_6+\frac{72\bar{u}_4\bar{u}_6}{5\pi}+\frac{4\bar{u}_4\bar{\tilde{u}}_6}{5\pi^2} -\frac{5\bar{u}_4^3}{18\pi} \,,}
\end{equation}
\begin{equation}
\boxed{\dot{\bar{\tilde{u}}}_6=- \mathrm{dim}_{6,n.l}\, {\bar{\tilde{u}}}_6\,,}
\end{equation}
Once again, let us remark that some numerical factors depending on $k$ but almost of order $1$ for $k$ small enough have been neglected. Taking into account these additional factor does not change our conclusions.
These equations can still be studied numerically, and the net result illustrated in Fig. \ref{FigureflowbehaviorfullUV1} has once again the appearance of singularities on a finite scale, becoming shorter with the increasing intensity of the disorder. This would allow in principle to reconstruct the phase space. We note however that the singularity scale $k_c \approx 0.72$ depends weakly on the choice of initial conditions, as we remain close to the Gaussian fixed point. 
\medskip

It is also important to move upstream from the Gaussian region, as shown in Fig. \ref{FigureflowbehaviorfullIR1}. We then observe an unexpected phenomenon. When disorder is zero, the behavior of the trajectory is almost what one would expect from ordinary field theory. After a phase of growth or decline, the trajectory reaches a plateau, in other words, the flow reaches an almost perfect scale regime. This regime ends in the deep UV, and the trajectory ends up diverging at the same time as the canonical dimensions. But this last phenomenon seems to be a pathology of our regularization. When disorder increases, a new phenomenon appears. The trajectories oscillate, more and more quickly, as the disorder increases. Furthermore, in the $(\bar{\Sigma},\bar{u}_4)$ plane, the flow quickly finds itself trapped in a narrow region, around $\bar{\Sigma} \sim 0.0015$ (see Fig. \ref{RGtrajectories}), before to be slowly repelled from this region in a spiral motion around the $\bar{\Sigma}$ axis. This oscillation takes place in the range of values $\bar{\tilde{u}}_6(k_0) \lesssim \tilde{\bar{u}}_{6,c}$, where $\tilde{\bar{u}}_{6,c}\approx 27$ on Fig. \ref{FigureflowbehaviorfullIR1}. For larger disorders, sharp singularities appear again, forbidding access to microscopic physics from the macroscopic states, for $k_c \approx 0.001$. Note that these oscillations appear only for a local sextic truncation and beyond. 
\medskip

At this stage, it is tempting to conclude that this behavior of the RG flow is a signature of the frozen regime, but a physical interpretation of these rapid oscillations remains to be understood.

\begin{subequations}
\begin{figure}
\begin{center}
\includegraphics[scale=0.35]{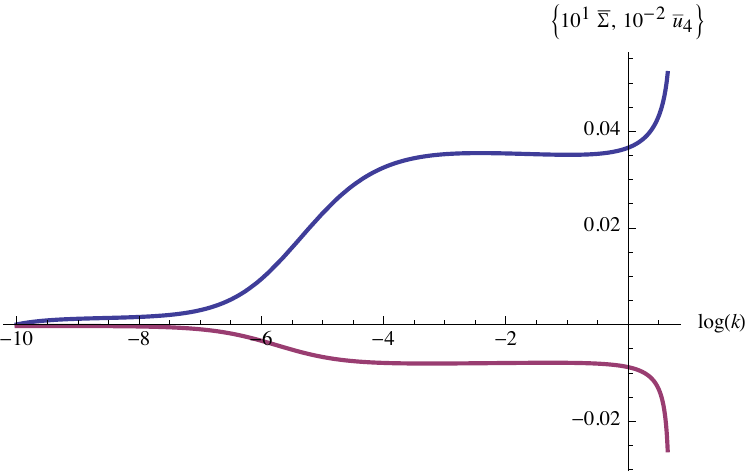}\qquad
\includegraphics[scale=0.35]{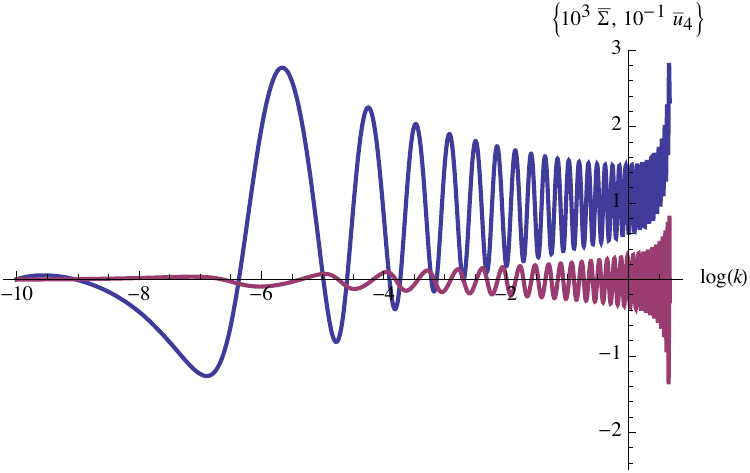}\qquad 
\includegraphics[scale=0.35]{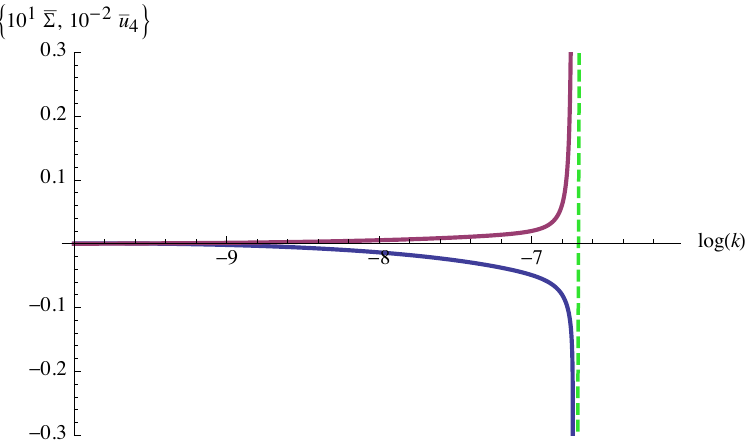}
\end{center}
\caption{Global behavior of the RG flow toward UV scales, in the vicinity of the critical surface. Initial values are for $(\bar{\Sigma}(k_0),\bar{u}_4(k_0),\bar{u}_6)=(0,-0.03,0.1)$. From the left to the right, we have respectively $\tilde{\bar{u}}_6(k_0)=0, 10, 27$. Red curve is for $\bar{u}_4$ and blue curve for $\bar{\Sigma}$.}\label{FigureflowbehaviorfullIR1}
\end{figure}
\end{subequations}

\begin{figure}
\begin{center}
\includegraphics[scale=0.5]{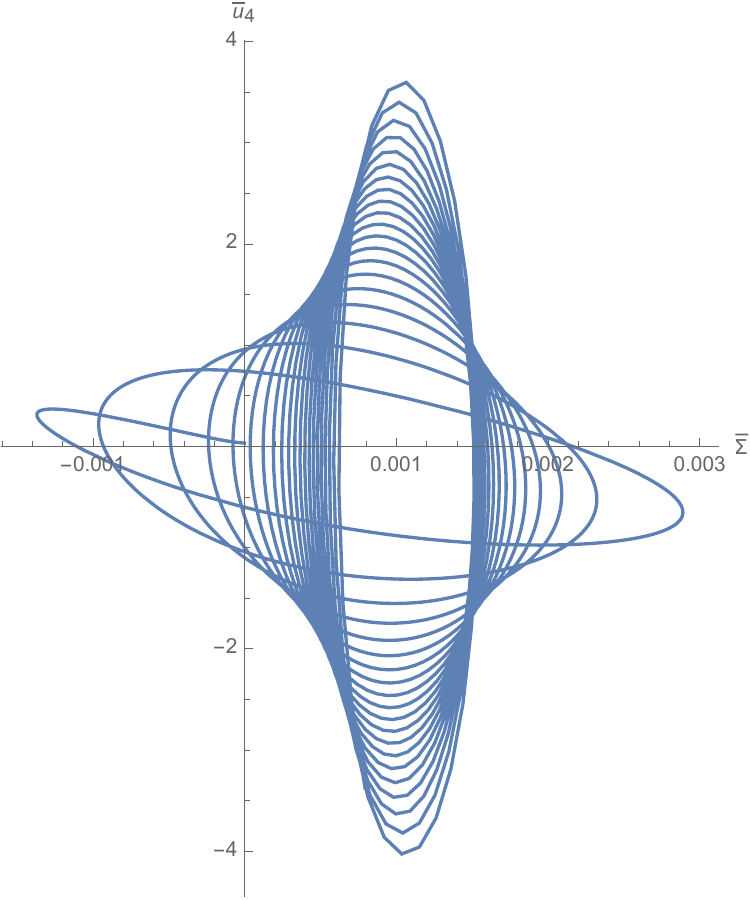}\qquad\includegraphics[scale=0.5]{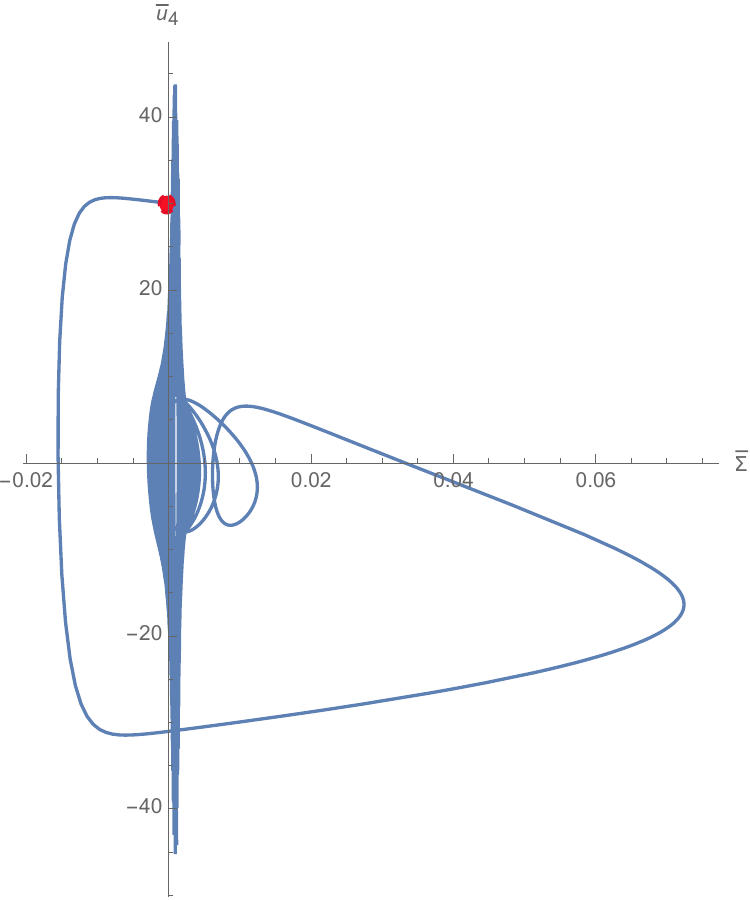}
\end{center}
\caption{Trajectories in the oscillating domain. The red dot on the trajectory on the right materializes the initial condition.}\label{RGtrajectories}
\end{figure}

\subsection{A discussion of the $p=\infty$ limit}

In this section, we are aiming to discuss a nontrivial large $p$ limit for the RG flow. Classically, the large $p$ limit of the Sherrington-Kirkpatrick is linked with the exactly solvable random energy model \cite{Derrida1,Derrida2} because for $p\to \infty$, spin configurations with different energy are essentially uncorrelated. Let us then consider what the RG contributes in this context. Obviously, our formalism is still limited, but we can perhaps hope to draw some conclusions from this limit. In the previous sections, we focused on the case $p=3$, and we could see that disorder couples to local truncations. In the case where $p\to \infty$ on the other hand, the disorder is decoupled from any finite truncation of the vertex expansion, since the flow equations only involve a single loop. Thus, we understand that the flow in the Gaussian point is always driven by local interactions, which therefore behaves as if no disorder was present. We were able to see in the previous sections and in the appendix \ref{AppC} that in this case, the flow behaves similarly to that of a  ‘‘masked" ferromagnetic system in dimension $4+\epsilon$. A phase transition towards a condensed state appears in the region of negative mass, controlled by the Gaussian fixed point. When the flow approaches the region of negative mass, we find the previous phenomenon, and the correlations between replicas increase.  However, while in the case $p=3$ the disorder influenced a part of the flow in the positive region, this is no longer the case here: the condensation induces correlation between replicas, but the concerned region of the space is essentially the same as for $p=0$, both in the deep IR and for the global flow.

\section{Large $N$ methods}\label{largeN}

\subsection{Large $N$ closed equations in the deep IR}

\begin{figure}
\begin{center}
\includegraphics[scale=0.5]{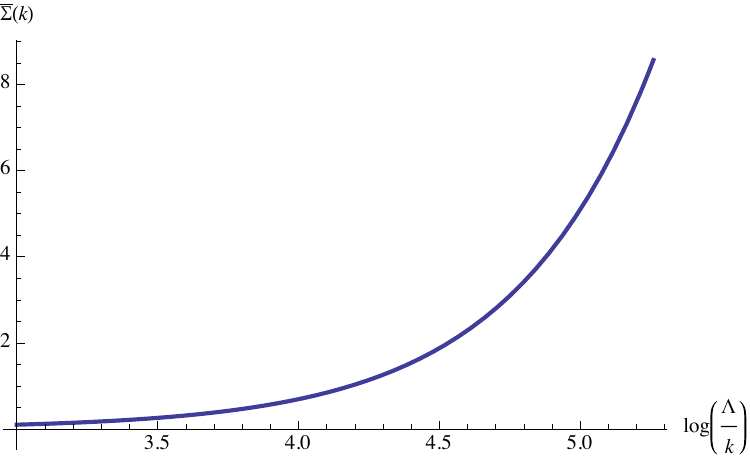}\qquad
\includegraphics[scale=0.5]{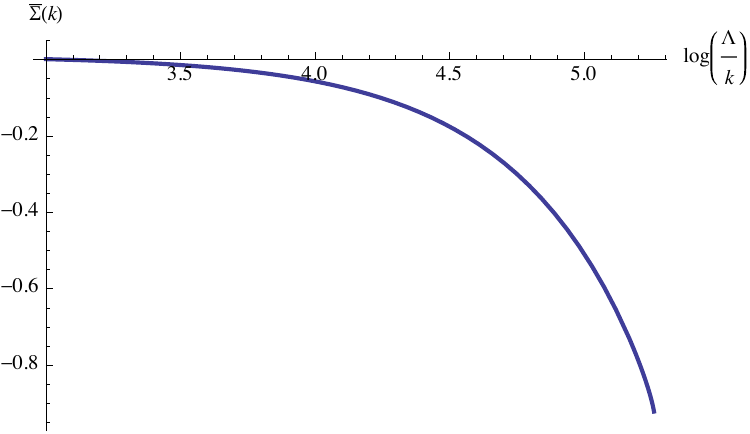}
\end{center}
\caption{Behavior of $\bar{\Sigma}(k)$ in the positive region, respectively below (on the bottom) and above (on the top) the critical temperature $\mu_{1c}$. }\label{behaviorclosedflow}
\end{figure}

\begin{figure}
\begin{center}
\includegraphics[scale=0.8]{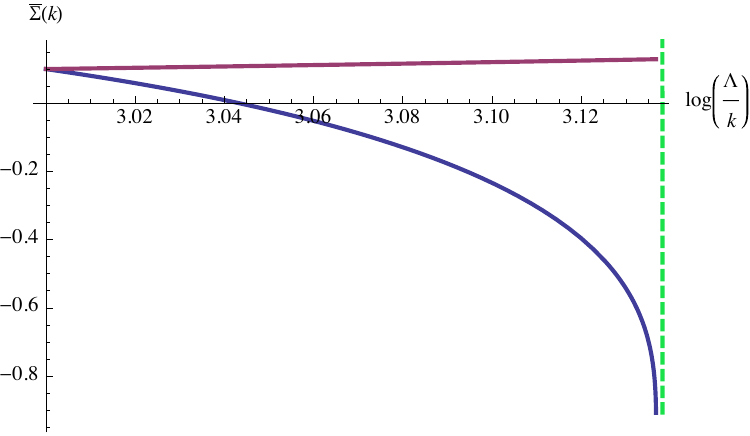}
\end{center}
\caption{Behavior of $\bar{\Sigma}(k)$ in the positive region above $\mu_{1c}$ respectively for $\lambda=0$ (purple curve) and for $\lambda=0.01$ (blue curve).}\label{behaviorclosedflow2}
\end{figure}

\begin{figure}
\begin{center}
\includegraphics[scale=0.8]{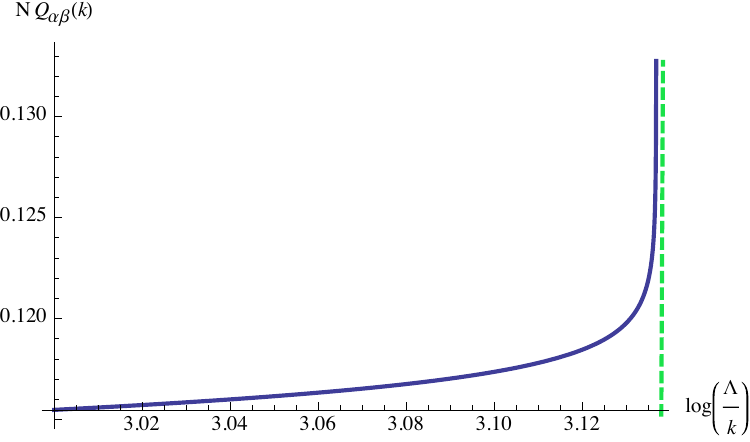}\qquad 
\includegraphics[scale=0.8]{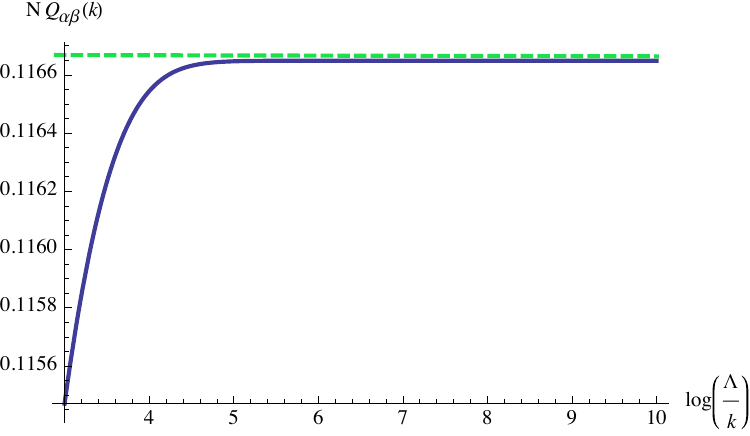}
\end{center}
\caption{Behavior of $N Q_{\alpha\beta}$ in the positive region and above  $\mu_{1c}$ for $\lambda=0$ (on the bottom) and for $\lambda=0.01$ (on the top). }\label{behaviorQabClosed}
\end{figure}

In section \ref{sectionPert}, we showed that the self-energy is diagonal and obeys a closed equation \eqref{closedeqation} in the large $N$ limit. This equation shows that a specific dependency on external frequency $\Omega$ appears due to the non-local couplings arising because of the disordered quenched averaging. Because of this dependency, the resulting closed equation fails to be algebraic and cannot be solved exactly. In the deep IR, for a small enough $k$, this complicated equation can be simplified since we expect this limit to be dominated by small frequencies (large time processes). We can therefore expand $\gamma(\Omega)$ in power of $\Omega$, 
\begin{equation}
\gamma(\Omega)=\gamma_0+\gamma_1\Omega+\gamma_2 \Omega^2+\cdots\,,
\end{equation}
and keep only the lowest term, of the expansion, the closed equation  \eqref{closedeqation} simplifies as:
\begin{align}
\nonumber \gamma_0\approx &  -i \int \frac{d\omega}{2\pi}\,\rho(p^2) dp^2 \bigg(\frac{\mu_2}{6}\, G_k^{(0)}(\omega,p,p,\alpha)\\
&+{12\lambda}\,\int  \rho(q^2)  dq^2 G_k^{(0)}(\omega,p ,p,\alpha) G_k^{(0)}(\omega,q,q,\alpha)\bigg)\,,\label{closedeqationIR}
\end{align}
where $G_k^{(0)}$ is:
\begin{equation}
G_k(\omega,p_\mu,p_\nu,\alpha):=\frac{\delta_{\mu\nu}}{\omega^2-p_\mu^2-\mu_1-\gamma_0-R_k(p_\mu^2)+i\epsilon}\,.\label{Gammak0}
\end{equation}
It is suitable to introduce the self-energy $\Sigma:=\mu_1+\gamma_0$. Computing the integral over frequency, we get:
\begin{align}
\nonumber \Sigma-\mu_1&= -\int\rho(p^2) dp^2 \bigg(\frac{\mu_2}{12}\, \frac{1}{\sqrt{p^2+\Sigma+R_k(p^2)}}\\
&+{6\lambda}\,\int  \rho(q^2)  dq^2 F(p^2+\Sigma+R_k(p^2),q^2+\Sigma+R_k(q^2))\bigg)\,,
\end{align}
where:
\begin{equation}
F(x,y):=\frac{1}{x\sqrt{y}+y\sqrt{x}}\,.
\end{equation}
This equation is again quite complicated, but differentiating with respect to $t:=\ln k$, we have:
\begin{align}
\dot{\Sigma}= \frac{\mu_2 I_1(k)+2\lambda I_2(k)}{1+\mu_2 J_1(k)+2\lambda J_2(k)}\,,
\end{align}
where:
\begin{equation}
I_1(k)=\frac{1}{24}\, \int \rho(p^2) dp^2\, \frac{\dot{R}_k(p^2)}{(k^2+\Sigma)^{\frac{3}{2}}}\,,
\end{equation}
\begin{eqnarray}
\nonumber I_2(k)&=&6 \int  \rho(q^2) dq^2  F^2(k^2+\Sigma,q^2+\Sigma+R_k(q^2))\\
&&\times \int  \rho(p^2) dp^2 \dot{R}_k(p^2) H(k^2+\Sigma,q^2+\Sigma+R_k(q^2))\,,
\end{eqnarray}
with:
\begin{equation}
H(x,y):= \sqrt{y}+\frac{1}{2} y \frac{1}{\sqrt{x}}\,.
\end{equation}
and 
\begin{equation}
J_1(k)=\frac{1}{24}\, \int \rho(p^2) dp^2\, \frac{1}{(p^2+\Sigma+R_k(p^2))^{\frac{3}{2}}}\,,
\end{equation}
\begin{eqnarray}
\nonumber J_2(k)&=&6 \int  \rho(q^2) dq^2  F^2(p^2+\Sigma+R_k(p^2),q^2+\Sigma+R_k(q^2))\\
&&\times \int  \rho(p^2) dp^2  H(p^2+\Sigma+R_k(p^2),q^2+\Sigma+R_k(q^2))\,.
\end{eqnarray}
These equations can be studied numerically and we focus on the positive region for simplicity. In Fig. \ref{behaviorclosedflow}, we show the large-scale behavior of the RG flow for $\lambda=0$ above and below the critical temperature $\mu_{1c}$. The behavior is exactly what we expect for ferromagnetic phase transition above the critical dimension $4$. Setting $\lambda\neq 0$, we essentially recover the result of the previous section, as can be seen for $\lambda=0.01$ in Fig. \ref{behaviorclosedflow2}. The presence of disorder increases the critical temperature and makes the singularity of the flow sharper compared to the case without disorder. The interpretation remains the same as in the previous section, and the behavior of $Q_{\alpha\beta}$ is shown in Fig. \ref{behaviorQabClosed}.
\medskip

\begin{figure}
\begin{center}
\includegraphics[scale=0.8]{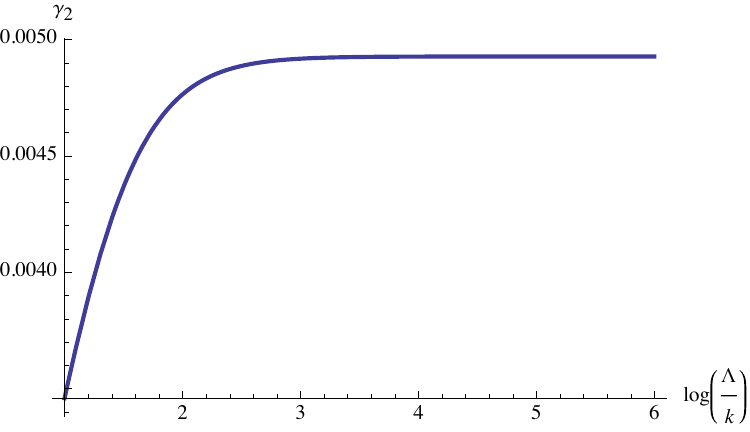}
\end{center}
\caption{Typical behavior of the field strength.}\label{figanomalous}
\end{figure}

To conclude, let us investigate the effect of the anomalous dimension induced by the coupling $\lambda$. For this purpose, let us expand both sides of equation \eqref{closedeqation} in powers of $\Omega$. The leading order imposes $\gamma_1=0$, and we get for $\gamma_2$ (which is dimensionless):
\begin{equation}
\gamma_2= \frac{2\lambda A(\Sigma)}{1+2\lambda B(\Sigma)}\,,
\end{equation}
where:
\begin{align}
 \nonumber A(\Sigma):= 6i\int \frac{d\omega}{2\pi}\,\rho(p^2) & dp^2 \rho(q^2) dq^2 G_k^{(0)}(\omega,p ,p,\alpha)\\
 &\times  \bigg((G_k^{(0)})^2(\omega,q,q,\alpha) -4\omega^2 (G_k^{(0)})^3(\omega,q,q,\alpha) \bigg)\,,
\end{align}
\begin{align}
\nonumber B(\Sigma):= 6i\int \frac{d\omega}{2\pi}\,\rho(p^2) dp^2 \rho(q^2) dq^2   G_k^{(0)}(\omega,p ,p,\alpha) (G_k^{(0)})^2(\omega,q,q,\alpha) \,.
\end{align}
Computing the integral over frequency, we get:
\begin{align}
\nonumber A(\Sigma)&:=6 \int \,\rho(p^2) dp^2 \rho(q^2) dq^2 \Big(f_1(p^2+\Sigma+R_k(p^2),q^2+\Sigma+R_k(q^2))\\
&+f_2(p^2+\Sigma+R_k(p^2),q^2+\Sigma+R_k(q^2))\Big)\,,
\end{align}
where, for $x\neq y$:
\begin{equation}
f_1(x,y):=\frac{\sqrt{x}+2 \sqrt{y}}{4 \sqrt{x} y^{3/2} \left(\sqrt{x}+\sqrt{y}\right)^2}\,, 
\end{equation}
\begin{equation}
f_2(x,y):=\frac{9 \sqrt{x y}+3 x+8 y}{16 \sqrt{x} y^{5/2} \left(\sqrt{x}+\sqrt{y}\right)^3}\,,
\end{equation}
and:
\begin{align}
\nonumber B(\Sigma):=6 \int \,\rho(p^2) dp^2 \rho(q^2) dq^2 f_1(p^2+\Sigma+R_k(p^2),q^2+\Sigma+R_k(q^2)\,.
\end{align}
In Fig. \ref{figanomalous}, we show the behavior of the field strength $\gamma_2$, for $\lambda=0.01$ and $\bar{\Sigma}(k_0)=\bar{u}_4(k_0)=1$ (in the positive region, outside the symmetry breaking region). Since the scale window in which the flow stabilizes is small and given the magnitude of the coupling we consider, we expect the anomalous dimension to be a less relevant effect. This justifies using the standard \textit{local potential approximation} as long as we remain in the symmetric phase.

\subsection{$2$-particle irreducible approaches for $N\to \infty$}

The previous sections provided us indications that in some regions of phase space, some interactions not explainable by perturbation theory could be in unstable equilibrium. Let us emphasize this point again, this situation is not specific to the problem that we study. In the theory of critical phenomena, singularities appear along the flow in the symmetric phase ($\bar{\Sigma}=-1$ with the Litim regulator). This is symptomatic of development around the ‘‘bad" vacuum, the vacuum ‘‘zero" becoming unstable in this region of phase space. In the case of the Ising model, the order parameter is identified with the $1$-point function, and the 1PI formalism is naturally imposed. Here it seems more natural to consider that the role of the order parameter is played by the $2$-point function, and the 2PI formalism seems more natural. This last formalism not being as well known as the more traditional 1PI formalism in theoretical physics, we recall the basics in the appendix \ref{AppE}, and the reader can also consult the \cite{Blaizot21,Dupuis,Ward} references for more details. 
\medskip

Assuming non-diagonal $2$-point couplings in the replica space become unstable in some regions of the full phase space, we introduce the truncation (again in the symmetric phase):
\begin{align}
\nonumber\Gamma_{k;\,\mu,\mu^\prime,\alpha,\beta}^{(2)}(\omega,\omega^\prime)&:=\delta_{\mu\mu^\prime}\Big[ (\omega^2-p_\mu^2) \delta_{\alpha\beta}\delta(\omega+\omega^\prime)-\mathbf{\Sigma}_{\alpha,\beta}(\omega,\omega^\prime)\Big]\,,
\end{align}
where the mass matrix has entries:
\begin{equation}
\mathbf{\Sigma}_{\alpha,\beta}(\omega,\omega^\prime)=\Big[\underbrace{\Sigma \, \delta_{\alpha\beta}\delta(\omega+\omega^\prime)}_{\text{diagonal}}\,+\, \underbrace{q_{\alpha \beta}  \delta(\omega+\omega^\prime)}_{\text{non-diagonal}} \Big]\,.
\end{equation}
Because $n$ is finite, we can focus on a replica symmetric solution, and choose :
\begin{equation}
q_{\alpha \beta} =-q (1-\delta_{\alpha\beta})\,.
\end{equation}
Note that the $q$ couplings remains local in time. One could also consider a more ordinary non-locality, with respect to time, but this adds subtleties that we will discuss in a later article.
\medskip 

\begin{figure}
\begin{center}
\includegraphics[scale=0.5]{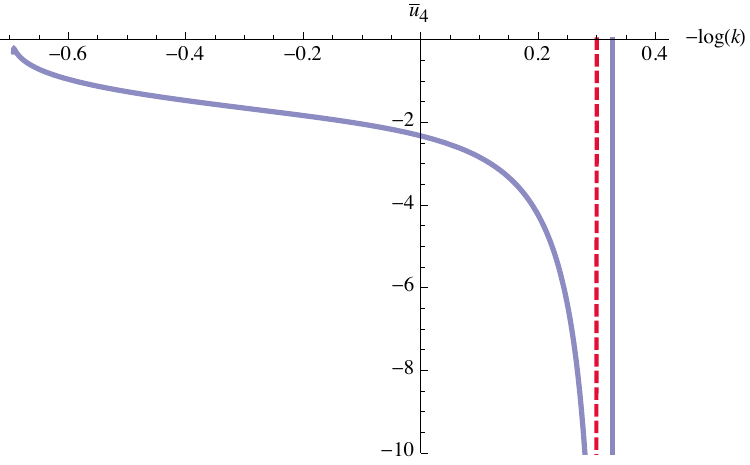}\qquad \qquad \includegraphics[scale=0.5]{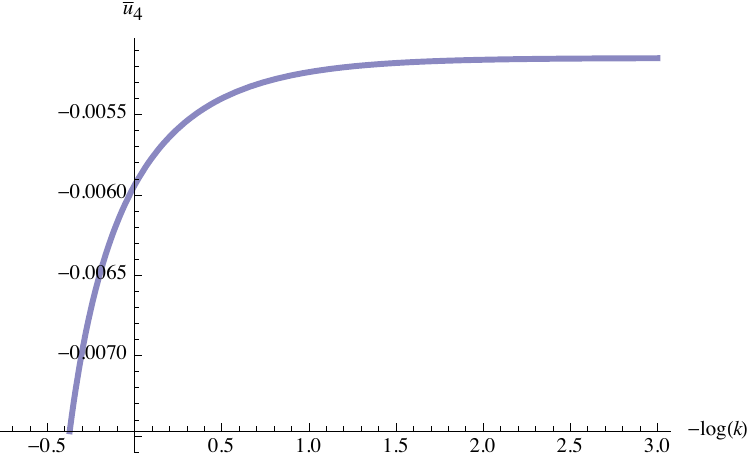}\\
\includegraphics[scale=0.5]{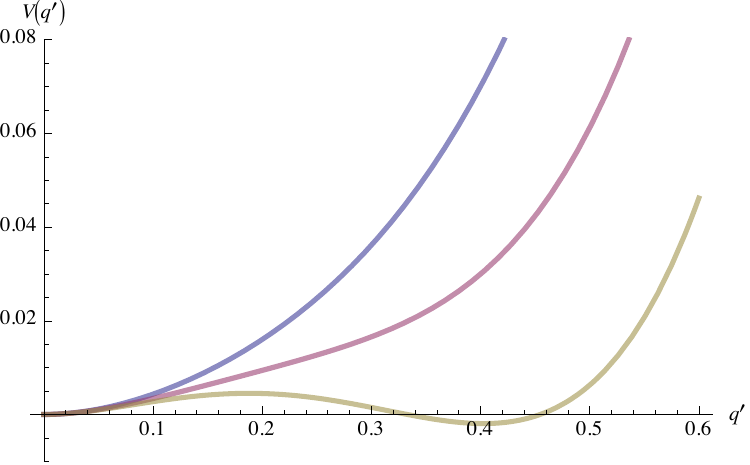}\qquad \qquad \includegraphics[scale=0.5]{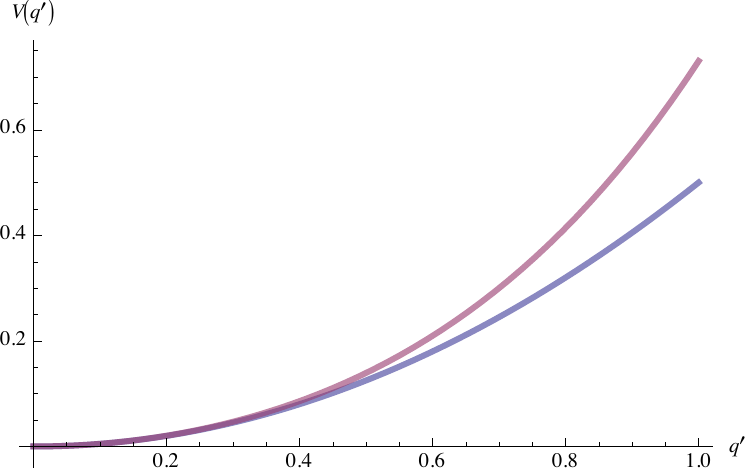}
\end{center}
\caption{On the top, the evolution of the quartic coupling along the RG flow for the initial conditions $(\bar{\Sigma}(k_0),\bar{u}_4(k_0),\bar{u}_6)=(-0.000045,-0.3,1)$, $k_0=1.99$, on the right for $\bar{\tilde{u}}_6(k_0)=10^5$ and on the left for $\bar{\tilde{u}}_6(k_0)=1$. On the bottom is the corresponding evolution for $V(q)$, on the right for $k\approx 1.82$ and $k\approx 0.13$ and $k=0.13$ (resp. blue and purple curves), and on the right for $k=1.50$, $k=1.45$ and $k=1.42$ (resp. blue, purple and yellow curves). The singularity is for $k_c=0.72$, but the value of the non zero vacuum invalidates the perturbative expansion near this point.}\label{potential2PI}
\end{figure}

In the 2PI formalism, $q$ is an order parameter, and we aim to find an equation of state to determine it. In the vicinity of the transition, and because of the power counting, it is suitable to consider a Ginzburg–Landau approach, and to expand the full effective propagator in power of $q$, which reads graphically:
\begin{align}
\nonumber G_k&=\vcenter{\hbox{\includegraphics[scale=1]{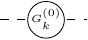}}}\,+\,\vcenter{\hbox{\includegraphics[scale=1]{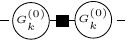}}}\\
&\,+\,\vcenter{\hbox{\includegraphics[scale=1]{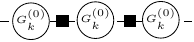}}}\,+\,\cdots \,,
\end{align}
Where only the term of order $q^0$, $G_k^{(0)}$, is local i.e. proportional to $\delta_{\alpha\beta} \delta(\omega+\omega^\prime)$, and we denoted insertions of the matrix with entries $q_{\alpha\beta}\delta(\omega+\omega^\prime)\delta_{\mu\mu^\prime}$. In the large $N$ limit, the LO 1PI 2-point function $\mathbf{\Sigma}$ can be closed accordingly with the results of the section \ref{sectionPert}, and the same equation can be deduced using 2PI formalism, equations, see Appendix \ref{AppE}, equations \eqref{largeNPhi} and \eqref{equationselfenergy}. We get graphically:
\begin{equation}
q=\vcenter{\hbox{\includegraphics[scale=0.8]{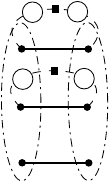}}}\,+ \,2\times \, \vcenter{\hbox{\includegraphics[scale=0.8]{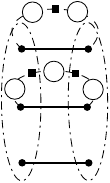}}}\,+\,\mathcal{O}(q^4)\,.
\end{equation}
Explicitly:
\begin{eqnarray}
\nonumber q&=&{-12i\lambda}\,\int \frac{d\omega}{2\pi}\,\rho(p^2_1) dp^2_1\,  \rho(p^2_2)  dp^2_2 \Big[ q^2 (G_k^{(0)}(\omega,p_1 ,p_1,\alpha))^2 (G_k^{(0)}(\omega,p_2,p_2,\alpha))^2\\\nonumber
&&-2 (n-1) q^3 (G_k^{(0)}(\omega,p_1 ,p_1,\alpha))^3 (G_k^{(0)}(\omega,p_2,p_2,\alpha))^2 \Big]+\,\mathcal{O}(q^4)\,.
\end{eqnarray}
This equation can be summarized as a stability condition:
\begin{equation}
V^\prime(q)=0\,,
\end{equation}
for the potential:
\begin{eqnarray}
\nonumber V(q)&=&\frac{q^2}{2}+{12i\lambda}\,\int \frac{d\omega}{2\pi}\,\rho(p^2_1) dp^2_1\,  \rho(p^2_2)  dp^2_2  \Big[ \frac{q^3}{3} (G_k^{(0)}(\omega,p_1 ,p_1,\alpha))^2 (G_k^{(0)}(\omega,p_2,p_2,\alpha))^2\\
&&- (n-1) \frac{q^4}{2} (G_k^{(0)}(\omega,p_1 ,p_1,\alpha))^3 (G_k^{(0)}(\omega,p_2,p_2,\alpha))^2 \Big]\,,
\end{eqnarray}
where the expansion stops at order $q^4$. At the leading order, we assume that the flow of $G_k^{(0)}$ can be computed from the zero-order equations in $q$. In Figure \ref{potential2PI} we compare the behavior of the effective potential along a typical RG trajectory, above and before the singular behavior (i.e. for large and low sextic initial disorder $\bar{\tilde{u}}(k_0)$). The result confirms our intuition of the previous section. The effective potential exhibits a typical shape reminiscent of a \textit{first order phase transition}: along the singular trajectories, the nonzero vacuum becomes deeper and deeper, and at the same time, the derivative at the origin of the potential becomes flatter and flatter. In this limited approximation scheme, the potential becomes singular before the stability of the ‘‘zero" vacuum changes. But the fact that the non-zero vacuum deepens more and more goes in the direction of the hypothesis of a phase transition, during which $q$ takes a non-zero value. Our assumption on the unstable nature of the truncation based on perturbation theory is therefore reinforced. Conversely, we observe that when the disorder is quite weak and the flow is not singular. The stability of the “zero” vacuum is reinforced in the IR. Once again, let us note in conclusion that it is at this stage difficult to construct a phase space in this formalism, which we have only been able to study in certain limited cases, and this remains an open technical problem for the future.

\section{Conclusion and outlooks}\label{conclusion}

In this article, we have constructed an unconventional renormalization group based on the effective kinetics due to the presence of rank two disorder. This method, already proven in the classical case, made it possible to study the long-time (or low temperature) behavior of a quantum particle subjected to a disorder of the ‘‘$2+p$" type. Although we have limited ourselves to the perturbative framework and vertex expansion, which in the vicinity of the Gaussian fixed point seem justified by the power counting, we were able to 1) Recover the second order phase transition occurring for $p=0$, asymptotically justified by the power counting itself, 2) to provide evidence in favor of a first-order phase transition toward a replica symmetry breaking phase occurring for $p>2$, explicitly checked in the case $p=3$ and $p=\infty$, which superpose to the broken phase. In the cases, $p=0,1$, a complementary analytical calculation presented in the appendix \ref{AppC} allows us to understand the nature of the transition, which is reminiscent of a ferromagnetic transition, exactly what we see in the classic case \cite{van2010second,Dominicis,Lahochep2}.
\medskip

The evidence for a transition manifests itself essentially in the form of finite-scale singularity, making it impossible to continue the flow in the deep IR from sufficiently UV scales. At this stage, our interpretation of these singularities, based on similar phenomena in the literature, is as follows:

\begin{itemize}
\item In the large time limit, they imply revising the power $N$ counting, because some superficially irrelevant contributions like $Q_{\alpha\beta}$ suddenly become of the same order, or even more important, than the superficially dominant effects at large $N$. In other words, the singularity could be the signal that something is missing in the parameterization of the flow, and that the presence of these missing interactions could make it possible to ‘‘continue" the flow through the singularity, as has been observed in other contexts \cite{Delamotte2,Tarjus}.

\item These observations also invite revision of the construction of the large time limit which should not be taken independently of the large $N$ limit. It is indeed suspected that these ‘‘missing" terms, absent from perturbation theory, could be generated dynamically. This moreover underlines the importance of considering a mixed renormalization group, both in the space of generalized moments and in the space of frequencies \cite{Lahoche2023functional}.
\end{itemize}
\medskip

These results were obtained at the lowest order of the perturbation theory (order $\hbar$) or for the minimal vertex expansion, a temporary limitation linked to the specific non-localities of the theory. Thus, this is above all a preliminary result that we consider encouraging, and complementary to a more advanced nonperturbative study which will be the subject of a series of forthcoming works. One avenue could be to combine the equations of the renormalization group with the non-trivial Ward identities coming from the formal breaking by the effective kinetics of the global $O(N)$ symmetry, as was recently studied in \cite{lahoche20241,lahoche20242,lahoche20244} and other symmetries of the path integral measure intrinsic to the model (like local $O(N)$ symmetries). Another path, continuing the work of \cite{Lahoche2022functional}, would consist of partially integrating on frequencies rather than on the spectrum of matrix disorder, which again is the subject of further work. Finally, and as pointed out in section \ref{largeN}, the order parameter being the 2-point function $Q_{\alpha\beta}$, the formalism $2PI$ \cite{Blaizot21} seems to impose itself naturally, and we will also consider it in our forthcoming works in continuation with our preliminary investigations of this work.
\medskip

 As mentioned in the introduction, our long-term objective is to develop reliable methods, in different contexts, allowing us to approach problems that still resist standard analytical approaches. Finally, this type of approach could be developed with the aim of revealing signals in a quantum channel, as was recently established for the detection of signals in the case of almost continuous spectra \cite{lahoche4}.

\section*{Acknowledgements} 
We would like to thank St\'ephane Dartois for fruitful discussions and comments in all stages of this work.

\appendix
\section{Formal sum of Gaussian integration}\label{AppA}

In this section, we consider the Gaussian averaging of some oscillating exponential. The typical integral we have to compute is of the form:
\begin{equation}
I[x]:=\frac{\int_{-\infty}^{+\infty}dJ\, e^{-\frac{J^2}{2\sigma^2}}\, e^{i J x^p}}{\int_{-\infty}^{+\infty}dJ\, e^{-\frac{J^2}{2\sigma^2}}}\,,
\end{equation}
where $x\in \mathbb{R}$. Expanding the oscillating exponential in power of $J$, we get:
\begin{equation}
I[x]=\sum_{n=0}^\infty\,\frac{i^n}{n!} (x^p)^n\,\int_{-\infty}^{+\infty}d\mu(J)\, e^{-\frac{J^2}{2\sigma^2}}\, J^n\,,
\end{equation}
where we have hidden the normalization of the Gaussian integral in the measure $d\mu(J)$. Each term can be computed using the standard Wick theorem, and noticing that the integral vanishes for $n$ is odd, we get:
\begin{equation}
I[x]=\sum_{q=0}^\infty\, (-1)^q \frac{(x^p)^{2q}}{(2q)!}\,\mathcal{N}(q)\,\left(\int_{-\infty}^{+\infty}d\mu(J)\, e^{-\frac{J^2}{2\sigma^2}}\, J^2\right)^q\,,
\end{equation}
Because of the normalization, the Gaussian averaging of $J^2$ equals $\sigma^2$. The numerical factor $\mathcal{N}(q)$ counts the number of independent Wick contractions, and can be easily computed recursively:
\begin{equation}
\mathcal{N}(q)=(2q-1)!!=\frac{(2q)!}{2^q q!}\,.
\end{equation}
Finally:
\begin{equation}
I[x]=\sum_{q=0}^\infty\, (-1)^q \frac{(x^p)^{2q}}{q!}\, \left(\frac{\sigma^2}{2}\right)^q =e^{-\frac{\sigma^2}{2}\, x^{2p}}\,.
\end{equation}

\section{Equivalence with large time behavior}\label{AppB}

In this section, we will show explicitly that choosing a diagonal regulator is equivalent to considering large-time behavior of the flow. Let us denote as $M_{\mu}(\omega)$ the Fourier component of the classical field (we consider the case $p=0$ for simplicity, and disregard replica). It is suitable in this case to work in a discrete setting, considering a finite time volume $T$ with periodic boundary conditions. In that way, frequencies are quantized:
\begin{equation}
\omega_n=\frac{2\pi n}{T}\,.
\end{equation}
In the large time limit, we expect that the Fourier modes decomposition is dominated by the component $\omega=0$, and:
\begin{equation}
M_{\mu}(\omega_n)\approx \delta_{n0} \mathcal{M}_{\mu}\,,\label{largeTvacuum}
\end{equation}
which is equivalent to consider a uniform vacuum $M_{\mu}(t)=\sum_n M_{\mu}(\omega_n) e^{i \omega_n t}=\mathcal{M}_{\mu}$. Projected on the large time vacuum \eqref{largeTvacuum}, the EAA  $\Gamma_k$ becomes:
\begin{equation}
\Gamma_k[M_{\mu}(t)=\mathcal{M}_{\mu}]=: \int_0^T dt\, U_k[\mathcal{M}_{\mu}]=TU_k[\mathcal{M}_{\mu}]\,,
\end{equation}
where at the leading order in the perturbation theory:
\begin{equation}
U_k[\mathcal{M}_{\mu}]:=-\frac{1}{2}\sum_\mu \, \mathcal{M}_{\mu} (p_\mu^2+\mu_1)\mathcal{M}_{\mu} -\frac{\mu_2}{4! N} \left(\sum_\mu \mathcal{M}_{\mu} ^2 \right)^2+\cdots\,.
\end{equation}
Then, projecting the flow equation \eqref{Wett} along $ \mathcal{M}_{\mu}$ on both sides, we get:
\begin{equation}
T\dot{U}_k=-\frac{T}{2\pi}\frac{i}{2}\sum_n \Delta \omega\, \sum_{p_\mu} \dot{R}_k(p_\mu^2) G_k(\omega_n,p_\mu,p_\mu)\,.\label{Wett3}
\end{equation}
where $\Delta \omega:=2\pi/T$ is the spacing between frequencies, and:
\begin{equation}
G_k^{-1}(\omega,p_\mu,p_\mu):=-\omega^2 + \frac{\partial^2 U_k}{\partial \mathcal{M}_\mu \partial \mathcal{M}_\mu} \,.
\end{equation}
In the continuum limit, the equation then reads:
\begin{equation}
\dot{U}_k=-\frac{i}{2}\int \frac{d\omega}{2\pi}\, \sum_{p_\mu}\frac{ \dot{R}_k(p_\mu^2)}{-\omega^2 + \frac{\partial^2 U_k}{\partial \mathcal{M}_\mu \partial \mathcal{M}_\mu}}\,,
\end{equation}
which is equivalent to the flow equation computed explicitly in section \ref{sectionscaling}, projecting both sides on the symmetric phase i.e. $\mathcal{M}_\mu=0$. 

\section{Solution for $p=0,1$}\label{AppC}

For $p=0$, the Schr\"odinger equation \eqref{schro} reduces to:
\begin{align}
\nonumber i \hbar \frac{\partial }{\partial t}\Psi(\textbf{x},t)&= \bigg(-\frac{\hbar^2}{2m}\frac{\partial^2}{\partial {\textbf{x}}^2}+\frac{1}{2}\sum_{i,j} J_{ij}\hat{x}_{i}\hat{x}_{j}+V(\hat{\textbf{x}}^2)\bigg)\, \Psi(\textbf{x},t)\,.\label{schro2}
\end{align}
It is suitable to make use of eigen-coordinates $x_\mu$ defined as: $x_\mu:=\sum_i x_i u_i^{(\mu)}$, where $u_i^{(\mu)}$ denotes some eigenvector of the random matrix $J$, as we have done in this paper. The Schr\"odinger equation then becomes:
\begin{equation}
 i \hbar \frac{\partial }{\partial t}\Psi(\textbf{x},t)= \bigg[-\frac{\hbar^2}{2m}\sum_\mu\frac{\partial^2}{\partial {{x}}^2_\mu}+\frac{1}{2}\sum_{\mu} (p_\mu^2+\mu_1)\hat{x}_{\mu}^2+\frac{\mu_2}{4! N} \bigg(\sum_{\mu} \hat{x}_\mu^2\bigg)^2\,\bigg]\, \Psi(\textbf{x},t)\,.\label{schro3}
\end{equation}
We will consider separately two different methods to find the energy vacuum of this equation. 
\medskip

\paragraph{Method 1: Schr\"odinger equation and variational method (p=0).} The first method is based on the observation that, for $N$ large enough, we expect that quantities like $\hat{\textbf{x}}^2$ weakly fluctuate around their average (self averaging) \cite{ZinnJustinBook2,ZinnJustinReview}. We therefore expect that the vacuum corresponds to a function locating the variance of the operator $\hat{\textbf{x}}^2$. Based on this observation, we will try to estimate the value of the ground state $E_0$ using the variational method; we try a vacuum of the form:
\begin{equation}
\Psi(\alpha_1,\alpha_2,{\textbf{x}},t)=e^{-\frac{i}{\hbar} E_0 t}\, \frac{e^{-\frac{N}{\hbar} Q(\textbf{x}^2/N)}}{\sqrt{A(\alpha_1,\alpha_2)}}\,,
\end{equation}
where $A(\alpha_1,\alpha_2)$ is the normalization:
\begin{equation}
A(\alpha_1,\alpha_2):= \int d\textbf{x}\, e^{-2\frac{N}{\hbar} Q(\textbf{x}^2/N)}\,,
\end{equation}
and we choose for $Q$:
\begin{equation}
Q(\textbf{x}^2)=\frac{1}{2}\alpha_1 \textbf{x}^2 +\frac{1}{4!}\alpha_2 (\textbf{x}^2)^2\,.
\end{equation}
The normalization $A(\alpha_1,\alpha_2)$ can be computed in the large $N$ limit using the steepest-descend method. Using radial and angular variables,
\begin{align}
\nonumber A(\alpha_1,\alpha_2)&:= \Omega_N \int_0^\infty\, r^{N-1} dr  e^{-2N Q(r^2/N)}\\
&\approx \Omega_N \int_0^\infty\, dr  e^{-\frac{N}{\hbar} (2Q(r^2/N)-\hbar\ln r)}\,,
\end{align}
where $\Omega_N$ is the result of the integration over angles:
\begin{equation}
\Omega_N:=\frac{2 \pi^{N/2}}{\Gamma(N/2)}\,.
\end{equation}
The saddle point is fixed by the condition:
\begin{equation}
4 y Q^\prime(y)=\hbar \to 2 y \left(6 \alpha _1+\alpha _2 y\right)-3 \hbar=0\,,
\end{equation}
explicitly:
\begin{equation}
y_{\pm}=\frac{-3\alpha_1\pm 3\sqrt{\alpha^2_1+\frac{1}{6} \alpha^2_2}}{\alpha_2}
\end{equation}
We assume $\alpha_2>0$ because the wave function is normalizable, and the only positive solution is $y_+$. We get:
\begin{equation}
A(\alpha_1,\alpha_2) \approx \Omega_N N^{-N/2}\, e^{- N(2Q(y_+)-\frac{1}{2}\ln y_+) }\,.
\end{equation}
We will compute the average energy. For the kinetic energy, we get, in the large $N$ limit:
\begin{equation}
\langle E_c \rangle \approx \frac{N \hbar^2}{m} \left(\frac{3}{16y_+}+2y_+ Q^{\prime\prime}(y_+)\right)\,,
\end{equation}
and in the same way, the averaging potential reads:
\begin{equation}
\langle V \rangle \approx V(N y_+)\,.
\end{equation}
In order to estimate the final piece of the averaging Hamiltonian, we use the trivial bound:
\begin{equation}
\sum_{\mu} p_\mu^2 x_\mu^2 \leq  \max(p_\mu) \sum_\mu x_\mu^2 \,.
\end{equation}
and we get the averaging of the remaining pieces:
\begin{equation}
\Big \langle \frac{1}{2}\sum_\mu p_\mu^2 x_\mu^2 \Big\rangle \leq N \sigma  y_+\,.\label{boundIntp}
\end{equation}
Now, we are aiming to find the values for $\alpha_1$ and $\alpha_2$ such that the average energy is stationary, namely:
\begin{equation}
\frac{\partial}{\partial \alpha_i} \, \langle E \rangle =0 \quad \forall \, i=1,2\,.
\end{equation}
After a tedious calculation, we get, at the leading order in $\hbar$:
\begin{equation}
\frac{E_0}{N} \approx -\frac{3 \left(\mu _1+2 \sigma \right){}^2}{2 \mu _2}+\mathcal{O}(\hbar)\,.
\end{equation}
This approximation is however bad enough, as we will see below, except in the limit $\sigma \to 0$. In this regime, the interaction $\frac{1}{2}\sum_\mu p_\mu^2 x_\mu^2$ can be treated as a perturbation (see \eqref{boundIntp}). It is especially interesting to focus on the limit $\mu_2 \gg 1$, and we get in the case of minimum energy estimate:
\begin{equation}
\frac{\mathcal{E}_0}{N} \approx \frac{1}{8} \left(\frac{3}{2}\right)^{2/3} \hbar^{4/3} \mu _2^{1/3}\,,
\end{equation}
furthermore, the leading order perturbation theory leads to the following expression for the first-order correction (using  the previous estimate for the ground state wave function):
\begin{equation}
\frac{\epsilon}{N} \approx \sigma \langle \hat{\textbf{x}}^2\,\rangle \simeq  \sigma \left({\frac{3}{2}}\right)^{1/3} \hbar^{2/3} \mu _2^{-1/3}\,.
\end{equation}
The behavior $ \mu _2^{1/3}$ is characteristic of the anharmonic oscillator \cite{ZinnJustinBook1}. It is instructive to introduce the frequency $\Omega_0:=(\hbar \mu_2)^{1/3}$, such that the estimated ground state energy reads also as:
\begin{equation}
\frac{E_0}{N}=\hbar\left( \frac{1}{8} \left(\frac{3}{2}\right)^{2/3} \Omega_0+\left({\frac{3}{2}}\right)^{1/3} \frac{\sigma}{\Omega_0}\right)\,.
\end{equation}
The first term dominates as soon as $\sigma/\Omega_0^2 \ll 1$, which defines the perturbative domain. The result of the next paragraph shows that $E_0\sim  \mu _2^{-1/3}$, which invalidates the perturbation calculation. 
\medskip

\paragraph{Method 2: Euclidean path integral (p=0,1).} We make use of the path integral formalism, and consider the Euclidean version of the path integral. For periodic paths with length $\beta$, this is equivalent to computing the partition function of a quantum particle in contact with a thermal bath having temperature $k_B T= \beta^{-1}$, and the corresponding partition function:
\begin{equation}
Z(\beta):= \mathrm{Tr} \, e^{-\beta\,  \hat{H}_{\text{SG}} }\,,
\end{equation}
becomes:
\begin{equation}
Z(\beta)=\int_{x(0)=x(\beta)} [dx]\, e^{-\mathcal{S}[x]}\,,\label{defpathint}
\end{equation}
where the Euclidean action $\mathcal{S}[x]$ is, in the eigenbasis for $J$:
\begin{equation}
S[x]=\int_0^\beta dt\, \left(\frac{1}{2}\sum_\mu \left(m\dot{x}^2_\mu+p^2_\mu x_\mu^2 \right) + V(\hat{x}^2)-2\sigma\hat{x}^2\right)\,.
\end{equation}
Note that the notation in \eqref{defpathint} means that random paths are periodic. In the large $N$ limit, the integral can be computed using the steepest descent method. We introduce two auxiliary function $\lambda(t)$ and $a^2(t)$, periodic, such that:
\begin{equation}
\int [d a^2 ] \,[d\lambda]\, e^{-\int_0^\lambda dt \frac{1}{2}\lambda(t) (\hat{x}^2(t)-N a^2(t)}= K(\beta)\,,
\end{equation}
where the integral over $\lambda$ is expected parallel to the imaginary axis and $K(\beta)$ is some normalization factor. The partition function then becomes, up to this irrelevant factor:
\begin{equation}
Z(\beta)=\int_{x(0)=x(\beta)} [dx]\,[d a^2 ] \,[d\lambda]\,  e^{-\mathcal{S}[x,a^2,\lambda]}
\end{equation}
with:
\begin{eqnarray}
 \mathcal{S}[x,a^2,\lambda]&=&\int_0^\beta dt\, \Big[\,\frac{1}{2}\sum_\mu \left(m\dot{x}^2_\mu+p^2_\mu x_\mu^2 +\lambda(t) x_\mu^2 \right) + V(Na^2)-2N\sigma a^2-\frac{1}{2}N \lambda(t) a^2(t)\,\Big]\,.\cr&&
\end{eqnarray}
The $N$ independent Gaussian integrals over $x_\mu(t)$ can be easily performed, and we have up to an irrelevant factor:
\begin{align}
\nonumber\int_{x(0)=x(\beta)}& [dx]\, e^{-\int_0^\beta dt\,\frac{1}{2}\sum_\mu \left(m\dot{x}^2_\mu+p^2_\mu x_\mu^2 +\lambda(t) x_\mu^2 \right)}=\prod_\mu( \mathrm{det} (\hat{A}_\mu))^{-1/2}=\prod_\mu e^{-\frac{1}{2}\mathrm{Tr} \ln \, (\hat{A}_\mu)}\,,\label{Gaussianxmu}
\end{align}
where we make use of the identity $\mathrm{Tr} \ln=\ln \mathrm{det}$, and the operator $\hat{A}_\mu$ is (we set $m=1$ for simplicity):
\begin{equation}
\hat{A}_\mu:=-\frac{\partial^2}{\partial t^2}+p_\mu^2+\lambda(t)
\end{equation}
The fact that $\lambda$ depends on time complicates the calculation of the determinant, but a moment of reflection shows that we must look for time-independent saddle points \cite{ZinnJustinBook1}. We denote as $\lambda_*$ the solution for $\lambda(t)$ at the saddle point, and we are faced with a classic problem, that of calculating the determinant of the operator:
\begin{equation}
\hat{A}_\mu^*:=-\frac{\partial^2}{\partial t^2}+p_\mu^2+\lambda_*\,.
\end{equation}
Because trajectories are periodic, they can be expanded in Fourier series:
\begin{equation}
x_\mu(t)=\sum_{n=1}^\infty\, q_\mu(\omega_n) e^{i \omega_n t}\,,
\end{equation}
where: $\omega_n:=2\pi n/\beta$. We are aiming to compute:
\begin{equation}
I(p_\mu^2+\lambda_*):=\mathrm{Tr} \ln \, (\hat{A}_\mu)= \sum_n\, \ln (\omega_n^2+p_\mu^2+\lambda_*)\,.
\end{equation}
Deriving the function $I(u^2)$ we have:
\begin{equation}
I^\prime (u^2)= \sum_n\, \frac{1}{\omega_n^2+u^2}\,,
\end{equation}
and the sum can be computed using the standard tricks of complex analysis, we find:
\begin{equation}
I^\prime (u^2)=\frac{\beta}{2u}\, \coth \left(\frac{u \beta}{2}\right)\,,
\end{equation}
we then get:
\begin{equation}
I(u^2)-I(\epsilon^2)=\ln \sinh  \left(\frac{u \beta}{2}\right)-\ln \sinh  \left(\frac{\epsilon \beta}{2}\right)\,.
\end{equation}
The second term is nothing but the normalization of the free path integral, which can be absorbed in the definition of the measure $[dx]$, then:
\begin{equation}
e^{-\frac{1}{2}\mathrm{Tr} \ln \, (\hat{A}_\mu)}=\frac{1}{\sqrt{\sinh  \left(\frac{\sqrt{p_\mu^2+\lambda_*} \beta}{2}\right)}}\,.
\end{equation}
The saddle point equations for $a^2_*$ then reads (note that $\lambda_*$ have to be real at the saddle point):
\begin{equation}
\frac{1}{2}\mu_1+\frac{\mu_2}{12} a^2_*-\frac{1}{2}\lambda_*=0 \to a^2_*=\frac{6}{\mu_2}(\lambda_*-\mu_1)\,,
\end{equation}
and the saddle point action becomes:
\begin{equation}
\frac{2\mathcal{S}_*}{N}=-\frac{3\beta}{\mu_2}(\lambda_*-\mu_1)^2+\int dp^2\, \rho(p^2) \,\ln \sinh  \left(\frac{\sqrt{p^2+\lambda_*} \beta}{2}\right)\,.
\end{equation}
From this saddle point solution, we then deduce the saddle point equation for $\lambda_*$:
\begin{equation}
-\frac{6\beta}{\mu_2}(\lambda_*-\mu_1)+\int dp^2\, \rho(p^2)\frac{\beta  \coth \left(\frac{1}{2} \beta  \sqrt{\lambda_* +p^2}\right)}{4 \sqrt{\lambda_* +p^2}}=0\,.
\end{equation}
In the low temperature limit, $\beta\to \infty$, and the saddle point equation reduces to:
\begin{equation}
-\frac{6}{\mu_2}(\lambda_*-\mu_1)+\int dp^2\, \frac{\rho(p^2)}{4 \sqrt{\lambda_* +p^2}}=0\,.
\end{equation}

\begin{figure}
\begin{center}
\includegraphics[scale=0.6]{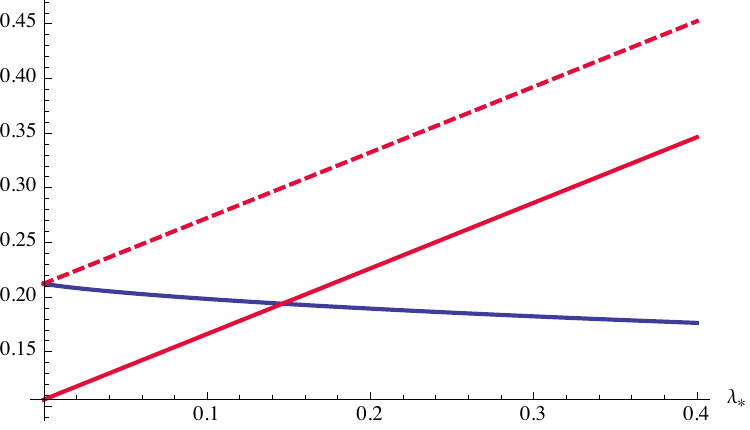}
\end{center}
\caption{Graphical solution of the equation. The blue line corresponds to the integral, and the red curves are the linear function $\frac{6\beta}{\mu_2}(\lambda_*-\mu_1)$, the dashed line is for the critical value $\mu_{1c}$.}\label{figGraphical}
\end{figure}

This equation can be solved graphically (see Fig. \ref{figGraphical}), and we find that this equation is always a single solution in the sector $\lambda_*\geq 0$, provided that\footnote{Classically, $\mu_c=0$, quantum fluctuations have the effect of opposing the phenomenon of condensation.}:
\begin{equation}
\mu_1 >\mu_{1c}:=-\frac{\mu_2}{6}\int dp^2\, \frac{\rho(p^2)}{4 \sqrt{p^2}}=-\frac{\mu_2}{9 \pi  \sqrt{\sigma }}\,.\label{conditionmu1}
\end{equation}
The ground state energy level is given by:
\begin{align}
&\frac{E_0}{N}=-\lim_{\beta\to \infty}\, \frac{\ln Z(\beta)}{\beta N}\label{EnergyPath}\\\nonumber
&=-\frac{3}{2\mu_2}(\lambda_*-\mu_1)^2+\int dp^2\, \rho(p^2) \,\left(\frac{\sqrt{p^2+\lambda_*}}{4}\right)-\frac{\ln 2}{\beta}\,.
\end{align}
The meaning of the condition \eqref{conditionmu1} can be understood as follows \cite{Dominicis}. Above the critical value $\mu_{1c}$, all the components $\vert x_\mu \,\vert$ are expected to be of order $1$; below in contrast, we expect that the component $p=0$ acquires a macroscopic number $\mathcal{O}(N)$ as a sort of Bose-Einstein condensation:
\begin{equation}
-\frac{6}{\mu_2} \mu_1=-\frac{6}{\mu_2} \mu_{1c}-\frac{6}{\mu_2} (\mu_{1}-\mu_{1c})\,.
\end{equation}
The first term can be identified with $\sum_{\mu \neq 0} \langle x_\mu^2 \rangle$ for $\lambda_*=0$:
\begin{equation}
-\frac{6}{\mu_2} \mu_{1c}=\frac{1}{N}\sum_{\mu \neq 0} \langle x_\mu^2 \rangle=\frac{2}{3 \pi  \sqrt{\sigma }}\,,
\end{equation}
and the second term identifies to:
\begin{equation}
\boxed{\langle x_0^2 \rangle = -\frac{6 N}{\mu_2} (\mu_{1}-\mu_{1c})\,.}\label{formulacompmacro}
\end{equation}
As announced, we find that the component $0$ becomes macroscopic below $\mu_{1c}$, and the transition is second order (continuous). Finally, let us return to the expression for energy \eqref{EnergyPath}. It is difficult to integrate directly the integral on the right-hand side, but it can be computed perturbatively, expanding the square $\sqrt{p^2+\lambda_*}$ in the power of $p^2$. Each term can be computed easily, and the resulting series looks as a hypergeometric function:
\begin{equation}
\int dp^2\, \rho(p^2) \,\left(\frac{\sqrt{p^2+\lambda_*}}{4}\right)=\frac{\sqrt{\lambda_*}}{4}\, \, _2F_1\left(-\frac{1}{2},\frac{3}{2};3;-\frac{4 \sigma }{\lambda_*}\right)\,.
\end{equation}
It is interesting to look at this function in different regimes. Let's first take $\lambda_*\to \infty$ (corresponding to $\mu_2\gg 1, \mu_1>0$), we then find:
\begin{equation}
_2F_1\left(-\frac{1}{2},\frac{3}{2};3;-\frac{4 \sigma }{\lambda_*}\right)\to 1\,,
\end{equation}
and:
\begin{equation}
\frac{E_0}{N}\simeq -\frac{3}{2\mu_2}\lambda_*^2+\frac{\sqrt{\lambda_*}}{4}-\frac{\ln 2}{\beta}\,.
\end{equation}
It is moreover easy to check that, in this case:
\begin{equation}
\lambda_*\approx \frac{(\mu_2)^{1/3}}{2 \times 3^{1/3}} \,,
\end{equation}
and $E_0 \sim (\mu_2)^{-1/3}$. 
\medskip

\begin{figure}
\begin{center}
\includegraphics[scale=0.6]{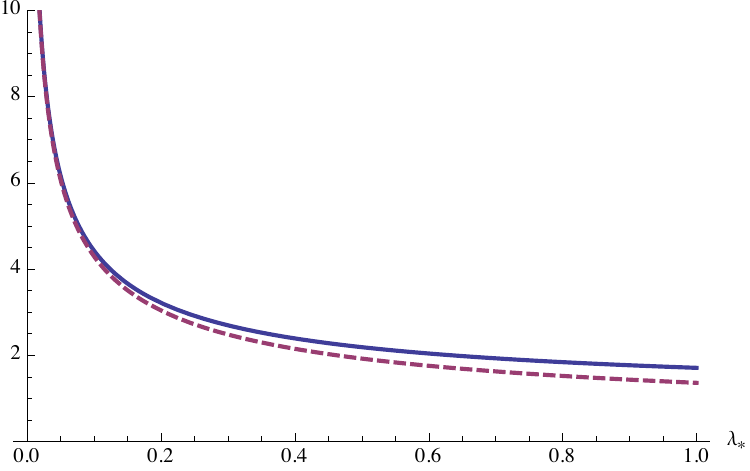}
\end{center}
\caption{Asymptotic behavior of the hypergeometric function $_2F_1$ (solid curve) v.s the asymptotic expression \eqref{asymptoticformF} (dashed curve).}\label{AsymptoticF}
\end{figure}

Now, let us consider the opposite limit, $\lambda_*\to 0$ i.e. close to the transition point. We have the asymptotic expression (see Fig. \ref{AsymptoticF}):
\begin{equation}
_2F_1\left(-\frac{1}{2},\frac{3}{2};3;-\frac{4 \sigma }{\lambda_*}\right)\approx\frac{32}{15 \pi} \left(\frac{4 \sigma }{\lambda_*}\right)^{1/2}\,,\label{asymptoticformF}
\end{equation}
and we get:
\begin{equation}
\frac{E_0}{N}\simeq-\frac{3\mu_1^2}{2\mu_2}+\frac{16}{15 \pi} \left(\sigma \lambda_*\right)^{1/2}-\frac{\ln 2}{\beta}\,,
\end{equation}
for:
\begin{equation}
\lambda_* \approx \mu_1+\frac{\mu_2}{9\pi \sqrt{\sigma}}\,.
\end{equation}
The same large $N$ approach can be considered for $p> 2$, but requires replica \cite{ANOUS1}. 
\medskip

The case $p=1$ is however easily calculable from the previous strategy. We simply need to add to the initial action a term in $\sum_\mu h_\mu x_\mu$, where $h_\mu$ is a Gaussian variable with zero mean and variance $\Delta$. The Gaussian integration on the field $x_\mu$ therefore becomes:

\begin{align}
&\int_{x(0)=x(\beta)} [dx]\, e^{-\int_0^\beta dt\,\frac{1}{2}\sum_\mu \left(m\dot{x}^2_\mu+p^2_\mu x_\mu^2 +\lambda(t) x_\mu^2 +h_\mu x_\mu \right)} =\prod_\mu e^{-\frac{1}{2}\mathrm{Tr} \ln \, (\hat{A}_\mu)} e^{\frac{\beta}{2}\,\sum_\mu \, h_\mu \left(\int_0^\beta dt \hat{A}_\mu^{*-1}\right) h_\mu}\,,\label{Gaussianxmu2}
\end{align}
where:
\begin{equation}
\int_0^\beta dt \hat{A}_\mu^{*-1}=\frac{\beta}{p_\mu^2+\lambda_*}\,,
\end{equation}
and the saddle point equation for $\lambda_*$ becomes:
\begin{align}
 &-\frac{6\beta}{\mu_2}(\lambda_*-\mu_1)+\int dp^2\, \rho(p^2)\frac{\beta  \coth \left(\frac{1}{2} \beta  \sqrt{\lambda_* +p^2}\right)}{4 \sqrt{\lambda_* +p^2}}+\frac{1}{2}\sum_\mu \, h_\mu \frac{\beta^2}{(p_\mu^2+\lambda_*)^2} h_\mu=0\,.
\end{align}
In the limit $\beta\to \infty$, the second term dominates, and its average reads:
\begin{equation}
\sum_\mu \,  \frac{1}{(p_\mu^2+\lambda_*)^2} = N \int_0^{4\sigma} dp^2 \, \frac{\rho(p^2)}{(p^2+\lambda_*)^2}\,.
\end{equation}
Then, we find that $\lambda_*=\mathcal{O}(\beta)$ is arbitrarily large, and the system does not have any finite equilibrium point.

\section{2PI Formalism and $1/N$ expansion}\label{AppE}

In this appendix, we provide a short presentation of the 2PI formalism and its $1/N$ expansion, focusing especially on the LO contribution. The reader may consult the recent reference \cite{Blaizot21} for more details. 
\medskip

\paragraph{General formalism.} Let us begin with a definition:
\medskip

\noindent
\textbf{Definition.} \textit{A 2PI diagram is a connected Feynman diagram $\mathcal{G}$ such that there now exist a pair of internal edges $(e,e^\prime)$ such that $\mathcal{G}/\{e,e^\prime\}$ is disconnected. }
\medskip

We denote as $\overline{S_{\text{cl},k}}$ the classical action \eqref{classicalaveraged} which we added the regulator $\Delta S_k$ defined by \eqref{defregulatorD}, and we introduce the (averaging) generating functional of connected correlation functions:
\begin{align}
\nonumber W^{(n)}[\mathcal{L},{\textbf{K}}]&=\ln \int \prod_{\alpha=1}^n [dx_\alpha]\, \exp \Big(- \overline{S_{\text{cl},k}}[\{\textbf{x}\}]+ \int dt \sum_{i,\alpha} x_{i,\alpha}(t) L_{i,\alpha}(t)\\
&+\frac{1}{2} \int dt dt^\prime \sum_{i,j,\alpha,\beta} x_{i,\alpha}(t) K_{i,\alpha;j,\beta}(t,t^\prime)  x_{j,\beta}(t^\prime) \Big) \,,\label{defW}
\end{align}
where $\textbf{K}$ is a bilocal (super) matrix with entries $K_{i,\alpha;j,\beta}(t,t^\prime)$ and is assumed to be symmetric, both in the variables $t$ and $t^\prime$ and on the pairs $(i,\alpha)$ and $(j,\beta)$. To simplify the notations, we introduce the short super indices $a=(i,\alpha)$, $b=(j,\beta)$. We Find:
\begin{equation}
2\frac{\delta W}{\delta k_{ab}(t,t^\prime)}=G_{ab}(t,t^\prime)+M_{a}(t)M_{b}(t^\prime)\,,
\end{equation}
where $G_{ab}$ is a short notation for the effective $2$ propagator and $M_{a}$ is the classical field. Now, we define the second Legendre transform $\Gamma_{2,k}[M,\textbf{G}]$ as:
\begin{align}
\nonumber &\Gamma_{2,k}[M,\textbf{G}]:=-W_k^{(n)}[\mathcal{L},\textbf{K}]+\int dt \sum_a L_a(t) M_a(t)\\
&+ \frac{1}{2}\int dt dt^\prime \sum_{a,b} M_a(t) [K_{ab}(t,t^\prime)-R_{k,ab}(t-t^\prime)]M_b(t^\prime) + \frac{1}{2} \mathrm{Tr} [\textbf{G}\textbf{K}]\,, \label{defLegendre2PI}
\end{align}
and we have:
\begin{align}
\frac{\delta \Gamma_{2,k}}{\delta M_a(t)}&=L_a(t)+\int dt^\prime \sum_b K_{ab}(t,t^\prime)  M_b(t^\prime)\,,\\
 \frac{\delta \Gamma_{2,k}}{\delta G_{ab}(t,t^\prime)}&=\frac{1}{2} K_{ab}(t,t^\prime)\,.\label{quantummoveeq}
\end{align}
Note that the super-matrix $\textbf{G}$ depends only explicitly on $k$ ‘‘on shell", as it identifies with the physical propagator. In \ref{defLegendre2PI}, it is viewed as the only variable of the functional $\Gamma_{2,k}$. Integrating out the quadratic contribution for fluctuation around the classical field, the definition \eqref{defLegendre2PI} leads to:

\begin{equation}
\boxed{\Gamma_{2,k}[M,\textbf{G}]=\overline{S_{\text{cl}}}[M]+\frac{1}{2}\mathrm{Tr} [\ln \textbf{G}^{-1}]+ \frac{1}{2}\mathrm{Tr}\, \textbf{G}\textbf{G}^{-1}_{0,k} + \Phi_k[M,\textbf{G}]  \,,}\label{equationG}
\end{equation}
where:
\begin{enumerate}
\item $\textbf{G}_{0,k}$ is the \textit{effective bare propagator} with entries:
\begin{equation}
(G_{0,k})_{ab}(t,t^\prime)=\frac{\delta^2 \overline{S_{\text{cl},k}}}{\delta x_{a}(t)\delta x_{b}(t^\prime)}\,.
\end{equation}
\item $\Phi_k[\Xi,\textbf{G}]$ is the Ward-Luttinger functional (WLF), whose expansion starts at two loops. Furthermore:
\begin{equation}
\boxed{\mathbf{\Sigma}:=2\frac{\delta \Phi_k}{\delta \textbf{G}}\,,} \label{equationselfenergy}
\end{equation}
is the self-energy, including only $1$PI contributions. 
\end{enumerate}

\paragraph{1/N expansion and LO.} As recalled in section \ref{sectionPert}, the vector field that we consider has a power counting, and the perturbative series can be organized accordingly with a non-trivial $1/N$ expansion given with the scaling law $\sim N^{-\omega+1}$ with $\omega$ given by \eqref{omegaindex}. Hence, the WLF expands in power on $N$, $\Phi_k=\sum_{\omega=1}^\infty \Phi_k^{(\omega)}$, where $\Phi_k^{(\omega)}$ expands in diagrams involving Feynman diagrams with scaling $\sim N^{-\omega+1}$. At leading order we therefore retain:
\begin{equation}
\Phi_k = \Phi_{k}^{(0)}+\mathcal{O}(N^{-1})\,.
\end{equation}
For instance, let us consider the quartic case. Accordingly with the statements of section \ref{sectionPert}, the LO diagrams are unrooted trees in the intermediate field formalism. But it is easy to check that a tree is not 2PI, except for the primary of them, made of a single edge and two nodes, namely:
\begin{equation}
\Phi_{k,(1)}\equiv \vcenter{\hbox{\includegraphics[scale=0.7]{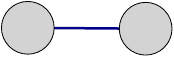}}}\,+\,\vcenter{\hbox{\includegraphics[scale=0.7]{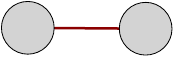}}}\,,
\end{equation}
where here grey discs are length one effective loops. For the quartic theory including non-local sextic interaction, we get in the same way:
\begin{equation}
\Phi_{k,(1)}\equiv  \vcenter{\hbox{\includegraphics[scale=0.7]{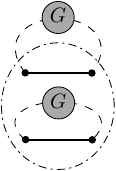}}}\,+\,\vcenter{\hbox{\includegraphics[scale=0.7]{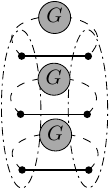}}}\,.\label{largeNPhi}
\end{equation}

\paragraph{RG equations} From the definition of $\Gamma_{2,k}$ given by equation \eqref{equationG}, we have:
\begin{align}
 \frac{d}{ds}\Gamma_{2,k}[M,\textbf{G}]&=\frac{1}{2}\mathrm{Tr}\, \textbf{G}\dot{\textbf{G}}^{-1}_{0,k} +\frac{d}{ds} \Phi_k[M,\textbf{G}]\,,
\end{align}
where one more time, $s=\ln(k)$. Note that we keep both $M$ and $\textbf{G}$ fixed in the derivation. On the other hand, from definition \eqref{defLegendre2PI}, we get straightforwardly:
\begin{align}
 \frac{d}{ds}\Gamma_{2,k}[M,\textbf{G}]:=-\frac{\partial}{\partial s}W_k[\mathcal{L},\textbf{K}]+ \frac{1}{2}\int dt dt^\prime \sum_{a,b} M_a(t) \frac{\partial R_{k,ab}}{\partial s}(t-t^\prime)M_b(t^\prime)\,. \label{defLegendre2bis}
\end{align}
The partial derivative of $W_k[\mathcal{L},\textbf{K}]$ can be computed exactly as:
\begin{align}
 \frac{\partial}{\partial s}W_k[\mathcal{L},\textbf{K}]&= \frac{1}{2} \mathrm{Tr}\frac{\partial}{\partial s} \textbf{R}_k (\textbf{G}+ M\otimes M)= \frac{1}{2} \mathrm{Tr} \dot{\textbf{G}}^{-1}_{0,k}(\textbf{G}+ M\otimes M)\,.
\end{align}
Therefore:
\begin{equation}
\frac{d}{ds}\Gamma_{2,k}[M,\textbf{G}]=\frac{1}{2}\mathrm{Tr}\, \textbf{G}\dot{\textbf{G}}^{-1}_{0,k}\,,
\end{equation}
and we conclude that the WLF is an RG invariant:
\begin{equation}
\boxed{\frac{d}{ds} \Phi_k[M,\textbf{G}]=0\,.}
\end{equation}
Hence, on shell, the dependence on $k$ of $ \Phi_k[M,\textbf{G}_{k}]$ is only through the dependency on $k$ of $\textbf{G}_{k}$:
\begin{equation}
\frac{d}{ds} \Phi_k[M,\textbf{G}]= \mathrm{Tr} \frac{\delta \Phi_k}{\delta \textbf{G}} \frac{d \textbf{G}_k}{ds} = \frac{1}{2}\mathrm{Tr} \mathbf{\Sigma}_k \frac{d \textbf{G}_k}{ds}\,,
\end{equation}
and generally:
\begin{equation}
\frac{d}{ds} \Phi_k^{(n)}[M,\textbf{G}]=\mathrm{Tr}  \frac{\delta  \Phi_k^{(n)}}{\delta \textbf{G}_k}\frac{d \textbf{G}_k}{ds}=-\mathrm{Tr} \frac{\delta  \Phi_k^{(n)}}{\delta \textbf{G}_k} \textbf{G}_k\frac{d \textbf{G}^{-1}_k}{ds}\textbf{G}_k\,,
\end{equation}
where $\Phi_k^{(n)}$ denotes the $n$-th functional derivative with respect to $\textbf{G}_k$.


\pagebreak
\printbibliography[title={Bibliography}]

\end{document}